\documentclass[a4paper, 12pt]{article}
\pdfoutput=1
\usepackage[a4paper, inner=3cm, outer=3cm, top=3cm, bottom=3cm]{geometry}
\usepackage{cite}
\usepackage{epsfig,amssymb,euscript,xspace,xcolor,setspace}
\usepackage{amsmath,mathtools,empheq,amsthm,hyperref,graphicx}
\usepackage{mathrsfs,float}  
\usepackage{pgf,tikz,pgfplots}
\usetikzlibrary{arrows}
\usepackage{tikz,caption,subcaption,marvosym} 
\usetikzlibrary{shapes.geometric}
\usetikzlibrary{decorations.markings,arrows,snakes}
\usepackage[belowskip=-20pt,aboveskip=0pt]{caption}
\usepackage[skip=10pt]{caption} 
\usepackage{comment}
\usepackage{wrapfig}

\def\bea{\begin{eqnarray}}
\def\eea{\end{eqnarray}}
\def\be{\begin{equation}}
\def\ee{\end{equation}}
\def\mW{\mathcal{W}}
\def\a{\alpha}
\def\m{\mu}
\def\mB{\mathcal{B}^{(\ell)}}

\makeatletter

\newcommand\blfootnote[1]{%
  \begingroup
  \renewcommand\thefootnote{}\footnote{#1}%
  \addtocounter{footnote}{-1}%
  \endgroup
}

\makeatother

\begin{document}
\baselineskip 24pt

\begin{center}

{\Large \bf  QFT, EFT and GFT} 

\end{center}

\vskip .5cm
\medskip

\vspace*{4.0ex}

\baselineskip=18pt

\centerline{\large \rm  Prashanth Raman$^{a*}$ \blfootnote{$^{*}$prashanth.raman108@gmail.com} and Aninda Sinha$^{a\dagger}$\blfootnote{$^{\dagger}$asinha@iisc.ac.in}  }
\vspace*{4.0ex}

\centerline{\it ~$^{a}$ Centre for High Energy Physics, Indian Institute of Science,}
\centerline{\it~ C.V. Raman Avenue, Bangalore 560012, India.}

\vspace*{1.0ex}
\centerline{\it ~} 
\vspace*{5.0ex}
\centerline{\bf Abstract} \bigskip

We explore the correspondence between geometric function theory (GFT) and quantum field theory (QFT). The crossing symmetric dispersion relation provides the necessary tool to examine the connection between GFT, QFT, and effective field theories (EFTs), enabling us to connect with the crossing-symmetric EFT-hedron. Several existing mathematical bounds on the Taylor coefficients of Typically Real functions are summarized and shown to be of enormous use in bounding Wilson coefficients in the context of 2-2 scattering. We prove that two-sided bounds on Wilson coefficients are guaranteed to exist quite generally for the fully crossing symmetric situation. Numerical implementation of the GFT constraints (Bieberbach-Rogosinski inequalities) is straightforward and allows a systematic exploration. A comparison of our findings obtained using GFT techniques and other results in the literature is made. We study both the three-channel  as well as  the two-channel crossing-symmetric cases, the latter having some crucial differences. We also consider bound state poles as well as massless poles in EFTs. Finally, we consider nonlinear constraints arising from the positivity of certain Toeplitz determinants, which occur in the trigonometric moment problem.

\vfill \eject

\baselineskip=18pt

\tableofcontents



\onehalfspacing
\section{Introduction}

Consider 2-2 scattering in quantum field theory. Suppose for simplicity that the external particles are identical massive scalars of mass $m$. In the low energy limit, the scattering amplitude admits an expansion
\be \label{Mdef}
{\mathcal M}(s,t)=\sum_{p,q=0} \mW_{pq} (st+su+tu)^p (stu)^q\,,
\ee
where $\mW_{pq}$ are the Wilson coefficients and $s,t,u$ are the standard Mandelstam invariants satisfying $s+t+u=4m^2$. In terms of some given scale, typically either the mass of the external particle or some cut-off, can the $\mW_{pq}$'s take on arbitrary values? The success of the Wilsonian picture suggests that the answer must be negative. There must exist two-sided bounds on these coefficients. If so, how do we go about showing this? This question has been investigated by several groups starting with the seminal works \cite{Aharonov,Pham, anant, Adams} followed more recently by \cite{RMTZ, tolley, rattazzi, SCH, SCH2, sasha, bern, meltzer, Miro:2021rof}. The typical starting point is to use a fixed-$t$ dispersion relation and examine the constraints imposed by crossing symmetry. In the geometric picture of \cite{nima}, we get the EFT-hedron which is the space of the $\mW_{pq}$'s constrained by positivity arising from dispersive arguments.

What are the mathematical apparatus available to us to tackle this question? The question posed above asks about Taylor coefficients of a certain series. Why should these be bounded? The two immediate answers a physicist would give are a) The series in question has a dispersive representation which implicitly makes certain assumptions about the high energy behaviour. The high energy behaviour feeds into the low energy properties. b) The amplitude satisfies additional constraints like unitarity and crossing symmetry. In \cite{HSZ} a different perspective was put forth where ingredients of Geometric Function Theory (GFT) were shown to have important restrictions on the Wilson coefficients. Unlike the fixed-$t$ dispersion relation, the starting point was a crossing symmetric dispersion relation recently resurrected in \cite{ASAZ}, building on an old work in the 1970s \cite{AK}. The topic of GFT is a vast one that has been examined for more than a 100 years by mathematicians. GFT studies geometric properties of complex analytic functions and includes famous results such as the Riemann mapping theorem, Schwarz's lemma. There are several more such as the Bieberbach conjecture (de Branges' theorem) from univalence, theorems about typically real functions, etc., which are not in the usual repertoire of a theoretical physicist. In different contexts, the constraints on physics from univalence have been briefly discussed in the literature, starting with the old work \cite{khuriuni} and more recently in \cite{saso}. In \cite{hebbar}, some elements of GFT have also been used. The area of GFT, which deals with Taylor expansion coefficients of meromorphic functions in the unit disk, will be the focus of this work.

 \begin{figure}[ht]
  \centering
 \includegraphics[width=0.7\textwidth]{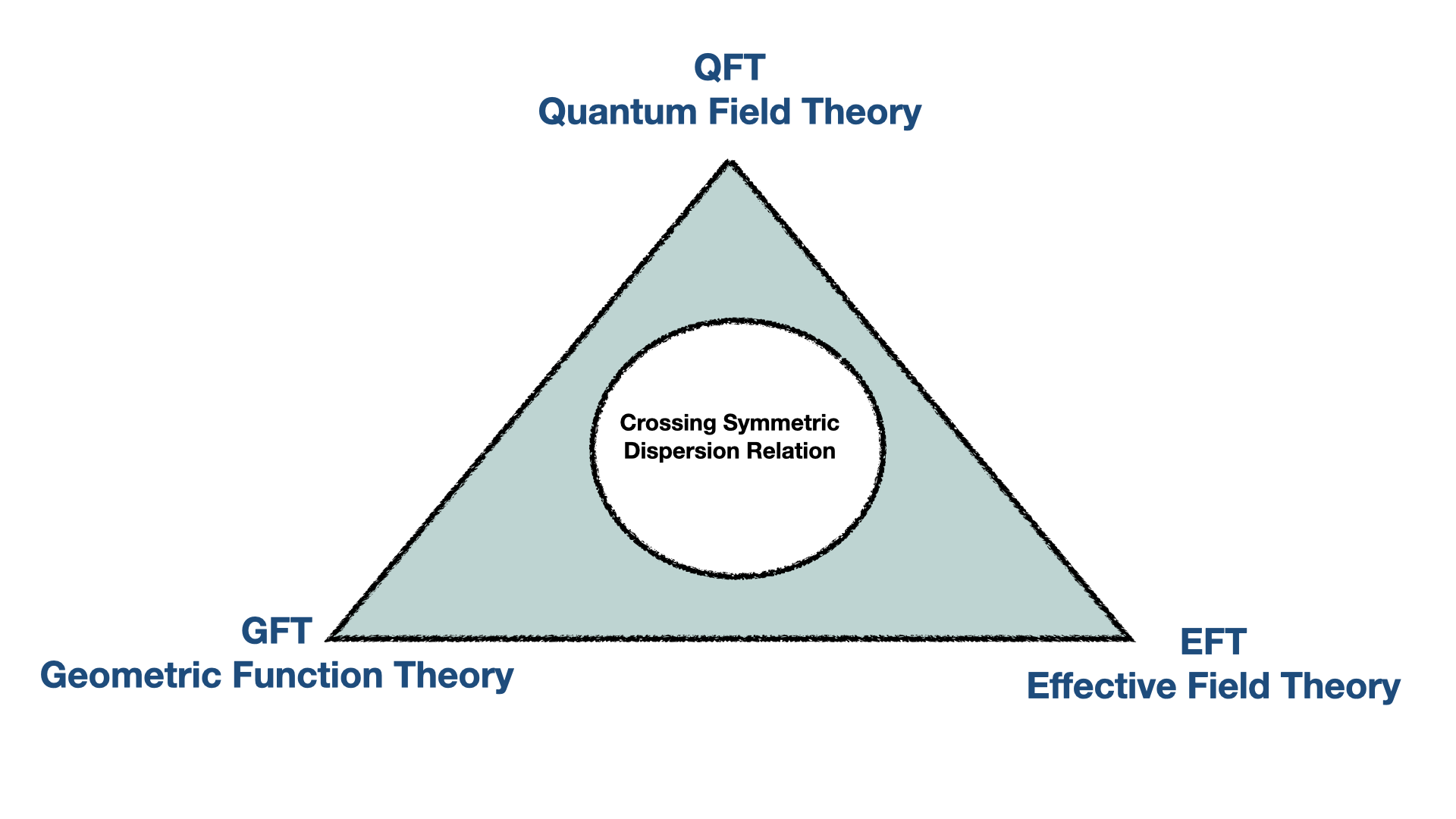}
 \caption{We examine the connections between QFT, EFT and GFT in this paper.}\vskip 0.5cm
\end{figure}

In \cite{HSZ}, it was shown how there is a close connection between the bounds arising from the Bieberbach conjecture and the bounds on the Wilson coefficients, which follow from a careful examination of the crossing symmetric dispersion relation. The findings in \cite{HSZ} can be summarized as follows. The crossing symmetric dispersion relation gives a representation for the amplitude, which involves an integration over the kernel times some positive measure factor. The kernel's property holds the key here. In the appropriate variables, the kernel is a univalent function in the unit disk. The Bieberbach conjecture applies precisely to this case! A preliminary study was carried out in \cite{HSZ}, and it was found that the leading order Wilson coefficients in all known cases respect all the consequences arising from the univalence of the kernel. Instead of the crossing symmetric invariants $x\sim (s t+ s u+ t u), y\sim s t u$, we work with complex variables $\tilde z, a$ which are related to $x,y$ via $x\sim a^2 \tilde z/(\tilde z-1)^2, y= a x$. We will give the precise formulae below. However, for now note that for fixed $a$, both $x,y$ are proportional to what are called Koebe functions defined as $k(z)=z/(z-1)^2$. Koebe functions, in the parlance of univalent functions, are extremal--they give the extremal values in the Bieberbach conjecture. If we ask what kind of amplitudes, which arise from a crossing symmetric dispersion relation, are Koebe extremal admitting up to simple poles, then there are just three possibilities. The amplitude could be proportional to $x$, $y$, or $x^2/y$. The other possibility $y^2/x$ is ruled out using the locality constraints discussed below. The amplitudes $x,y$ could be thought of as the leading order expansion in some EFT while $x^2/y$ is special. $x^2/y$ is nothing but the leading order 2-2 dilaton amplitude with a graviton exchange!

 However, there were a few important drawbacks that prevented us from extending this correspondence further. Namely, 
\begin{itemize}
\item While the kernel was univalent, the amplitude itself was shown to be a convex sum of univalent functions. To our knowledge, there appear to be no theorems which would allow us to conclude that the amplitude itself is univalent. Then what is the mathematical property of the full amplitude analogous to what the kernel satisfies?
\item What happens in cases where there are bound state poles, corresponding to singularities in the unit disk? What are the correct mathematical tools to use in such a situation?
\end{itemize}

In this paper, we will show that the kernel is also a Typically Real function (also called Herglotz function). 
These are functions satisfying
$$
\Im z \Im f(z)>0 ~{\rm} {\rm for} ~\Im z \neq 0                 \,,
$$
where $\Im$ is the imaginary part. A considerable amount of results exist for this class of functions, although in a scattered way (both spatially and temporally; for a book see \cite{Goluzin})\footnote{The Bieberbach conjecture was proved for this class of functions by Rogosinski \cite{rogo}, whose work we will use in what follows. We could not find an English translation of his beautiful paper and landed up doing a translation for ourselves, which we will be happy to share. Another historical note of interest is that Bieberbach was a Nazi and was instrumental in driving out several Jewish German mathematicians. Rogosinski, being a Jewish mathematician, was forced to leave Germany and found refuge in Cambridge, thanks to Hardy and Littlewood.}, which appear to be ideally suited to address all the shortcomings in \cite{HSZ}. {Typical-realness has been used in physics in the context of scattering in both quantum mechanics and relativistic quantum field theory and we briefly review some of these applications in  (\ref{trp})}. We will find novel applications of this property in the context of 2-2 scattering. In the course of our work, we will show how we can connect up very nicely with existing literature using techniques that appear to be ideally suited to tackle the question framed in the beginning. Amongst many other things, we will show
\begin{itemize}
\item The convex sum of typically real functions can be shown to be typically real! Thus, under restrictions spelt out in \cite{HSZ} and to be discussed in this paper, the amplitude in eq.(\ref{Mdef}) is a typically real function.
\item We will make use of  existing mathematical literature which deal with the Taylor coefficients of functions not just inside the disk but also inside the annulus. This enables us to include in our discussions bound states as well as a massless pole.
\end{itemize}
The findings in \cite{HSZ} are, of course, valid, and considerations made in this paper make them stronger. We have the Bieberbach-Rogosinski bounds for typically real functions (eq.(\ref{rogo1})), which are stronger than the bounds in the Bieberbach conjecture. For typically real functions, the derivation of the bounds is very simple, unlike the proof of the de Branges theorem. We will review the proof below. 
In the course of our investigations, we will find a sound mathematical argument as to why $\mW_{pq}$'s in eq.(\ref{Mdef}) should be bounded on both sides (sec.\ref{marko}). The surprisingly simple argument we will present uses the Markov brothers' inequality\footnote{Named after A. Markov and V. Markov, Russian brothers, and students of P. Chebyshev. A. Markov of the ``Markov chain'' fame, proved the simplest case in 1890, which was generalized in 1892 by his brother V. Markov.} which was proved in the late 1800s! The techniques we will develop in this paper, after getting over the lack of mathematical familiarity, will turn out to be very simple to implement in Mathematica. This enables us to compare in detail our bounds with those in \cite{SCH}--see tables (\ref{mint}),(\ref{maxt}). The bounds in \cite{SCH, tolley} were obtained, imposing crossing symmetry constraints on the usual fixed-$t$ dispersion relation; these constraints were termed as null constraints. In \cite{ASAZ}, it was shown that from the perspective of the crossing symmetric dispersion relation, these null constraints are identical to the locality constraints; as such, we will use the two terminologies interchangeably. An important point is that in our approach, we do not impose the locality constraints\footnote{Imposing these constraints should reduce the theory space further, but this is something we will not attempt in this paper.} (yet). Furthermore, our constraints will not depend on the spacetime dimensions as we will only use the positivity properties of Gegenbauer polynomials and the positivity of the absorptive part of the partial wave amplitudes. Nevertheless, we will not only find close agreement with the results of \cite{SCH}, in some cases, we will get tighter bounds. Considering the vast differences in our toolboxes, this is quite fascinating. We will elaborate more on this in this paper. Our punchline is:

\begin{center}
{\it Typically Real-ness is intimately tied with positivity of amplitudes. }
\end{center}

We will then consider amplitudes with two channel symmetry and explain how the entire machinery works in a similar manner in that situation as well. The essential mathematical step is to identify the correct analog of the complex variable which enables us to map to a disk. Rather than the cube roots of unity which played an important role in the 3-channel case, it turns out that the square root of unity plays the same role in this situation. This hints at a unifying framework to tackle such dispersion relations. We will draw parallels between our 2-channel analysis and the open string analysis in \cite{yutin}. An important point that we should emphasise here is that for technical reasons, we were led to consider the large-$s$ behaviour of the amplitude to be $o(s)$ rather than $o(s^2)$ used in the fully crossing symmetric case. There are some important differences between the 2-channel and 3-channel cases. The main one is that instead of two-sided bounds, here the analogous methods lead to one-sided bounds. Nevertheless, both 2-channel and 3-channel cases respect certain determinant conditions which enable us to draw parallels with the EFT-hedron picture in \cite{nima}. These determinant conditions demand the positivity of certain Toeplitz determinants and were written down in \cite{rogo}. We provide an argument using the trigonometric moment problem. We demonstrate that the nonlinear constraints on the Wilson coefficients, arising from the Cauchy-Schwarz inequality in the fixed-$t$ dispersion relation considerations in \cite{tolley} follow from these conditions. Moreover, the Toeplitz positivity gives rise to many more systematic nonlinear inequalities which can be used to constrain the allowed theory space further.

The paper is organized as follows. In section 2, we consider the 3-channel symmetric amplitudes from the perspective of the crossing symmetric dispersion relation in QFT and EFT's then we motivate the need for typically-real functions and prove why the 2-2 crossing scattering amplitude is typically real. After briefly reviewing the necessary results from the GFT of typically-real functions \cite{rogo,NS,Goodman} we proceed to get bounds for the Wilson coefficients in the low-energy expansion of the amplitude and compare with several other results known in the literature. We also spell out the role of extremal functions and prove why the space of low-energy Wilson coefficients is bounded. In section 3, we will turn to the 2-channel symmetric amplitudes. We will set up the 2-channel symmetric dispersion relation along the lines of \cite{AK} and find similarities as well as important differences compared to the 3-channel symmetric analysis. In section 4, we elaborate on nonlinear conditions, which are positivity of certain Toeplitz determinants, and explore the connection to  EFT-hedron \cite{nima}. We conclude with a brief discussion of future directions. The appendices have several useful mathematical results and review material referred to in the main text.

{\bf Notation:} We warn the reader in advance for our notation. In keeping with \cite{AK}, we had started using $\mu=4m^2$ in \cite{ASAZ} and this has percolated in \cite{HSZ}. We will keep using this notation. Unless explicitly specified, we will frequently work in units where $m^2=1$.

\section{The fully crossing symmetric case}
\subsection{Set-Up}
We consider the 2-2 scattering amplitude $\mathcal{M}(s,t)$ of identical scalar particles \footnote{The generalization to particles with spin shall be considered in \cite{CGHRS}.} which we assume has the following properties:
\begin{itemize}
\item {\bf Causality:} $\mathcal{M}(s,t)$ is analytic modulo poles and branch cuts on the real axis.
\item {\bf Polynomial boundedness:} For a fixed $t$ and  $|s|\rightarrow \infty$, $|\mathcal{M}(s,t)|= o(s^2)$.
\item{\bf Unitarity:} The amplitude admits a well defined partial wave expansion 
\be
\mathcal{M}(s,t)= \Phi(s)\sum_{\ell=0}^{\infty}\left(2\ell+2 \a\right) f_\ell(s)C^{(\a)}_{\ell}\left(\cos \theta \right) \,,
\ee
with $0\le |f_{\ell}(s)|^2 \le 2 \Im f_{\ell} \le 1$. In the above $\a =\frac{d-3}{2}$ and $\Phi(s)=2^{4\alpha+3}\pi^\alpha \Gamma(\a)\frac{\sqrt{s}}{(s- \mu)^\a}$. 
\item {\bf Crossing symmetry:} This follows from the above assumptions if we assume there is a mass gap.
\end{itemize}
The first couple of assumptions above enable one to write down twice-subtracted fixed-$t$ dispersion relations for the amplitude. We shall follow \cite{AK,ASAZ} to write a crossing symmetric dispersion that keeps $a=\frac{s t u}{st +t u +u s}$ fixed.
The method we use in this paper can be broadly summarised in the following 3-steps:
\begin{enumerate}
\item We begin with the parametric crossing symmetric dispersion relation for the 2-2 scattering  amplitude $\mathcal{M}(s,t)=\mathcal{M}(s(z,a),t(z,a))$ derived in \cite{AK} and reviewed recently in \cite{ASAZ} which we shall use to connect low-energy and high energy physics.
\item In the crossing symmetric variable $\tilde z=z^3$ for a fixed range of the parameter $a$ the amplitude can be shown to have the special analytic property of Typical-realness, which we shall discuss below.
\item As we shall show several results from the geometric function theory of typically-real functions can be used to get linear and nonlinear constraints on the Wilson coefficients $\mW_{p,q}$ in the low energy expansion ${\mathcal M}(s,t)=\sum_{p,q=0} \mW_{pq} (st+su+tu)^p (stu)^q$ of the amplitude.
\end{enumerate}
We shall also obtain several other qualitative results along the way such as the proving that all the $\mW_{pq}$ are bounded, role of extremal functions and connection with the EFT-hedron. We shall now start with crossing symmetric dispersion.
\subsection{Dispersion relation in QFT}
In \cite{ASAZ} the crossing symmetric dispersion relation was derived 
\be
\label{disp}
\mathcal{M}_{0}(s_1, s_2)=\alpha_{0}+\frac{1}{\pi} \int_{\frac{2\m}{3} }^{\infty} \frac{d s_1^{\prime}}{s_1^{\prime}} \mathcal{A}\left(s_1^{\prime} ; s_2^{(+)}\left(s_1^{\prime} ,a\right)\right) H\left(s_1^{\prime} ;s_1, s_2, s_3\right)\,,
\ee
where $\mathcal{A}\left(s_1; s_2\right)$ is the s-channel discontinuity and the kernel is given by
\be
\begin{split}\label{Hs2p}
&H\left(s_1^{\prime} ; s_1, s_2,s_3\right)=\left[\frac{s_1}{\left(s_1^{\prime}-s_1\right)}+\frac{s_2}{\left(s_1^{\prime}-s_2\right)}+\frac{s_3}{\left(s_1^{\prime}-s_3\right)}\right]\\
&s_{2}^{(+)}\left(s_1^{\prime}, a\right)=-\frac{s_1^{\prime}}{2}\left[1 - \left(\frac{s_1^{\prime}+3 a}{s_1^{\prime}-a}\right)^{1 / 2}\right]\,.
\end{split}
\ee
Here $s_1=s-\mu/3, s_2=t-\mu/3, s_3=u-\mu/3$, with $s,t,u$ being the usual Mandelstam variables satisfying $s+t+u=\mu$ so that $s_1+s_2+s_3=0$. The crossing symmetric combination $a=\frac{s_1 s_2 s_3}{s_1 s_2+s_2 s_3+s_3 s_1}$ was held fixed instead of $t$ in the fixed $t$-dispersion relations and this made crossing symmetry manifest. The absorptive part is given by:
\be\nonumber \label{partw}
\begin{split}
&\mathcal{A}\left(s_1,s_2^{(+)}(s_1,a)\right)=\Phi(s_1)\sum_{\ell=0}^{\infty}\left(2\ell+2\a\right)a_\ell(s_1)C^{(\a)}_{\ell}\left(\sqrt{\xi(s_1,a)}\right)\,,\\
&\xi(s_1,a)=\xi_0+4\xi_0\left(\frac{a}{s_1-a} \right),~\xi_0=\frac{s_1^2}{(s_1-2\mu/3)^2}\,.
\end{split}
\ee

\noindent The above dispersion relation is a non-perturbative representation of the  2-2 amplitude $\mathcal{M}_0(s_1,s_2)$ and is valid for any QFT with a mass gap. If we are considering a low energy effective field theory then we can shift the lower limit \footnote{This can be done because we can subtract out $\int_{\frac{2\m}{3} }^{\Lambda} \frac{d s_1^{\prime}}{s_1^{\prime}} \mathcal{A}\left(s_1^{\prime} ; s_2^{(+)}\left(s_1^{\prime} ,a\right)\right) H\left(s_1^{\prime} ;s_1, s_2, s_3\right)$ from the amplitude as this is a quantity that is computable in the EFT.}of the dispersion integral to $\Lambda^2=\delta+ \frac{2 \mu}{3}$ where $\Lambda$ is the cutoff of the EFT above which one expects to see new physics.
 We remain agnostic about physics above the cut-off $\Lambda$ except the assumption that the parent UV theory for which the EFT provides an IR description is unitary, causal and local. We make the following assumptions regarding the EFT:

\noindent $\bullet$ There is no exchange of massless particles in the loops. \\
$\bullet$ There is a parameter in the original theory with respect which tree-level amplitudes in the EFT provide the leading contribution and can be approximated by higher derivative operators.\\

\noindent Under these assumptions the low energy expansion of the 2-2 identical particle amplitude with the Wilson coefficients $\mW_{pq}$ is given by 
\be
\label{Wdef}
\mathcal{M}_{0}(s_1,s_2)=\sum_{p, q=0}^{\infty} {\mathcal W}_{p q} x^{p} y^{q}\,,
\ee
with $x=-\left(s_1 s_2 + s_2 s_3+s_3 s_1\right)$, $y=-s_1 s_2 s_3$. %
Next we Taylor expand this expression around $a=0$ and match powers to obtain\footnote{Compared to \cite{ASAZ} we have pulled out a $2\pi s_1^{2n+m}$ factor from $\mB_{n,m}$. }
\be
\begin{split}\label{Belldef1}
&\mW_{n-m,m}=\int_{\delta+\frac{2\m}{3}}^{\infty}\frac{d s_1}{2\pi s_1^{2n+m+1}}\Phi(s_1)\sum_{\ell=0}^{\infty}\left(2\ell+2\a\right)a_\ell(s_1)\mathcal{B}_{n,m}^{(\ell)}(s_1)\,,\\
&\mathcal{B}_{n,m}^{(\ell)}(s_1)=2\sum_{j=0}^{m}\frac{(-1)^{1-j+m}p_{\ell}^{(j)}\left(\xi_{0}\right) \left(4 \xi_{0}\right)^{j}(3 j-m-2n)\Gamma(n-j)}{ j!(m-j)!\Gamma(n-m+1)}~~{\rm for} ~ n\ge 1 \,, 
\end{split}
\ee
where $p_{\ell}^{(j)}(\xi_0)=\partial^j C_\ell^{(\a)}(\sqrt{\xi})/\partial\xi^j|_{\xi=\xi_0}$. Furthermore, it can be checked that $p_\ell^{(j)}(\xi_0)\geq 0$ for $\xi_0\geq 1$. This is an important property since the sign of $\mathcal{B}^{(\ell)}_{n,m}$ which in-turn determines the sign of the term in $\mW_{p,q}$ is then governed by the other factors in eq.(\ref{Belldef1}). This has the following important implications: 
\begin{enumerate}
\item We first note that for $m=0$ we have:
\bea \label{posm0}
\mW_{n,0} =\int_{\delta+\frac{2\m}{3}}^{\infty}\frac{d s_1}{\pi s_1^{2n+1}}\Phi(s_1)\sum_{\ell=0}^{\infty}\left(2\ell+2\a\right)a_\ell(s_1) C_\ell^{(\a)}(\sqrt{\xi_0}) >0
\eea
Since $C_\ell^{(\a)}(\sqrt{\xi_0})>0$ for $\sqrt{\xi_0} = \frac{s_1}{s_1-\frac{2\mu}{3}} > 1+\frac{2 \mu}{3 \delta}>1$.\\

\item {\bf Locality/Null constraints:} As alluded to earlier the expression \eqref{Belldef1} for $\mW_{n-m,m}$ is valid for $n\ge 1$ and any $m$. However, in a local theory we know that we cannot have negative powers of $x$ \footnote{ A simple pole in $y$ is however allowed as this corresponds to the massless pole such as the graviton pole $\frac{1}{s t u}$.}. We thus need need to impose the following {\it Locality constraints} which we denote by $N_c$.\\

\vspace*{-40 pt}
\begin{centering}
\bea \label{nc}
\mW_{n-m,m}=0, ~ \forall ~ n <m
\eea
\end{centering}
This the price we pay for making crossing symmetry manifest, locality is now lost and has to be imposed as an additional set of constraints. Unlike in the case of the fixed-$t$ dispersion relations where locality was manifest but full crossing symmetry was lost (as fixing $t$ which explicitly broke crossing symmetry) and had to be imposed as an additional set of constraints called the {\it Null constraints} recently in the literature \cite{SCH}. It was argued in \cite{GSZ} that both these methods are equivalent. In other words:\\

\begin{centering}
Crossing symmetric ~~~~~~~~~~~~~~~~~~~~~~~~~~~~~~~~~~~~~~~~~~~~~~~~~~~~Fixed-$t$ \\
~~~~~~~dispersion relation~~~~~~~~~~~~~~~~~~~~~~~~~~~~~~~~~~~~~~~~~~~~~~~~~~dispersion relation \\      
~~~~~~$\oplus $~~~~~~~~~~~~~~~~~~~~~~~~~~~~$\equiv$~~~~~~~~~~~~~~~~~~~~~~~~~~~~~~~~   $\oplus$ \\
~~~~~~~~~~~~~~~~Locality constraints~~~~~~~~~~~~~~~~~~~~~~~~~~~~~~~~~~~~~~~~~~~~ Null constraints
\end{centering}

The locality constraints have several non-trivial consequences for the theory. We shall list a few of them in the context of EFT where $\Lambda \gg \mu$, here and refer the interested reader to Appendix \ref{nullapp} for further details. In this regime the $\mathcal{B}^{(\ell)}_{n,m}$'s \eqref{Belldef1} simplify considerably and we get:
\begin{itemize}
\item The first non-zero contribution to $\mW_{n-m,m}$ is from $\ell=2n$.
\item An infinite number of higher-spin partial waves are non-zero in the partial wave expansion.
\item All spin up-to $\ell=28$ are necessarily present and we will have an infinite number of spins present.
\end{itemize}
The locality/null constraints can also be solved and this can also be used to obtain bounds for $\mW_{p,q}$'s as was done in \cite{SCH}. We present results of this analysis in section \ref{schtables} both in the EFT context where the particle masses are negligible compared to the cutoff of the theory i.e.,$\delta >> \mu$ which implies $\xi_0 \sim1$ as was considered in \cite{SCH} as well as when we cannot neglect $\mu$ compared to the EFT cutoff scale $\Lambda^2$ which implies $\xi_0 >1$ which is the case we consider in this paper.\\
\noindent However, the main focus of this work will be on obtaining constraints/bounds on $\mW_{p,q}$'s  using a different method involving the Geometric function theory(GFT) of typically real functions. 
In our approach, we {\it do not} solve the Locality/Null constraints but implicitly assume that $\mW_{-n,m}=0$ by starting with a local low energy expansion \eqref{Wdef} without negative powers.
We shall compare our results using this approach to the ones obtained by solving the Locality constraints in both the above cases as well other known results in the literature in \ref{schtables}.

\item {\bf Positivity constraints:} We can also take linear combinations of different $\mW_{p,q}$'s with specific coefficients such that all the negative terms in the $\ell$-expansion cancel out and we get a result that is a positive combination of $p_{\ell}^{(j)}(\xi_0) \ge 0$ and hence manifestly positive \cite{ASAZ}. We call these conditions $PB_C$:
\bea \label{pbc}
\sum_{r=0}^m \chi_n^{(r,m)} (\mu,\delta) \mW^{\delta}_{n-r,r} &\ge& 0  \\
0\le \mW^{\delta}_{n,0}&\le& \frac{1}{\left(\delta + \frac{2 \mu}{3}\right)^2} \mW^{\delta}_{n-1,0}\,,  ~~~~~~~~~~~~~~~~~~~~~~~~~~~~~~~n\ge2\,.
\eea
The $\chi_n^{(r,m)} (\mu,\delta) $ satisfy the recursion relation:
\bea
\chi_n^{(m,m)} (\mu,\delta) &=&1\nonumber\\
\chi_n^{(r,m)} (\mu,\delta) &=& \sum_{j=r+1}^m (-1)^{j+r+1} \chi_n^{(j,m)} \frac{{\mathscr U^{\alpha}}_{n,j,r} (\delta + \frac{2 \mu}{3})}{{\mathscr U^{\alpha}}_{n,r,r}(\delta + \frac{2 \mu}{3})}
\eea
with  ${\mathscr U}^{\alpha}_{n,m,k}(s_1)= -\frac{4^k \left(\alpha \right)_k (3 k-m-2 n) \Gamma (n-k) \,
   _4F_3\left(\frac{k}{2}+\alpha,\frac{k}{2},k-m,k-\frac{m}{3}-\frac{2
   n}{3}+1;k+1,k-n+1,k-\frac{m}{3}-\frac{2 n}{3};4\right)}{ s_1^{2n+m} \Gamma (k+1) (n-m)!
   \Gamma (-k+m+1)}$ and $\alpha=\frac{d-3}{2}$.
\end{enumerate}
\subsection{GFT: The need for typically real functions} \label{gft}
In \cite{HSZ}, an intriguing correspondence between the crossing symmetric dispersion relation of 2-2 scattering and geometric function theory (GFT) was pointed out. In order to exhibit the full 3-channel symmetric \cite{AK} introduced the variable $z,a$ via $s_k=a-a(z-z_k)^3/(z^3-1)$ where $z_k$'s are the cube-roots of unity and $a= s_1 s_2 s_3/(s_1 s_2+s_1 s_3+s_2 s_3)$. We note that
\be \label{xydef}
y\equiv -s_1 s_2 s_3=-27\frac{a^3 z^3}{(z^3-1)^2}\,,\quad x\equiv=-(s_1 s_2+s_1 s_3+s_2 s_3)=-27\frac{a^2 z^3}{(z^3-1)^2}\,.
\ee
In \cite{HSZ}, it was observed that the kernel as a function of $\tilde z=z^3$ is a {\it univalent function} in the unit disk\footnote{An analytic one-one mapping inside the disk $|\tilde z|<1$.}. Namely, writing
\be\label{betan}
H(s_1,\tilde z)= \frac{27 a^2 \tilde z (2 s_1-3a)}{27 a^3 \tilde z-27 a^2 \tilde z s_1-(\tilde z-1)^2 s_1^3}=\sum_{n=0}^\infty \beta_n(a,s_1)\tilde z^n\,,
\ee
it was found that $H(s_1,\tilde z)$ is a Mobius transformation of the Koebe function $\tilde z/(\tilde z-1)^2$ and is hence univalent  \footnote{This follows since the Koebe function is univalent as $z_1/(z_1-1)^2 =z_2/(z_2-1)^2 $ for any $z_1,z_2$ inside the disk necessarily implies $z_1 =z_2$  as can be easily checked and a M\"{o}bius transformation preserves 1-1 mappings.} in the unit disk provided $|27 (a/s_1)^2(1-a/s_1)-2|<2$. The last condition is needed to avoid singularities in the unit disk and restricted the range of the parameter $a$. This was the key step in relating the Bieberbach bounds $\bigg| \frac{\beta_n(a,s_1)}{\beta_1(a,s_1)}\bigg| \le n ~ \forall ~n\ge 2$
 to the bounds on the Wilson coefficients \cite{HSZ}. We note that $\beta_1(a,s_1)=\frac{27 a^2}{s_1^3}(3a-2 s_1)$ so if $a$ is real and in the dispersion relation $s_1>8/3$ we have $\beta_1<0$ if $a<16/9$. This restriction is important to note as we do not want $H$ to change sign; typical realness will need a positive measure as we will see. Together with the singularity-free condition inside the disk mentioned above, we get the range of $a$ as $$a \in \left(-\frac{8}{9},0\right)\cup \left(0,\frac{16}{9}\right).$$
\begin{wrapfigure}{H}{0.3\textwidth}
 \centering
  \includegraphics[width=0.25 \textwidth] {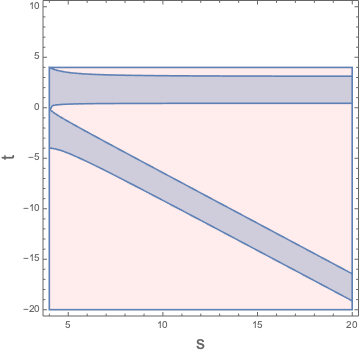} 
 \label{sta}
 \end{wrapfigure}
 
When we write $-8/9<a<16/9$ this is what we mean. $a=0$ gives a trivial constant amplitude. As noted in \cite{HSZ}, in the original Mandelstam variables, the $t\in (-\infty,4)$ for $s>4$ contains this region. In the figure, the $s>4,-\infty<t<4$ are depicted in pink and the $a$ domain above in blue.  We know that the partial wave expansion converges here, as this is inside the Martin ellipse.

\noindent If we are interested in real $a, s_1$, then we can do more. This is what we will explain now. In fig.2 we have plotted several $\beta_n/\beta_1$ as a function of $s_1$ . The interesting thing to note is that while for $n$ even the upper and lower bound for $\beta_n/\beta_1$ appears to be $\pm n$, for $n$ odd, the lower bound is stronger. We will explain how this happens in the next section using the concept of Typically Real functions--see also appendix (\ref{appA}). 
  \begin{figure}[H]
 \centering
  \includegraphics[width=0.8 \textwidth] {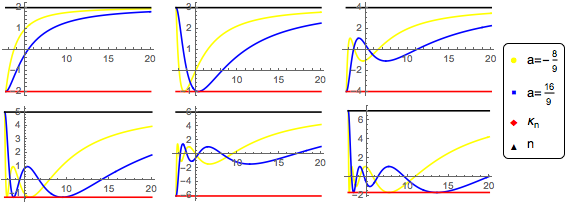} 
 \caption{$\beta_n/\beta_1$ vs $s_1$ for $n=2,3,4$ and $n=5,6,7$, from left to right. The above plots show that the Rogosinksi bounds are respected the amplitude for the full range of $a$ namely $-\frac{8}{9} \le a\le \frac{16}{9}$.}
 \label{fig22}
 \vskip 0.5cm
 \end{figure}
 
\subsubsection{Typically-real functions}
When $a, s_1$ are real, then $\beta_n$'s are also real. This motivates us to consider a restricted class of functions. We consider functions of the type
\be \label{fdef}
f(z)=\sum_{n=-\infty}^{\infty} a_n z^n\,,\quad  |z|<1\,,
\ee
and satisfying 
\be
\Im z \Im f(z)>0 ~{\rm for}~\Im z \neq 0\,,
\ee
where $\Im$ is the imaginary part, 
define Typically Real functions inside the unit-disk $|z|<1$. It is easier to introduce typically real functions with poles since we would making use of these later. But we would like to emphasise that the notion of typically real is quite general and does not require the existence of poles. From the physics point of view, we are interested in Laurent series and the annulus since we want to enlarge the analysis in \cite{HSZ} to include poles on the real axis. 

\noindent We shall first argue why this class of functions is relevant for us. The reasons are two fold firstly the kernel $H(s_1,\tilde z)$ being a univalent function with real coefficients is actually a typically real  function \footnote{We caution the reader here that univalence does {\bf not} imply typically-realness in general, only univalence with real coefficients $a_n$ in \eqref{fdef} translates to typical realness (see \eqref{counterex} for a counter example). } of $\tilde z$ inside the disk as can be seen by the following argument:

\noindent If $f(z)$ in eq.(\ref{fdef}) is a univalent function with real coefficients, then the $\Im z \Im f(z)>0$ restriction automatically follows. This can be seen as follows. $\Im f(z)=\frac{1}{2 i}(f(z)-f^*(z))$ so a change of sign will need $f(z)=f^*(z)$ at some $z$. This gives $f(z)=f(z^*)$ for that $z$ which is forbidden by univalence unless $z=z^*$. In other words, the change of sign of $\Im f(z)$ happens only on the real axis. By choosing either $f(z)$ or $-f(z)$ we can always satisfy $\Im z \Im f(z)>0$. The kernel $H(s_1,\tilde z)$ in QFT is precisely of this kind for real $a, s_1$.

\noindent Secondly a positive sum of typically real functions such as $f(z)=\lambda f_1(z) +\mu f_2(z)$ is typically real as  can be readily seen from $\Im (\lambda f_1(z) +\mu f_2(z)) \Im z  =  \lambda \Im f_1(z) \Im z +\mu \Im f_2(z) \Im z>0$ for $\Im z \neq 0$ whenever $\lambda,\mu >0$ for any pair of typically-real function $f_1(z),f_2(z)$. Since the amplitude $\mathcal{M}_0$ is infinite positive sum (when the absorptive part is postive) of such functions from the dispersion relation \eqref{disp} we could expect that the amplitude would also be typically real provided the integral converges , as we shall argue in the next section it turns this is indeed the case.
 
We shall now list the important results from the geometric function theory of typically real functions that we will use in our analysis. We refer the reader to appendix \ref{appA} for further details. 
There are four important properties that typically real functions \eqref{fdef} in the disk have: 
\begin{enumerate}
\item All poles lie on the real line. 
\item All poles are first order.
\item Poles have negative residues.
\item They are closed under convex linear combinations.
\end{enumerate}
 These statements are proved in appendix (\ref{appA}).The above results mean we have 3-cases to consider since $a_{-n}=0$ for all $n\ge 2$ .
\begin{enumerate}
\item {\bf {$T_R$ :}} $f(z)=z+ \sum_{n=2}^\infty a_n z^n$ and $f(z)$ is regular inside the unit disk.
\item {\bf {$TM$ :} }$f(z)=z+ \sum_{n=2}^\infty b_n z^n$ and $f(z)$ has poles inside the unit disk except at $z=0$.
\item {\bf {$TM^*$ :}} $f(z)=-\frac{1}{z}+ \sum_{n=0}^\infty \beta_n z^n$ and $f(z)$ has a pole at $z=0$ with possibly other poles inside the unit disk.
\end{enumerate}
\noindent From the EFT perspective the classes $T_R$,$TM$ and $TM^*$ are suited to describe dispersive part of the amplitude, massive poles away from the origin and a massless pole respectively.

We uses the following theorems \cite{Goodman,Wigner,Robertson} for our analysis: 
\begin{enumerate}
\item {\bf Schiffer-Bargmann representation:} Every typically real function in $TM$, $f(z) = z+\sum_{n=2}^{\infty} b_n z^n$ inside the unit disk $|z|<1$ can be expanded into an absolutely convergent Mittag-Leffler series:
\bea\label{SB1}
f(z) =\mu~g(z) +\sum_{i} m_i \frac{(1-p_i^2)}{p_i^2} \left(\frac{z}{1-(p_i+p_i^{-1})z+z^2}\right)
\eea
where $\mu \ge 0$ and $g(z)$ is an regular typically real function inside the unit disk, where the sum extends over all poles $p_i $( note $0<|p_i |<1$) of $f(z)$ with $-m_j<0$ being the corresponding residues\footnote{One can think of $a(z)$ as the dispersive part and the remaining sum as the bound state poles.}. Furthermore
\bea \label{mu1}
\mu + \sum_i m_i  \frac{(1-p_i^2)}{p_i^2}=1.
\eea
If $f(z)$ is in $TM^*$, in other words $f(z)$ also has a pole at $z=0$ i.e. $f(z) =-\frac{1}{z}+\sum_{n=0}^{\infty} \beta_n z^n $ then the above result gets modified as:
\bea\label{SB2}
f(z) =-\frac{1}{z}+ \beta_0 -z + \mu~g(z) +\sum_{i} m_i \frac{(1-p_i^2)}{p_i^2} \left(\frac{z}{1-(p_i+p_i^{-1})z+z^2}\right)
\eea
and
\bea\label{mu2}
 \mu +\sum_{i} m_i  \frac{(1-p_i^2)}{p_i^2} =1+\beta_1. 
\eea
For the proof see appendix \ref{SBproof}.

\item {\bf Robertson representation (important):} A function $f(z)$ in $T_R$ is typically real  {\it if and only if}  it has the following representation:
\bea\label{robrep}
f(z) = \alpha_0+ \frac{1}{\pi} \int_{-1}^{1} d\alpha(\xi) \frac{z}{1- 2 \xi z+z^2} 
\eea
where $\alpha(z)$ is a non-decreasing function such that $\alpha(1)-\alpha(-1)=1$ (See also (\ref{A22})).

We shall in-fact show that our dispersion relation can be directly recast into the Robertson provided the absorptive part is positive in the next section. For now assuming this we can immediately see by combining the \eqref{SB2} and \eqref{robrep} that typically real functions have the usual analytic structure that usual scattering amplitude do namely with bound state-poles only on the real axis and branch cuts.
\be
f(z)= \underbrace{\sum_{i} m_i \frac{(1-p_i^2)}{p_i^2} \left(\frac{z}{1-(p_i+p_i^{-1})z+z^2}\right)}_{\text{bound state poles}}
+\underbrace{\frac{1}{\pi} \int_{-1}^{1} d\alpha(\xi) \frac{z}{1- 2 \xi z+z^2} }_{\text{Dispersive part} } \, .
\ee

We note though we have written the above in the $z$-plane, the statement is true even in the $s_1$-plane due to the nature of the map \eqref{xydef}. In particular for $|p_i|<1$ by using the inverse map it can be easily checked that $\frac{z}{1-(p_i+p_i^{-1}) z+z^2}$ is equal to the crossing symmetric bound state poles $\frac{1}{(s_1-m^2)(s_2-m^2)(s_3-m^2)}$ or $(\frac{1}{s_1-m^2}+\frac{1}{s_1-m^2}+\frac{1}{s_1-m^2})$ modulo an additive constant.\footnote{The massless pole corresponds to a $\frac{1}{z}$ term in the above expansion and can be readily seen to correspond to the pole $s_i=0$ since $z=0$ corrsponds to $s_i=0$ in \eqref{xydef}.} We shall remark on the significance of this fact in section \ref{SEG}. 
We now state a few theorems that give us bounds on the Taylor-coefficients of typically real functions.
\item {\bf Coefficient Bounds:}
The coefficient bounds for regular class and classes with poles are:

{\it Bieberbach-Rogosinski bounds:} For $f(z) \in T_R$ in the disk we have the following:
  \bea \label{rogo1}
 \begin{cases}
-n~~~~~~~~~~~~~~~~~~~~~~~~~~~~~~~~ & , n~even \\
\\  ~~~~~~~~~~~~~~~~~~~~~~~~~~~~~~~~  \le a_n \le n  \\
-n<\kappa_n = \frac{\sin n~\vartheta_n}{\sin \vartheta_n} & ,n ~odd
\end{cases} \,,
\eea
where $\vartheta_n$ is the smallest solution of $\tan n \vartheta = n \tan \vartheta$ located in $( \frac{\pi}{n},~\frac{3 \pi}{2n} )$ for $n>3$ and $\kappa_3=3$ (see \eqref{rog1} for details).

For cases where the function has poles we could consider the function is analytic either inside an annulus or inside the punctured disk (see \ref{boundswithpoles} for details) and have the following bounds:

 {\it Goodman bounds:} For functions $f(z)$ with poles at $p_j$. If we define $p$ such that $|p_j|\geq p>0, j=1,2,3\cdots$, and
\be
B(n,p)=\frac{p(p^{2n}-1)}{p^n(p^2-1)}\,,
\ee
we have for $TM$
\bea \label{good}
-B(n,p)&\leq& b_n \leq B(n,p) \,, \qquad n~{\rm even}\,,\\
-\kappa_n &\leq& b_n \leq B(n,p)\,, \qquad n~{\rm odd}\,.
\eea
Finally for $TM^*$ we have $\beta_1\geq -1$ and for $n\geq 2$,
\bea\label{good1}
-(1+\beta_1)B(n,p)&\leq& \beta_n \leq (1+\beta_1)B(n,p)\qquad n~{\rm even}\,,\\
-(1+\beta_1)\kappa_n&\leq& \beta_n \leq (1+\beta_1)B(n,p)\qquad n~{\rm odd}\,.
\eea

{\it RNS bounds:} For functions $f(z)$ with poles at $p_j$. If we define $\rho$ such that $\rho \ge |p_n|\ge \cdots \ge |p_1|>0, j=1,2,3\cdots n$, and
\be
\alpha_1(r)= (a_1-a_{-1} r^{-2})\,,
\ee
then
\bea \label{rns}
-\frac{n}{1-\rho^{2n}}\left(\frac{\kappa_n}{n}\a_1(1)+\a_1(\rho)\rho^{n+1}\right)&\leq& a_n \leq \frac{n}{1-\rho^{2n}}\left(\a_1(1)+\frac{\kappa_n}{n}\a_1(\rho)\rho^{n+1}\right)
 \eea
\end{enumerate}
We note that the RNS and Goodman bounds are complementary approaches as we recover the Bieberbach-Rogsinski bounds by taking $\rho \rightarrow 0$ and $p \rightarrow 1$ respectively and the bounds get weaker in the opposite limit $\rho\rightarrow 1$ and $p\rightarrow 0$. We shall use only the Goodman bounds and Bieberbach-Rogosinski bounds in this paper.

\noindent We shall argue the crossing symmetric amplitude is typically-real and proceed to apply the above coefficient bounds to constrain the Wilson coefficients in the rest of the section.
\subsection{Amplitude and Typically Real-ness}
We can recast the crossing symmetric dispersion relation \eqref{disp} in the Robertson form using the following steps.
If we set $-\frac{27 a^3}{(s_1^{'})^3}+\frac{27 a^2}{(s_1^{'})^2} -2 =-2 \xi$ and make a change of variable to $\xi$ in \eqref{disp} we get\footnote{Note that $d\xi=\frac{27 a^2}{2(s_1')^4}(3a-2s_1')ds_1'$ so when $a \in \left(-\frac{8}{9},0\right)\cup \left(0,\frac{16}{9}\right)$, the outside factor has a definite sign for $s_1'\geq 8/3$. Further note that in this range of $a$, $-1\leq \xi\leq 1$ and $\xi(s_1')$ is a single valued function which is needed to cover the full range of $s_1'$. This is an alternative argument to getting the range $a \in \left(-\frac{8}{9},0\right)\cup \left(0,\frac{16}{9}\right)$.}:
\bea
{\mathcal {\overline M}}_0(\tilde z,a) =\alpha_0 + \frac{2}{ \pi}  \int_{-1}^{1} d \xi
{\bar{\mathcal A}}\left(\xi,s_2(\xi,~a)\right)  \frac{\tilde z}{1-2 \xi \tilde z+\tilde z^2}\,.
\eea

  The range of the integral follows by noting that the range of $a$ was chosen to avoid singularities inside the unit disc which gives $|\xi|< 1$. Notice that without the absorptive the part the measure of the integral is a probability measure. 
By defining the $d\mu(\xi) = \frac{{\bar{\mathcal A}}\left(\xi,s_2(\xi,~a)\right) d\xi}{ \int_{-1}^1 d\xi {\bar{\mathcal A}}\left(\xi,s_2(\xi,~a)\right)} $, we easily see that $\mu(\xi)$ is a non-decreasing function\footnote{$\mu(x)=\frac{\int_{-1}^x\,{\bar{\mathcal A}}\left(\xi,s_2(\xi,~a)\right) d\xi}{ \int_{-1}^1 d\xi {\bar{\mathcal A}}\left(\xi,s_2(\xi,~a)\right)}$ which is obviously non-decreasing since ${\bar{\mathcal A}}>0$.} and we can recast the above integral as:
\bea \label{reprob}
\frac{{\mathcal {\overline M}}_0(z,a) -\alpha_0}{2\int_{-1}^1 d\xi {\bar{\mathcal A}}\left(\xi,s_2(x,~a)\right)} =\frac{1}{\pi} \int_{-1}^1 d\mu(\xi)  \frac{\tilde z}{1-2 \xi \tilde z+\tilde z^2}\,.
\eea
which is in the Robertson form, which is a necessary and sufficient criterion for being typically real. 
\vskip 0.5cm
{\it This proves that the fully crossing symmetric representation of 2-2 scattering of massive particles is typically real, when the absorptive part is positive.} 
\vskip 0.5cm
We contrast this with \cite{HSZ} where {\bf only the kernel $H(\tilde z,a)$ was univalent not the whole amplitude} and the coefficients bounds had to transferred for the kernel onto the amplitude to get bounds on $\mW_{pq}$. In our case we have shown that {\bf the full amplitude including the bound state poles is typically-real} and we can thus directly use the bounds \eqref{rogo1} on amplitude $\mathcal{M}$. As alluded to earlier we can systematically explore the presence of massive bound states or massless poles by using \eqref{good},\eqref{good1} in EFTs as well. 
\subsection{Scalar EFTs and GFT} \label{SEG}
In \cite{SCH}, it was suggested that that most scalar EFTs compatible with unitarity and crossing symmetric are {\it infinite positive} linear combinations of 
\bea\label{M0M2}
{\mathcal M}^{(M_h^2)}_0(s_1,s_2,s_3)&=&\left(\frac{1}{M_h^{2}-s_1}+\frac{1}{M_h^{2}-s_2}+\frac{1}{M_h^{2}-s_3} \right) \,,\\
{\mathcal M}^{(M_h^2)}_2(s_1,s_2,s_3)&=&  \frac{M_h^4}{(M_h^{2}-s_1)(M_h^{2}-s_1)(M_h^{2}-s_3)} -\gamma_d{\mathcal M}^{(M_h^2)}_0(s_1,s_2,s_3)\,,
\eea 
with varying $M_h$. 
Here the $\gamma_d {\mathcal M}_0$ subtracts off the spin-0 contribution, with $\gamma(d)=\frac{4}{9}{}_2F_1(\frac{1}{2},1,\frac{d-1}{2},\frac{1}{9})$.
Let us examine this using GFT arguments. Notice that
\bea\label{zMsch}
{\mathcal M}^{(M_h^2)}_0-\frac{3}{M_h^2}&=& \frac{\tilde \beta_1}{M_h^2} \frac{\tilde z}{\tilde z^2-2\xi \tilde z+1}\,,\\
{\mathcal M}^{(M_h^2)}_2 +\gamma_d{\mathcal M}^{(M_h^2)}_0&=& \frac{1}{M_h^2}\frac{(\tilde z-1)^2}{\tilde z^2-2\xi \tilde z+1}=\frac{1}{M_h^2}\left(2(1-\xi) \frac{\tilde z}{\tilde z^2-2\xi \tilde z+1}-1\right)\,.
\eea
Here $-2\xi=27 a^2 (1-a/M_h^2)/M_h^4-2$ and $\tilde \beta_1=27a^2(3a-2M_h^2)/M_h^6$. For absence of singularities inside the unit disk, for all $M_h>\Lambda$ we have $|\xi|<1$ which gives $-\Lambda^2/3<a<\Lambda^2$.
 According to \cite{SCH} the SDPB results support the fact that almost all EFTs are convex sums of the above two amplitudes except for a tiny sliver in the region plots. This means (removing the constant pieces, indicated by $\sim$) we have a general amplitude to be 
\be\label{chul}
k_1 {\mathcal M}_0+k_2 {\mathcal M}_2\sim \left[k_1 \frac{\tilde \beta_1}{M_h^2}+k_2\left(\frac{2(1-\xi)}{M_h^2}-\gamma_d \frac{\tilde \beta_1}{M_h^2}\right)\right] \frac{\tilde z}{\tilde z^2-2\xi \tilde z+1} \,,
\ee
with $k_1,k_2>0$ with $k_1+k_2=1$. Now the factor inside the brackets is non-zero, else the amplitude vanishes so this enables us to divide by it so that the coefficient of $\tilde z$ is unity. At this stage, this simply implies that the factors multiplying $k_1, k_2$ should have definite signs since setting either $k_1=0$ or $k_2=0$ should still allow us to normalize the amplitude to put in the Robertson form.  Now notice that $\gamma_d>0$, and for $\tilde \beta_1<0$,  we find the coefficient of $k_2$ is greater than zero, so that if we set $k_1=0$, we can still normalize. Thus the necessary and sufficient condition is that $\tilde \beta_1<0$ which implies $a<2M_h^2/3$ for all $M_h^2>\Lambda^2$. In all we get  the $a$-range 
\be
-\frac{\Lambda^2}{3}<a<\frac{2\Lambda^2}{3}\,.
\ee
This is an important condition that we will come back to frequently in due course. 
\subsubsection{Extremal functions}
${\mathcal M}_0$ and ${\mathcal M}_2$ provide the extremal values for some of the Wilson coefficients in \cite{SCH}, where it was empirically observed that the convex hull of ${\mathcal M}_0$ and ${\mathcal M}_2$ appear to generate most of the allowed space of theories (at least in some of the coupling constants space) that arise from SDPB considerations---except for a small sliver, most of which gets ruled out by our considerations as discussed below. Operationally, what this means is that we put a cutoff $\Lambda$ in the theory and dial $M_h>\Lambda$ as well as $k_1, k_2$ in eq.(\ref{chul}) and examine the span of the Wilson coefficients. Under these circumstances the range of $a$ becomes $-\frac{\Lambda^2}{3}<a<\frac{2\Lambda^2}{3}$. Now in our language, the Robertson representation suggests that the full theory space allowed by positivity and typically real-ness should be generated by $\tilde z/(1-2\xi \tilde z+\tilde z^2)$. Further, this is extremal for typically real functions as discussed in (\ref{A22}). Both ${\mathcal M}_0$ and ${\mathcal M}_2$ are of this form.\\

\noindent {\it Using well known properties of extremal functions in GFT and the Krein-Milman theorem (\ref{A22}), we have thus provided a proof why the theory space should be generated by such extremal functions.} \\

However, it is not always straightforward to write a general extremal function in the integral representation given in (\ref{A22}) in the conventional Mandelstam variables since both $\xi$ as well as the measure factor $\mu(\xi)$ implicitly depend on the parameter $a$. In some cases, we can write an interesting amplitude. 

\noindent For instance, suppose we ask the question: Which amplitude (namely, what value of $\xi$) saturates $a_n=n$ as defined in eq.(\ref{fdef}). 

Taylor expanding $\tilde z/(1-2\xi \tilde z+\tilde z^2)$, we readily find the answer to be $\xi=1$. As pointed out in the introduction, this is just the Koebe function. We can use $-27 a^2 \tilde z/(1-\tilde z)^2=x, a x=y$, to rewrite this in terms of $x,y$, thereby obtaining $-x^3/(27 y^2)$. Now in our typically real considerations, we normalized the function such that $a_1=1$--so we are allowed to multiply by suitable powers of $a$ since the measure is independent of $\tilde z$. Thus, if we were interested in local theories with singularities being simple poles only, the actual amplitude can only be proportional to:
\begin{itemize}
\item $x$ on multiplying by $a^2$ 
\item $y$ on multiplying by $a^3$
\item $x^2/y$ on multiplying by $a$. 
\end{itemize}
Curiously, the latter amplitude is the leading dilaton scattering with a graviton exchange! 

 \noindent Consider another example: $a_n=-n$ (in the Bieberbach case $|a_n|\leq n$) for all even $n$'s, then the amplitude is proportional to $x^3/(4 x^3-27y^2)$. However, it does not appear to arise from a local QFT.  If we wanted to write a single amplitude that gives $a_n=-\kappa_n$, i.e., the Rogosinski lower bounds, then the answer is more complicated and  involves Elliptic functions see \eqref{wp} and appendix \ref{boundswithpoles} for more details.

\subsubsection{A toy problem}
Now let us investigate the implications of the Goodman bounds \eqref{good3} discussed in appendix \ref{boundswithpoles} in a toy example.  Here for simplicity let us set $\gamma(d)=0$. We will take $M_h^2=8/3$ in ${\mathcal M}_2$ and choose $-8/9\leq a\leq 8/3$ so that this pole gets mapped to the boundary of the unit disk\footnote{Any choice of $a$ outside this range will give two real poles one inside the disk and one outside.}.  Now consider the amplitude
\be
{\mathcal M}(\tilde z)=g^2 \left({\mathcal M}^{(M_h^2)}_0(\tilde z)-\frac{3}{M_h^2}\right)+\left({\mathcal M}^{(\frac{8}{3})}_2(\tilde z)-\frac{27}{512}\right)\,.
\ee
$M_h$ here will be real.
The constant shifts have been chosen to set the $z^0$ term to zero. To apply the Goodman bounds we have to consider
\be
f(\tilde z)=\frac{{\mathcal M}(\tilde z)}{\partial_{\tilde z}{\mathcal M}(\tilde z)|_{\tilde z=0}}\,.
\ee
Now consider eq.(\ref{good3}). This tells us to calculate the negative residue for a pole inside the unit disk. If $M_h^2<8/3$, we expect that for a range of $a$, there will be real pole inside the disk. It is easy to check that the condition for this is
\be
a<-\frac{M_h^2}{3}\,, \quad {\rm or~~} a>M_h^2\,.
\ee
Eq.(\ref{good3}) tells us that the negative residue has to be positive. If we do not demand a unitary ${\mathcal M}_0$ at this stage, i.e., we do not impose $g^2>0$, then eq.(\ref{good3}) leads to the following necessary conditions
 \bea
 g^2<0\,, \quad {\rm for ~~} M_h^2<a<\frac{8}{3}\\
 g^2>0\,,\quad {\rm for ~~} -\frac{8}{9}<a<-\frac{M_h^2}{3}\,.
 \eea
 This means that demanding $g^2>0$ will select out only the first range of $a$ where the combination is typically real. There is no further restriction on $g^2$ using just eq.(\ref{good3}). Using the Goodman bounds in eq.(\ref{good}) it is not possible to get further constraints on $g^2$ in this toy example in a unitary theory as we have checked. We can easily generalize this analysis to the following case:
 \be
{\mathcal M}(\tilde z)=g^2 \left({\mathcal M}^{(M_h^2)}_0(\tilde z)-\frac{3}{M_h^2}\right)+\sum_{p,q=0} \mW_{p,q}x^p y^q-\mW_{0,0}\,,
\ee
where we have added the dispersive contribution to the pole term. The previous case is a sub-case of this. Now imposing the Rogosinski bounds (namely eq.(\ref{rns}) with $\rho=0$)  on the dispersive part, we can ask if $g^2$ is bounded as a function of $M_h^2$. Using the Goodman bounds, the analysis is similar to what we wrote above and the conclusions are identical. This leads us to the following important conclusion: {\it Typically Real-ness, being analogous to positivity, is not enough to give us bounds on $g^2$. We need the unitarity condition $a_\ell<1$, possibly including the non-linear condition on the full partial wave amplitude, to get bounds on $g^2$.} 
\subsection{$\mW_{pq}$ bounds}
We now turn to Wilson coefficient bounds.
Let us first collect some useful formulae. Writing the amplitude in terms of the $\tilde z$ variable we have \cite{HSZ}
\be\label{Malpha}
{\mathcal M}(\tilde z, a)=\sum_{n=0}^{\infty} \tilde z^n a^{2n} \alpha_n(a)=\sum_{p,q} \mW_{pq}x^p y^q\,.
\ee
Comparing the powers of $\tilde z$ and {\bf assuming the locality constraints to hold}, which means we have thrown away $\mW_{pq}$'s with $p<0$, we have
\be\label{alphadef}
\alpha_p(a)=\sum_{n=0}^p \sum_{m=0}^n \mW_{n-m,m} a^{2n+m-2p}(-27)^n\frac{\Gamma(n+p)}{\Gamma(2n)(p-n)!}\,,\quad p\geq 1\,.
\ee
This means that $\alpha_p(a) a^{2p}$ is a polynomial of degree $3p$.
For instance
\be
\alpha_1(a)=-27 (a \mW_{0,1}+\mW_{1,0})=-27 \mW_{1,0}(a w_{0,1}+1)\,,
\ee
where in the last equality we have pulled out the positive $\mW_{1,0}$ and defined $w_{0,1}\equiv \mW_{0,1}/\mW_{1,0}$. In terms of the $\beta_n$'s defined in eq.(\ref{betan}), we have
\be\label{alphadef}
a^{2n}\alpha_n(a)=\frac{1}{\pi}\int_{\frac{2\mu}{3}}^\infty \frac{ds_1'}{s_1'} {\mathcal A}(s_1';s_2^+(s_1',a))\beta_n(a,s_1')\,,
\ee
where $s_2^+$ is defined in eq.(\ref{Hs2p}). 
The Bieberbach-Rogosinski bounds \cite{HSZ} read
\be \label{bbound}
-\kappa_n \leq \frac{\alpha_n(a) a^{2n}}{\alpha_1(a)a^2}\leq n, \forall n\geq 2\,.
\ee
We will discuss stronger bounds arising from typically real-ness below. Note at this stage that eq.(\ref{bbound}) is of the form of a degree $3n-2$ polynomial in $a$ divided by $27 |a \mW_{0,1}+\mW_{1,0}|$. We will make use of this later on. We will also find it convenient to introduce normalized $w_{pq}$'s via
\be
w_{pq}\equiv \frac{\mW_{p,q}}{\mW_{1,0}}\,.
\ee

\subsection{Why we get two-sided bounds: a general proof}\label{marko}
We will give an argument that the Bieberbach type inequalities will lead to two-sided bounds on the Wilson coefficients. The argument relies on the so-called Markov brothers' inequality which we will summarize below.
First, recall that as we noted near eq.(\ref{bbound}),  all Rogosinski/Bieberbach inequalities we use, are always of the form:
\bea
\Bigg|\frac{P_{3n-2}(\{ w_{p,q}\} ,a)}{(a w_{01}+1)}\Bigg| \leq n
\eea  
where $P_{3n-2}(w_{p,q},a)$ are degree $3n-2$ polynomials in $a$ with coefficients that are $w_{p,q}$'s. Now noting that $-9/16\leq w_{01}\leq 9/8$ and $8/9\leq a \leq 16/9$ we have $|a w_{01}+1|\leq 3$. This means
\bea
|P_{3n-2}(\{w_{p,q}\} ,a)| \leq 3 n\,.
\eea  

The coefficients of such polynomials are bounded by the Markov brothers' inequality \cite{Markov1,Markov2} which states the following:\\

\noindent{\bf Theorem:} Let $p(x)= \sum_{k=0}^n a_k x^k$ be a real polynomial  of degree at most $n$ that satisfies $|p(x)| \le 1$ for all $-1\le x\le 1$. Then we have:
 \be \label{markov}
|a_k| \leq \frac{1}{k !} |p^{(k)}(0)| \leq  
\frac{1}{k !} |T_n^{(k)}(1)| =\frac{1}{k!}\frac{\prod_{\nu=1}^k (n^2-(\nu-1)^2)}{(2k-1)!!}, 
\ee
where, $T_n^{k}(x)$ is the $k$'th derivative of the Chebyshev Polynomial of the first kind. 
We first begin transforming \eqref{markov} from the interval $[-1,1]$ to $[b_1,b_2]$ using $x\rightarrow \frac{2 x-(b_2+b_1)}{b_2-b_1}$ \cite{Markov1,Markov2} to get:
\bea\label{markov2}
|a_k| &\le& \frac{1}{k!}\left( \frac{2}{b_2-b_1} \right)^k \Bigg|\left( \frac{d^{k}}{dx^{k}} T_n\left( \frac{2 x-(b_2+b_1)}{b_2-b_1}\right) \right){\bigg |}_{x=b_2}\Bigg | \nonumber\\
&\le& \frac{1}{k!}\frac{\prod_{\nu=1}^k (n^2-(\nu-1)^2)}{(2k-1)!!} \frac{2^{2k}}{ (b_2-b_1)^{2k}},
\eea
where, $a_k$ are the coefficients of polynomial in $[b_1,b_2]$ now.
We can now apply this directly to various cases:\\
{\bf n=2 worked out:}

As before we first rewrite the $n=2$ inequality by rearranging factors of $aw_{01}+1$ and using the fact that it has an maximum value of $|a w_{01}+1| \le 3$:
\bea
|27 a^4 w_{02}+27a^3 w_{11}+27 a^2 w_{20}| \le 12\,.
\eea
Applying \eqref{markov2} with $b_2=16/9, b_1=-8/9$ we get,
\bea
|w_{02}| \le0.356, ~~~|w_{11}|\le 2.53125,~~~|w_{20}|\le 5.625 \,.
\eea
{\bf n=3 worked out:}
\bea
|729 w_{03} a^7+729 w_{12} a^6+729 w_{21} a^5+(729 w_{30} -108 w_{02} )a^4-108 w_{11} a^3-108 w_{20} a^2| \le 18
\eea
which gives
\bea
|w_{03}| &\le& 0.02815,~|w_{12}|\le 0.3503,~|w_{21}|\le 1.713,~\Bigg|w_{30}-\frac{108}{729} w_{02}\Bigg|\le 4.153, \nonumber \\
|w_{11}|&\le& 34.88,~~~~|w_{20}|\le 20.67 \,.
\eea
Combining the $n=2$ result we find
\bea
|w_{02}| &\le&0.356, ~~~|w_{11}|\le 2.53125,~~~|w_{20}|\le 5.625,~~~ |w_{30}| \le 4.206,\nonumber \\
 |w_{03}| &\le& 0.02815,~~|w_{12}|\le 0.3503,~|w_{21}|\le 1.713 \,.
\eea

As should be clear from the above examples, we will always get two-sided bounds on the $w_{pq}$'s which are all $O(1)$ numbers. We have checked to very high orders and find that for {\it any} $p,q$, $$|w_{pq}|\leq 5.625$$ holds (recall this is in units where $m^2=1$). We would like to emphasise that the bounds we get by this method are an {\it overestimate}. However, they still prove the finite extent of the $w_{pq}$-space and explain why we get bounded regions.  

\subsection{Numerical tighter bounds}
We would now like to bound  Wilson coefficients  using the inequalities we get from typically realness. 
We will begin with the RNS-bounds and first consider the case where $\rho=0$, i.e., the Bieberbach-Rogosinski bounds where we have $TR_U$ defined in the full unit disk.
 Note that \eqref{rns} gives us a stronger lower bound than the one we get from univalence alone. We would like to use these results to bound the Wilson coefficients.  We implement this as follows:

We use 2 sets of inequalities to constrain the Wilson coefficients:
\begin{enumerate}
\item {\bf Bieberbach-Rogosinski:} We first impose linear inequalities \eqref{rogo1} . We call these conditions $TR_U$. The inequalities on the $\mW_{p,q}$'s are obtained by using the \eqref{rogo1} bounds on $\alpha_n(a)$'s in eq.(\ref{alphadef}). 
\item{\bf Positivity Constraints:}  We also consider the  $PB_C$ conditions \eqref{pbc} with $\delta=0$ and $\Lambda^2= \frac{2 \mu}{3}$. 
\end{enumerate}
From here onwards we work with\footnote{Since we will work with low values of $p,q$ we will drop the comma between $p,q$ to make the notation slightly simpler.} $w_{pq}=\frac{\mW_{p,q}}{\mW_{1,0}}$.
\subsection*{Restriction on the parameter $a$--summary:}
Before we begin the numerical analysis, we will remind the reader one more time about the restriction on the parameter $a$ that enters the dispersion relation since it plays a very important role. We will work in units where $m^2=1$. Recall that the objective is to use Typically Real-ness and in order to do so, we observed that the kernel is univalent and typically real in the unit disc provided it does not have a singularity inside the disc. This gave rise to the restriction on $\gamma$ defined as $\gamma\equiv 27 (a/s_1)^2(1-a/s_1)-2$ to satisfy $|\gamma|<2$. Since the range of the integration variable is $s_1\geq 8/3$ we have $a\in (-8/9,0)\cup (0,8/3)$. We further note that $\beta_1$ in eq.(\ref{betan}) is negative in the integration domain if $a<16/9$ which is needed to have a positive measure in the Robertson representation.  Next note that the argument of the Gegenbauer polynomials in the partial wave expansion in eq.(\ref{partw}) is $\sqrt{\xi}$ where $\xi=\frac{9 s_1^2(s_1+3a)}{(8-3s_1)^2 (s_1-a)}$. It can be checked that in the integration domain $s_1\geq 8/3$ and when $a\in (-8/9,0)\cup (0,16/9)$, $\xi>1$. As such the Gegenbauer polynomials are positive. The partial wave expansion converges for this range of $a$; in fact it converges for a bigger range of $a$ which was derived in \cite{AK}.

We will also compare with the results in \cite{SCH}. In that reference, SDPB methods and fixed-$t$ dispersion relations were used to obtain two-sided bounds on $w_{pq}$'s. These results were somewhat stronger than \cite{tolley}.  In scalar EFTs, one assumes that the dispersive integral starts at some $s_1=\Lambda^2\gg \mu$ so that the external scalars can be taken to be massless. As discussed earlier, \cite{SCH} observed that all scalar EFTs are convex sums of the two amplitudes in eq.(\ref{zMsch}). This in turn is in the Robertson form (in other words in the form used in the crossing symmetric dispersion relation) and will be Typically Real provided $-\Lambda^2/3<a<2\Lambda^2/3$. To convert our results into these units we replace $\mu\rightarrow 3\Lambda^2/2$ or conversely $\Lambda^2\rightarrow 2\mu/3=8/3$ in the \cite{SCH} results. 
\section*{$n=2$ results:}
We impose the following conditions for $n=2$ as alluded to:\\ 
1) There is a single  constraint at $n=2$ in $TR_U$ and this reads:
\bea 
-2 ({a w_{01}+1})\le 2 ({a w_{01}+1})- {27 a^2 \left(a^2 w_{02}+a w_{11}+w_{20}\right)}\le 2({a w_{01}+1})\,.
\eea
2) From $PB_C$ for $n=2$ we have:
\bea
-\frac{9}{16}\leq w_{01}\leq \frac{9}{8},~~w_{11}+\frac{15 w_{20}}{16}\geq 0,\nonumber \\ w_{02}+\frac{9 w_{11}}{16}+\frac{81 w_{20}}{256}\geq 0,~~0\leq w_{20}\leq \frac{9}{64} \,.
\eea

\noindent Imposing these conditions gives us finite regions as shown in the figures below.

 \begin{figure}[H]
  \centering
    \includegraphics[width=\textwidth]{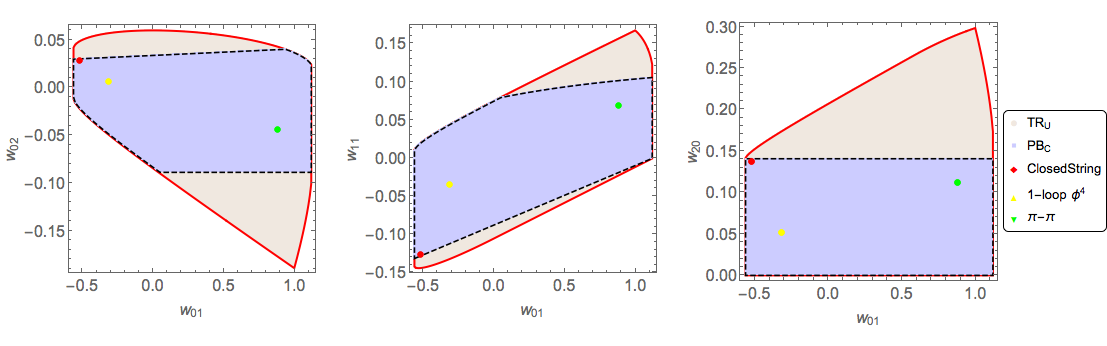}
\caption{The comparison of the regions obtained for $w_{02}$, $w_{11}$ and $w_{20}$ with $w_{01}$.} 
\end{figure} 
\vspace{10 pt}
We have indicated the closed string, 1-loop $\phi^4$ and pion results (see appendix of \cite{HSZ} for these values as well as what is used below) in the plot above as special points. Note that the string solution is very close to boundary of the region and this confirms the validity our bounds. We could also project in the $w_{01}$ and $a$ directions to see get a finite 3d region as shown below: \\
\vspace*{-25 pt}
 \begin{figure}[H]
  \centering
 \includegraphics[width=0.3\textwidth]{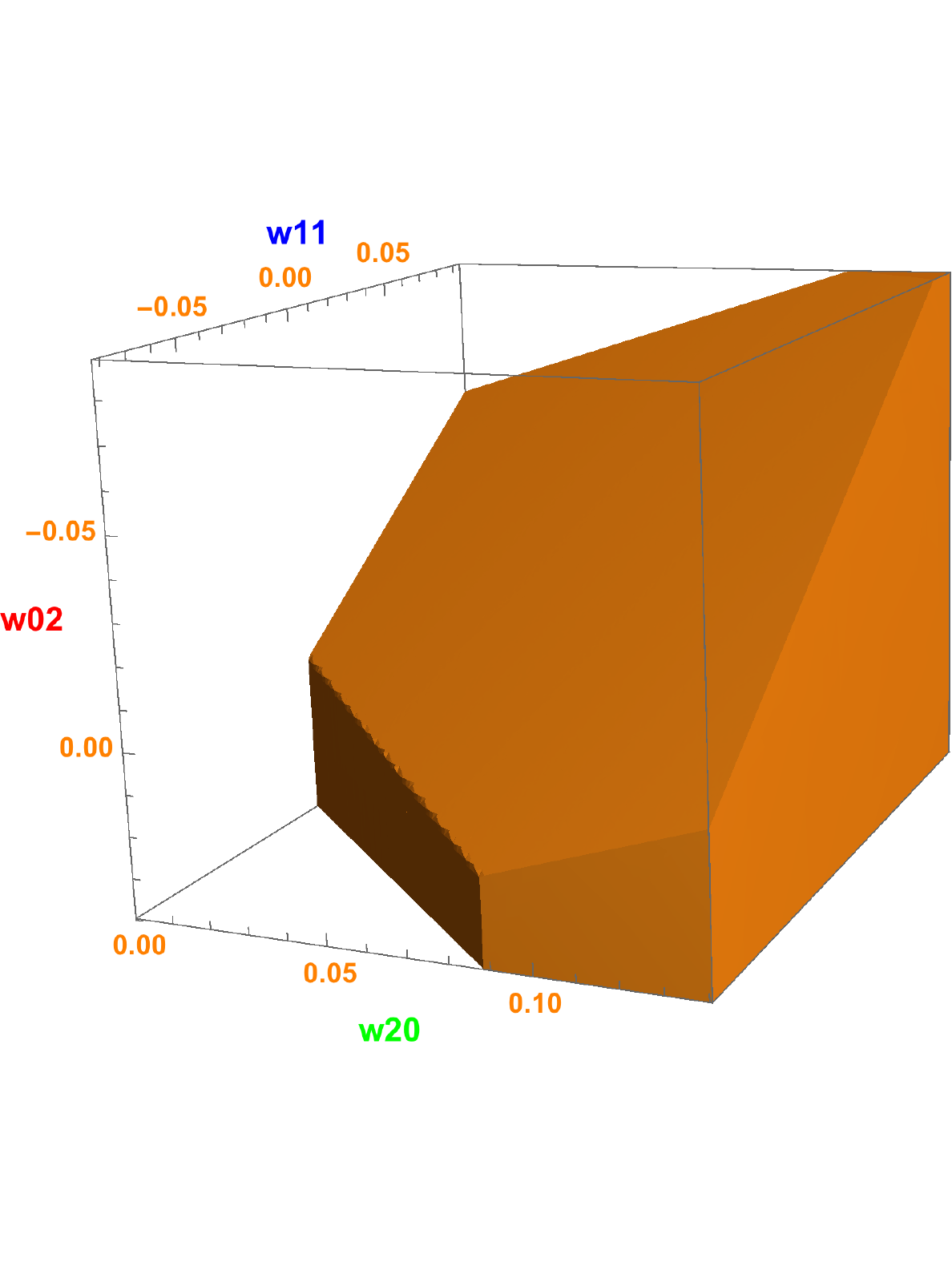}
  \caption{The 3d region obtained when we project down to  $w_{02}$, $w_{11}$ and $w_{20}$.}\vskip 0.5cm
\end{figure}
The results we get at $n=2$ are:
\bea
-0.08898&\leq& w_{02}\leq 0.0396,\quad-0.1318\leq w_{11}\leq 0.1054, \nonumber\\
0 &\leq& w_{20}\leq 0.140625,\quad -0.5625\leq w_{01}\leq 1.125 \,.
\eea
We can repeat the analysis for higher $n$ easily and we present the analysis till $n=5$ in the appendix \ref{n35}.We now proceed to compare our results with the ones with other known answers from the literature.

\subsection{Comparsion with known results} \label{schtables}
We can compare our results with the pion bootstrap \cite{ST} building on \cite{pionworks}, the results of \cite{SCH}. We also contrast these with bounds obtained from implementing Locality constraints $N_c$ defined in eq.\ref{nc} namely $\mW_{n-m,m} =0$ for $n<m$ for the following two scenarios:

\noindent $\bullet$ $N_c |_{\xi_0=1}$: The EFT cutoff $\Lambda \gg \mu$ i.e., $\xi_0= \frac{s_1}{(s_1-8 \mu/3)}~1$.\\
$\bullet$ $N_c |_{\xi_0>1}$: The EFT cutoff $\Lambda$ is comparable to $\mu$ i.e., $\xi_0>1$.\\
\noindent We do the above for all $m>n,~n$ up to $2n+m\le 8$. The results for $d=4$ (upto 2-significant digits) are listed in the tables below. We note that in the pion bootstrap \cite{ST} nonlinear unitarity constraints were used and the results respect the bounds obtained. We find good agreement between the results of \cite{SCH} and the first case $N_c |_{\xi_0 =1}$ and this is expected since the results of \cite{SCH} were obtained assuming $\Lambda \gg \mu$ and treating the particles as effectively massless. We also find reasonable agreement of our results with the second case $N_c |_{\xi_0>1}$.

We note that to compare with the results \cite{SCH} we need to make the following identifications to match conventions :
\bea\label{conv}
\tilde{g}_3&\to& -2 w_{0,1},~\tilde{g}_4\to \frac{w_{2,0}}{2},~\tilde{g}_5\to -w_{1,1},~\tilde{g}_6\to \frac{w_{3,0}}{4}, \nonumber\\
\tilde{g}'_6&\to& 2w_{0,2},~\tilde{g}_7\to -\frac{w_{2,1}}{2},~\tilde{g}_8\to \frac{w_{4,0}}{8} ,~ \tilde{g}'_8\to w_{1,2}, \nonumber \\
\tilde{g}_9&\to&-\frac{w_{3,1}}{4},~\tilde{g}'_9\to-2 w_{0,3},~\tilde{g}_{10}\to \frac{w_{5,0}}{16},~\tilde{g}'_{10}\to \frac{w_{2,2}}{2} 
\eea
We also need to multiply with suitable powers of $\frac{8}{3}$ to match the EFT scale conventions; concretely,  we multiply the $w_{pq}$ values in \cite{SCH} by $(3/8)^{2p+3q-2}$. 
\begin{table}[h!]
   \centering
$\begin{array}{|c|c|c|c|c|c|}
   \hline
    w_{p,q}=\frac{\mW_{p,q}}{\mW_{1,0}} & TR_U &EFT &Pion&N_c|_{ \xi_0>1}&N_c|_{\xi_0=1}\\
   \hline
w_{01} & -0.5625 &-0.5625 &-0.335& -0.5625& -0.5625\\
 \hline
 w_{11} & -0.1318 & -0.1318 &-0.0846 & -0.1318&-0.1318\\
  \hline
 w_{02} & -0.089 &-0.1268  &-0.056&-0.0664 &-0.127 \\
  \hline
 w_{20} & 0 &  0 &0.020& 0&0\\
  \hline
 w_{21} &-0.026 &  -0.02595  &-0.0157& -0.02595&-0.02595\\
  \hline
 w_{12}& -0.019& -0.02789&-0.0121 & -0.014&-0.0278\\
  \hline
   w_{30} & 0 &  0 & 0.001236 &0 &0\\
  \hline
 w_{03} & -0.0049& -0.00156 &-0.000891& -0.00156&-0.0015\\
  \hline
 w_{31}& -0.0047 &-0.0047 &-0.00270 &-0.0047 &-0.0047 \\
 \hline
 w_{40}& 0& 0 &0.0000905& 0&0\\
 \hline
 w_{50}& 0&0& 8.32\times 10^{-6} & 0&0 \\
 \hline
\end{array} $
\caption{A comparison of minimum values obtained using our results with \cite{SCH},\cite{ST} and Locality constraints $N_c$ for $\xi_0=1$ and $\xi_0>1$ cases.}\label{mint}\vskip 0.5cm
\end{table} \\
\begin{table}[h!]
   \centering
$\begin{array}{|c|c|c|c|c|c|}
  \hline
  w_{p,q}=\frac{\mW_{p,q}}{\mW_{1,0}} & TR_U & EFT & Pion&N_c|_{ \xi_0>1}&N_c|_{\xi_0=1} \\
   \hline
w_{01} &   1.125 &1.939&1.07& 1.125&1.94 \\
 \hline
 w_{11} &   0.106 & 0.216 & 0.0918& 0.106&  0.216\\
  \hline
 w_{02} &  0.039& 0.0296 & 0.0182& 0.0296&0.029 \\
  \hline
 w_{20} &  0.140625& 0.140625 & 0.128&0.140625 &0.140625 \\
  \hline
 w_{21} & 0.011  & 0.023 & 0.005995& 0.00741&0.0229\\
  \hline
 w_{12}&  0.011 &  0.0111 & 0.06525&0.011 &0.011\\
  \hline
   w_{30} &0.01977 &0.01977 & 0.01706& 0.01977& 0.01977 \\
  \hline
 w_{03} & 0.0065 &0.0071  & 0.0028&0.0036 &0.0071\\
  \hline
 w_{31}&  0.0022 &0.0022 &0.0000378&0.0022 &0.00218  \\
 \hline
 w_{40}&  0.00278& 0.00278& 0.00228& 0.00278&0.00278\\
 \hline
 w_{50}&0.00039  &0.00039&0.000306 & 0.00039&0.00039\\
 \hline

\end{array} $
\caption{A comparison of maximum values obtained using our results with \cite{SCH}, \cite{ST} and Locality constraints $N_c$ for $\xi_0=1$ and $\xi_0>1$ cases.}\label{maxt}\vskip 0.5cm
\end{table} 

\noindent A few key points that we would like to emphasise are as follows:
\begin{enumerate}
\item {\bf Constraining:} The inequalities in $TR_U$ are really constraining despite being only linear conditions due to the fact that they correspond to an {\it infinite} set  of inequalities for each value of $-\frac{8}{9}<a<\frac{16}{9}$ and this is why they are both powerful enough to strongly constrain the space of $\mW_{p,q}$'s and also simple enough to be implemented on Mathematica without a need for more sophisticated computational algorithms like SDPB.
\item {\bf Faster Convergence:} In the usual Null constraints approach one has to discretise $s_1$, spin $\ell$-sum truncation and also the number of Null constraints used. In our method using $TR_U$ and $PB_c$ we did not have to worry about the spin truncation or the number of constraints  and we only had discretisation of $a$. 
Furthermore using $TR_U$ constraints till level $n+1$ we get bounds for all $w_{p,q}$ s.t $p+q\le n$ which are convergent and including higher $n$ values does not improve these bounds.
\item {\bf Massive vs massless bounds:} We obtained analytic bounds for $w_{01}$ that followed from the range of $a$ directly. This gave $-\frac{9}{16}<w_{01}<\frac{9}{8}$ for the massive case ($\xi_0>1$). The massless case ($\xi_0=1$) requires consideration of the phenomenon of low-spin dominance (LSD) which changes the range of $a$ and shall be discussed in detail in an upcoming work \cite{CGHRS} for now we merely state that this gives us $-\frac{9}{16}<w_{01}<1.94$ in agreement with $N_c|_{\xi_0=1}$. The fact that several values of $w_{pq}$ agree between the massless and massive cases suggests that the EFT scenario ($\Lambda^2\gg \mu^2$) provides a good approximation. 
\item{\bf Dimension-dependent bounds:} The key parameter in our analysis is $a$ its range of determines the bounds. It might seem like our methods give dimension independent bounds in the massive case since the range of was fixed to be $-\frac{8}{9}<a<\frac{16}{9}$. In fact we can obtain dimension dependent bounds by doing a more careful analysis of the positivity of the absorptive part and our bounds are in-fact the ones corresponding to infinite-dimension limit in the massive case as we explain in the next section. The massless case is more subtle involving both positivity and LSD and shall be discussed in \cite{CGHRS} as alluded to earlier.

\item {\bf Nonlinear-constraints:} We have seen that the bounds we get from typical real-ness constrain the theory space quite strongly despite being {\it only linear} in the $w_{pq}$'s due to the nature of the constraints which hold true for a continuous range of the parameter $a$. There are also several non-linear inequalities imposed by typical real-ness \footnote{For univalent functions analogous non-linear constraints called the {\emph {Grunsky inequalities}} were discussed in \cite{HSZ}.However, since only the kernel was univalent these were valid only around $a\sim 0$ for the full amplitude $\mathcal{M}$.In our case \eqref{det} conditions are valid for the full range of $a$}analogous to the Hankel determinant conditions in the EFT-hedron. We will come back to this in section 4.
\item{\bf Full Unitarity:} We have not fully used the non-linear unitarity of the partial-waves.We have only used the positivity of the spectral function $0\le a_{\ell}= \Im f_{\ell}$. However some of the results obtained from S-matrix bootstrap which use the full-unitarity such the the ones for pion scattering \cite{ST,pionworks} which we have shown in the tables and the figure below are close to certain upper/lower bounds despite not saturating any of them. This suggest that some of the bounds obtained using our methods are already close to the allowed boundary of $w_{pq}$-space. It will be interesting to see if we can find amplitudes that satisfy non-linear unitarity and also saturate our linear bounds.
\end{enumerate}

We can directly compare the plots of \cite{SCH} for $w_{20}$ vs $w_{01}$ yielding the following \footnote{Note that the parabolic region here has been reflected about the $w_{20}$ axis compared to the one in \cite{SCH}. This is due to the difference in conventions $w_{01}$ in \cite{SCH} differs from ours by a sign apart from factors of 2 and $\frac{3}{8}$ as explained above \eqref{conv} and we have carefully accounted for these before making the comparison. } :
 \begin{figure}[H]
  \vspace*{-10 pt}
  \centering
 \includegraphics[width=0.6\textwidth]{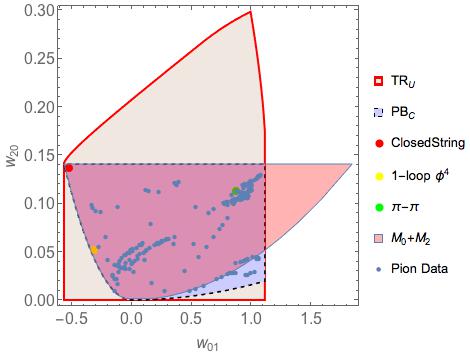}
 \vspace*{-10 pt}
  \caption{The comparison of plots for $w_{20}$ vs $w_{01}$ for our case and the one in\cite{SCH} .} \label{schcomp} \vskip 0.5cm
\end{figure}
The pink parabolic region is the convex hull of the models in eq.(\ref{M0M2}). There is also an extra sliver which is not captured by this, which we have not shown.
We see that our region overlaps with the ones in \cite{SCH} but excludes a part of the parabolic region and hence will exclude part of the sliver as well unlike in the massless case considered in \cite{SCH}. The pion bootstrap data \cite{pionworks} are consistent with our bounds. Note that some pion data lies outside the convex hull of $\mathcal{M}_0+\mathcal{M}_2$. The bounding parabolic dashed lines were obtained using \cite{tolley} $-3/2\sqrt{w_{20}}\leq w_{01}\leq 8\sqrt{w_{20}}$. We will discuss nonlinear inequalities further in sec.4.
\subsection{Comments on low spin dominance and the $d\rightarrow \infty$ limit}\label{low}
Our bounds arose essentially using only Typically Real-ness which needed unitarity and positivity of the absorptive part in a range of $a$--they did not depend on spacetime dimensions. Furthermore, it is quite coincidental that our $w_{01}, w_{11}$ bounds agree with the infinite spacetime dimension limit bounds in \cite{SCH}--the $w_{01}$ observation was also pointed out in \cite{HSZ}. We will now explain these observations.
We compared with the results in \cite{SCH} where the calculation was done in an EFT set up with a lower cutoff in the dispersive integral $\Lambda^2\gg m^2$ so that one could take the external scalar to be massless. Our calculations were for external massive scalars with the $s$-channel cut starting at $s=4m^2$.  As such, we have to be careful in making the comparison. As explained above, when we made the comparison, we put $\Lambda^2\rightarrow 2\mu/3$ in the \cite{SCH} $w_{pq}$ bounds. This was motivated by the observation in \cite{SCH} that all consistent scalar EFTs are positive combinations of the amplitudes in eq.(\ref{zMsch}), which in turn is in the Robertson form. This enabled us to bypass asking the question about the positivity of the Gegenbauer polynomials. Is the absorptive part necessarily positive? For the massive case, we had explicitly performed this check and showed that $-2\mu/9<a<4\mu/9$ is compatible with this positivity. Let us examine this point a bit more carefully now for the EFT or massless case.

Let us begin with the massive case. The argument of the Gegenbauer polynomial in the dispersion relation, eq.(\ref{partw}), is $$\frac{s_1}{s_1-\frac{2}{3}\mu}\sqrt{\frac{s_1+3a}{s_1-a}}.$$ Demanding that this is greater than unity for $s_1>\Lambda^2$ leads to
\be
\frac{\mu(\mu-3\Lambda^2)\Lambda^2}{\mu^2-3\mu \Lambda^2+9 \Lambda^2}<a<\Lambda^2\,.
\ee
Setting $\Lambda^2=2\mu/3$ as the lower limit of the dispersive integral, we recover our previous results, namely that Gegenbauer positivity needs $-2\mu/9<a<2\mu/3$ so that the $a$-range we consider in deriving the bounds is compatible with this restriction.
If we consider the massless limit, $\mu=0$, we would simply find $0<a<\Lambda^2$. Using $a w_{01}+1>0$, we would conclude that $w_{01}$ has  no upper bound, leading to a tension with the findings in \cite{SCH, tolley}. However, this is too restrictive as we will see now. Denoting the largest zero of the Gegenbauer polynomial $C^{\lambda}_\ell(x)$ by $x^{(0)}_\ell$, it is known \cite{elbert} that
\be
x^{(0)}_\ell\leq \zeta_\ell\equiv \frac{\sqrt{\ell^2+2(\ell-1)\lambda-1}}{\ell+\lambda}<1\,.
\ee
Beyond $x>x^{(0)}_\ell$, $C^{\lambda}_\ell(x)>0.$

If we assume that the partial wave expansion is dominated by contributions of spins below some $\ell=L_c$, such that the sign of the absorptive part cannot change from contributions from $\ell>L_c$ called {\it Low-spin dominance}, we should demand instead that the argument of the Gegenbauer ($\mu\rightarrow 0$) should satisfy
\be
\frac{s_1+3a}{s_1-a}>\zeta_{L_c}^2\,,
\ee
which together with $\beta_1<0$ in eq.(\ref{betan}) gives 
\be\label{recal}
\frac{\zeta_{L_c}^2-1}{\zeta_{L_c}^2+3}\Lambda^2<a<\frac{2}{3}\Lambda^2\,.
\ee
If we had naively set $\zeta_{L_c}=1$ then we would only get positive allowed values of $a$ for Typically Real-ness. But this is too restrictive in the situation where we have low spin dominance.
Now notice that the lower bound is spacetime dimension dependent as well as dependent on the $L_c$ parameter. In the large dimension limit, $\lambda\rightarrow\infty$ (and also when $L_c=1$ in $d=4$) we get 
\be
-\frac{\Lambda^2}{3}<a<\frac{2}{3}\Lambda^2\,.
\ee
 In our case we were considering $-2\mu/9<a<4\mu/9$.
When we compared our results with \cite{SCH}, we replaced $\mu$ in our bounds by $3\Lambda^2/2$ so that the lower limit of the dispersive integral started at $\Lambda^2$ as in \cite{SCH}. These considerations give the same range of $a$ we have been using once we identify $\mu\rightarrow 3\Lambda^2/2$.  It is because of this that the $w_{01},w_{11}$ bounds coincided with \cite{SCH} in the infinite dimension limit. 
A careful analysis of the massless case using Typical-realness will be presented in an upcoming work \cite{CGHRS}. A proper consideration of the positivity of the absorptive part to find $\zeta_{L_c}$ and the  Locality constraints to find $L_c$ gives a different dimension dependent range of $a$ \footnote{If we were to use the recalibrated range of $a$ in eq.(\ref{recal}), for massless scalar EFTs, then some of the $w_{pq}$ bounds are expected to get weaker depending on $L_c$ as well as the spacetime dimension. The simplest result is
\be
-\frac{3}{2\Lambda^2}<w_{01}<\frac{3+\zeta_{L_c}^2}{\Lambda^2(1-\zeta_{L_c}^2)}\,.
\ee
For $d=4$ where $\lambda=1/2$ and $L_c=4$, the RHS of the bound is $35/\Lambda^2$, while for $L_c=1$, it is $3/\Lambda^2$. }. \\
\bea
-0.192 ~\Lambda^2<a<\frac{2}{3} \Lambda^2 \, ,
\eea
in $d=4$. Using our methods with this improved range of $a$ gives good agreement with \cite{SCH}.\\
{\bf Key point:} What the reasonable agreement supports is the fact that positivity of the absorptive part for EFTs ($\Lambda^2\gg\mu$) exists for a bigger range of $a$ values than what the fixed-$d$ Gegenbauer argument gives. This in turn suggests that physical theories living at the boundary yielded by the lower limit of the $a$-range must have few spins contributing. These observations find support from what happens in string theory, as explained in \cite{nima}. We elaborate further on these issues in \cite{CGHRS}.



\subsection{Bounds in presence of poles}
In this section, we turn to examining the effect of parameters $\rho, p$ in the RNS and Goodman bounds. To remind the reader, $\rho$ in eq.(\ref{rns}) was a parameter to indicate the size of the annulus. It signifies that there could be unknown poles up to radius $\rho$ and the amplitude becomes typically real after that. $p$ in the Goodman bounds in eq.(\ref{good}), on the other hand was a complementary parameter which signified that we know of the existence of some poles but beyond some radius $p$. We will start with the more interesting case of $TM^*$ which enables us to consider a massless pole. 
\subsubsection{Maximal Supergravity $g_0$ bound}
It is of great recent interest to obtain bounds for theories with a massless pole along with usual branch cuts and in particular the bound on the constant term in the expansion of such an amplitude has been considered in the literature. For instance \cite{SCH2} consider the Maximal supergravity amplitude in $D=10$ dimensions: 

\bea
{\mathcal M}_{susy} &=& \frac{8 \pi G}{s_1 s_2 s_3} + g_0+ g_2 (s_1^2+s_2^2+s_3^2) + g_3 s_1 s_2 s_3 +\cdots\nonumber\\
&=& -\frac{8 \pi G}{y} + g_0 - 2 g_2 x- g_3 y+\cdots \nonumber\\
-\frac{{27 a^3\mathcal M}_{susy}}{8 \pi G}&=& -\frac{1}{ \tilde z} +\left(27 a^3 \hat g_0-2\right) +\left(1458 a^5 \hat g_2 -1-729 a^6 \hat g_3 \right) \tilde z+\cdots \nonumber
\eea
where $\hat g_i =\frac{g_i}{8 \pi G}$. In order for us to be able to normalize the pole to be $-1/\tilde z$, we need $27a^3/(8\pi G)$ to be nonvanishing. This means we can have either $a>0$ or $a<0$.  From the Schiffer-Bargman representation of $TM^*$ as in eq.(\ref{SB2}), if we assume no poles on the real axis, we must have a representation of the form $-1/z-z+\beta_0+\mu g(z)$ with $g(z)$ being $T_R$ and $\mu>0$. This last $g(z)$ can have a dispersive representation with the lower limit related to the location of the first massive pole. Since this means that $g(z)$ is in the Robertson form, we must have a positive measure. This part is in the form in eq.(\ref{reprob}) but now dressed with the $a^3$ factor that we have pulled out. Thus we must have $a>0$. 

\begin{figure}[ht]
 \centering
 \includegraphics[width=0.4\textwidth]{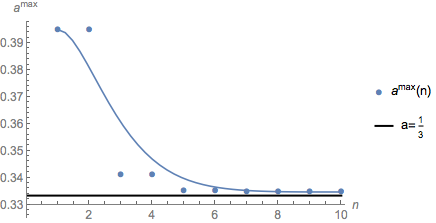}
  \caption{The plot above shows the maximum value of $a$ for which the string amplitude is typically real by imposing Rogosinski conditions at higher $n$ we see that the maximum value of $a$ asymptotes to $\frac{1}{3}$.}\vskip 0.5cm \label{chkk}
\end{figure}

We note that though the above is in Goodman class $TM^{*}$ with $\beta_0=\left(27 a^3 \hat g_0-2\right)$ and $\beta_1=\left(1458 a^5 \hat g_2 -1-729 a^6 \hat g_3\right)$ we cannot bound $\beta_0$ directly in this form since typical-realness involves looking at the imaginary part for which $\beta_0$ decouples due to being a real constant. However we can bound $\beta_0$ by  using the following simple argument which can be found in \cite{
Goodman}:

{\it If $f(z)= -\frac{1}{z}+\beta_0 +\beta_1 z+\cdots$ is in $TM^{*}$ then  $g(z)=-\frac{1}{f(z)} = z+\beta_0 z^2+(\beta_0^2+\beta_1) z^3 +\cdots$  is in $TM$.}
Thus we can use the above by applying the Rogosinksi bounds \cite{rogo}, namely eq.(\ref{rns}) with $\rho=0$, for $g(z)$ to bound $\beta_0, \beta_1$. Doing this for the case discussed above gives:
\bea
-2&\le&\left(27 a^3 \hat g_0-2\right)\le 2 \nonumber\\
-1&\le& \left(729 a^6 \hat g_0^2-729 a^6 \hat g_3-1458 a^5
   \hat g_2-108 a^3 \hat g_0+3\right)\le 3
\eea
 which give ($a>0$)
 \bea
 0&\le& \hat g_0 \le \frac{4}{27 a^3} \nonumber\\
 \frac{27 a^3 \hat g_0^2-27 a^3 \hat g_3-4
   \hat g_0}{54 a^2}&\leq& \hat g_2\leq \frac{729
   a^6 \hat g_0^2-729 a^6 \hat g_3-108 a^3
   \hat g_0+4}{1458 a^5}\,.
 \eea
 What we need to find now is the maximum positive value $a$ can take. 
Numerical checks using the full string amplitude (see fig. \ref{chkk}), including the massless pole,  suggest $0< a\lesssim 1/3$ for typically-realness to hold. The checks involved taking the full string amplitude ${\mathcal M}(z,a)$ and asking for what range of $a$ is $-1/{\mathcal M}(z,a)$ is in $TM$. We do not have an axiomatic argument for the maximum value of $a$ which would lead to $-1/{\mathcal M}$ being in $T_R$ as it relies on the properties of the zeros of ${\mathcal M}$.  

 This gives
\bea
0&\leq& \hat g_0 \leq 4\,, \nonumber\\
0&\leq& \hat g_2\le 10.23 \nonumber\\
-63.2&\leq& \hat g_3 \le 4.24\nonumber
\eea
which is somewhat weaker than \cite{SCH2} where they find $0\leq \hat g_0\leq 3$ in $d=10$ but stronger than \cite{rastelli2021} where they find $0\leq \hat g_0\leq 10.9$ in $d=5$. As usual, our bounds do not depend on the spacetime dimension. The values of $\hat g_2$ and $\hat g_3$ shown here for illustrative purposes are very weak due to the fact that there are other terms involving logarithms which contribute at this order in $z$ and we would have to do the maximization/minimization once we include these terms. 


\subsubsection{Impact parameter representation}
In \cite{rastelli2021}, it was pointed out that bounds in the presence of a massless pole in EFTs come from functions that are positive in impact parameter space and have compact support in momentum space. In the case where $\Lambda^2\gg m^2$, we have for $d=4$, \cite{bg} 
\be\label{impa1}
A(s_1,b)=\int_0^\infty dy\, y J_0(\sqrt{s_1} b y) {\mathcal M}(s_1,-s_1 y^2)\,,\quad {\mathcal M}(s_1,-s_1 y^2)=s_1\int_0^\infty db\, b J_0(\sqrt{s_1} b y) A(s_1,b)\,,
\ee
where $J_0$ is the zero-th order ordinary Bessel function, $b$ is the impact parameter and $A(s,b)$ is the corresponding representation.  It can be shown that our parameter $a$ works out to be
\be
a=s_1 \frac{y^2(y^2-1)}{y^2(y^2-1)+1}\,,
\ee
so to cover the entire $y$ range in the integral we will need $-1/3<a/s_1<1$. Since for $T_R$ we need $-1/3<a/s_1<2/3$, it would appear that $A(s_1,b)$ depends on the non-$T_R$ part of the amplitude. Further, for fixed $s_1$, we can trade $y$ for $a$ and hence $a$ can be interpreted as the conjugate variable for the impact parameter $b$. In \cite{rastelli2021}, it was pointed out that in order to probe the graviton pole, it is beneficial to work in impact parameter representation and measure $g_0$ using small impact parameters. In our discussion above, the upper bound on $\hat g_0$ arises from the lower bound on $a$ (since following \cite{Goodman} we made a $TM$ function from $1/{\mathcal M}$). Now when we go from the $a$ variable to the $y$ variable, the minimum value of $a/s_1=-1/3$ occurs when $y^2=1/2$. Since  $J_0(\sqrt{s_1}b y)$ peaks when $\sqrt{s_1}b y \sim 0$, we expect that for $y^2=1/2$, it is peaked when $\sqrt{s_1} b\sim0 $ or in other words,  $a/s_1\sim-1/3$ corresponds to small impact parameters.

It will be interesting to study our sum rules in impact parameter space, which we will leave for future work.  In order to exploit the $T_R$ properties, it would appear that we need to truncate the $y$ integral at $y^2=2$. We are not sure of the physical interpretation of this, but one should be able to reconstruct the typically real part of the amplitude from such a truncated representation in impact parameter space, at least in the high energy limit. To see that this is true, we start with the second equation in eq.(\ref{impa1}) and use \cite{bg}
\be
\int_0^{k} dx\, x J_0 (x b) J_0 (x b')=\frac{1}{b}\delta(b-b')+O(\frac{1}{k})\,,
\ee
to find that
\be
A(s_1,b)=\int_0^{\sqrt{2}} dy\, y J_0(\sqrt{s_1} b y) {\mathcal M}(s_1,-s_1 y^2)+O(\frac{1}{\sqrt{s_1}})\,.
\ee
This enables us to consider the impact parameter representation in terms of only the typically real part of the amplitude in the high energy limit. One can in fact use this result and the bound on $|\mathcal M|$ which follows for typically real functions to bound $|A|$. We proceed as follows. First consider the situation without any massless pole. We use $|{\mathcal M}|\leq |\alpha_1(a)||x|/27$ which follows from the distortion theorem in eq.(\ref{disto}) on ignoring the constant term in the high energy limit. We use $|\alpha_1(a)|=27\mW_{10}|a w_{01}+1|\leq 81 \mW_{10}$. Further we can check that $|J_0(x)|\leq \sqrt{\frac{2}{\pi x}}$. These lead to
\be
|A(s_1,b)|\leq 3 \mW_{10} s_1^2 \int_0^{\sqrt{2}} dy\, y (1-y^2+y^4) \sqrt{\frac{2}{\pi \sqrt{s_1} b y}}\lesssim 3.31 \mW_{10} \frac{s_1^{\frac{7}{4}}}{\sqrt{b}}\,.
\ee 
In the presence of the massless pole (we are considering here $d=4$), we will have an extra contribution due to $8\pi G/(s_1 s_2 s_3)$. This term will be logarithmically divergent at $y=0$ and we will need to regulate it by putting $y=\epsilon$. Using the triangle inequality, we get an extra contribution $8\pi G \ln \epsilon/s_1^3$, symptomatic of the infrared divergences,  to the above result. 
\section{Two channel symmetric case}
We will now analyse a situation where we do not have 3-channel symmetry but only 2-channel symmetry. The reason for looking at this is to identify the correct analog of the $\tilde z$ variable that played a crucial role in our earlier analysis. We want to examine what changes in the analysis happens compared to the 3-channel case. We will find certain interesting and important differences. 
\subsection{Set-up}
We consider 2-2 scattering amplitude $\mathcal{M}(s,t)$ for $a b \rightarrow a b$ scattering. We assume the amplitude has the following properties:
\begin{itemize}
\item {\bf Causality:} $\mathcal{M}(s,t)$ is analytic modulo poles and branch cuts on the real axis.
\item {\bf Polynomial boundedness:} For a fixed $t$ and  $|s|\rightarrow \infty$, $|\mathcal{M}(s,t)|= o(s)$. \\
This is technical choice we make to simplify our analysis and the $o(s^2)$ case can also be treated similarly and also leads to typical-realness as we show in appendix \eqref{appF}.
\item{\bf Unitarity:} The amplitude admits a well defined partial wave expansion 
\be
\mathcal{M}(s,t)= \Phi(s)\sum_{\ell=0}^{\infty}\left(2\ell+2 \a\right) f_\ell(s)C^{(\a)}_{\ell}\left(\cos \theta \right) \,,
\ee
with $0\le |f_{\ell}(s)|^2 \le 2 \Im f_{\ell} \le 1$. In the above we have used $\a =\frac{d-3}{2}$ and $\Phi(s)=2^{4\alpha+3}\pi^\alpha \Gamma(\a)\frac{\sqrt{s}}{(s- \mu)^\a}$. 
\item {\bf 2-channel crossing symmetry:} This follows from the above assumptions if we assume there is a mass gap.
\end{itemize}
 The assumptions allow us to write down a {\it once}-subtracted dispersion relation for $\mathcal{M}$ which will serve as link between UV and IR physics. We will follow the same procedure as the fully-crossing symmetric case to get bounds.  
\subsection{Dispersion relation in QFT}
In appendix (\ref{appB}) we derive a parametric dispersion relation for the two-channel crossing symmetric amplitudes following closely \cite{AK, ASAZ}. 
The two-channel symmetric dispersion relation with $s_3=u-\frac{\mu}{3}$ held fixed is given by:
\be
\label{disp2}
{\mathcal  M}_0(s_1,s_2) =\alpha_0 + \frac{1}{\pi}  \int_{\frac{2 \mu}{3}}^{\infty} \frac{ds'}{s'}
{\bar{\mathcal A}}\left(s_1',s_2(s_1^{'},a)\right) H(s_1',s_1,s_2)\,,
\ee
where $\mathcal{A}\left(s_1; s_2\right)$ is the s-channel discontinuity and the kernel is given by
\be
\begin{split}\label{Hs2p2}
&H\left(s_1^{\prime} ; s_1, s_2\right)=\left[\frac{s_1}{\left(s_1^{\prime}-s_1\right)}+\frac{s_2}{\left(s_1^{\prime}-s_2\right)}\right],\\
&s_{2}\left(s_1^{\prime}, a\right)=\frac{ a~s_1'}{s_1'-a} \,.
\end{split}
\ee
As alluded to earlier in deriving this dispersion relation we have assumed the amplitude to behaves like $o(s)$ rather than $o(s^2)$. In the two-channel symmetric dispersion, we use the absorptive part
\be\nonumber \label{partw2}
\begin{split}
&\mathcal{A}\left(s_1,s_2^{(+)}(s_1,a)\right)=\Phi(s_1)\sum_{\ell=0}^{\infty}\left(2\ell+2\a\right)a_\ell(s_1)C^{(\a)}_{\ell}\left(\xi(s_1,a)\right)\,,\\
&\xi(s_1,a)=\xi_0\left(\frac{s_1+a} {s_1-a}\right),~\xi_0=\frac{s_1}{(s_1-2\mu/3)}\,.
\end{split}
\ee
As in fully-symmetric case we assumed a EFT with cutoff $\Lambda^2=\delta+2 \mu/3$ then we have a low-energy expansion of the amplitude given by 
\be
\label{Wdef2}
\mathcal{M}_{0}(s_1,s_2)=\sum_{p, q=0}^{\infty} {\mathcal W}_{p q} x^{p} y^{q}\,,
\ee
with $x=\left(s_1+ s_2\right)$, $y=s_1 s_2 $. Here $s_1=s-\mu/3, s_2=t-\mu/3$ with fixed $s_3=u_0-\mu/3$ (with $u_0 <\mu$ to avoid crossing the $u$-channel cut), with $s,t,u$ being the usual Mandelstam variables satisfying $s+t+u=\mu$ so that $s_1+s_2+s_3=0$.

As before,  we can Taylor expand this expression around $a=0$ and match powers to obtain\footnote{$\Phi(s_1)=2^{4\alpha+3}\pi^\alpha \Gamma(\a)\frac{\sqrt{s_1+\frac{\mu}{3}}}{(s_1-\frac{2\mu}{3})^\a}$. }
\be
\begin{split}\label{Belldef}
&\mW_{n-m,m}=\int_{\delta+\frac{2\m}{3}}^{\infty}\frac{d s_1}{2\pi s_1^{n+m+1}}\Phi(s_1)\sum_{\ell=0}^{\infty}\left(2\ell+2\a\right)a_\ell(s_1)\mathcal{B}_{n,m}^{(\ell)}(s_1)\,,\\
&\mathcal{B}_{n,m}^{(\ell)}(s_1)=2\sum_{j=0}^{m}\frac{(-1)^{1-j+m}p_{\ell}^{(j)}\left(\xi_{0}\right) \left(2 \xi_{0}\right)^{j}(2 j-m-n)\Gamma(n-j)}{ j!(m-j)!\Gamma(n-m+1)}\, ~{\rm for}~ n\ge 1.
\end{split}
\ee
Here odd spins also contribute unlike the fully symmetric case and $p_{\ell}^{(j)}(\xi_0)=\partial^j C_\ell^{(\a)}(\xi)/\partial\xi^j|_{\xi=\xi_0}$ is much simpler than the fully symmetric case  due the absence of $\sqrt{\xi}$ argument for the Gegenbauers. Furthermore  $p_\ell^{(j)}(\xi_0)\geq 0$ for $\xi_0\geq 1$, since $C_{\ell-j}^{(\a+j)}(\xi_0)>0$ for $\xi_0 = \frac{s_1}{s_1-\frac{2\mu}{3}}>1$. \\\\\\
This implies that the signs of other factors in $\mathcal{B}_{n,m}^{(\ell)}$ determine the sign of a particular term in the $\ell $-sum and this has the following important consequences similar to the once observed for the fully symmetric case which we list below:
\begin{enumerate}
\item \noindent We first note as in 3-channel case that for $m=0$ we have:
\bea \label{posm0}
\mW_{n,0} =\int_{\frac{2\m}{3}}^{\infty}\frac{d s_1}{\pi s_1^{n+1}}\Phi(s_1)\sum_{\ell=0}^{\infty}\left(2\ell+2\a\right)a_\ell(s_1) C_\ell^{(\a)}(\xi_0) >0\,,
\eea
\item {\bf Locality/Null constraints:} The expression \eqref{Belldef1} for $\mW_{n-m,m}$ is valid for $n\ge 1$ and any $m$. However, in a local theory we know that we cannot have negative powers of $x$ .We thus need need to impose the following {\it Locality constraints} which we denote by $N_c^{(2)}$.\\

\vspace*{-40 pt}
\begin{centering}
\bea \label{nc}
\mW_{n-m,m}=0, ~ \forall ~ n <m
\eea
\end{centering}
We do not attempt to solve these constraints and obtain bounds in this work. 
We merely list a few of them in the context of EFT where $\Lambda \gg \mu$, here and refer the interested reader to Appendix \ref{nullapp2} for further details. In this regime the $\mathcal{B}^{(\ell)}_{n,m}$'s \eqref{Belldef} simplify considerably and we get:
\begin{itemize}
\item The first non-zero contribution to $\mW_{n-m,m}$ is from $\ell=n$.
\item An infinite number of higher-spin partial waves are non-zero in the partial wave expansion.
\item $\ell=1$ is necessarily present.
\end{itemize}
\item{\bf Positivity constraints:} We can take linear combinations of different $\mW_{p,q}$ with specific coefficients to a generate manifestly positive result. These yield:
\bea \label{ps2}
\sum_{r=0}^m \chi_n^{(r,m)} (\mu,\delta) \mW^{\delta}_{n-r,r} &\ge& 0,  \\
0\le \mW^{\delta}_{n,0}&\le& \frac{1}{\left(\delta + \frac{2 \mu}{3}\right)} \mW^{\delta}_{n-1,0} ~~~~~~~~~~~~~~~~~~~~~~~~~~~~~~~{\rm for}~n\ge2.
\eea
We work with only the $\delta =0$. The dictionary for the EFT case is to replace $\mu\rightarrow 3\Lambda^2/2$ as in the 3-channel case.
The $\chi_n^{(r,m)} (\mu,\delta) $ satisfy the recursion relation:
\bea
\chi_n^{(m,m)} (\mu,\delta) &=&1,\nonumber\\
\chi_n^{(r,m)} (\mu,\delta) &=& \sum_{j=r+1}^m (-1)^{j+r+1} \chi_n^{(j,m)} \frac{{\mathscr U^{\alpha}}_{n,j,r} (\delta + \frac{2 \mu}{3})}{{\mathscr U^{\alpha}}_{n,r,r}(\delta + \frac{2 \mu}{3})},
\eea
with  ${\mathscr U}^{\alpha}_{n,m,k}=  \frac{2^j (2 \xi_0)^j (\alpha)_k (m+n-2j)\Gamma(n-j)}{(\frac{2}{3})^{m+n+1}j!(m-j)!(n-m)!}$ and $\alpha=\frac{d-3}{2}$. For a derivation of these see app\ref{appH}.
\end{enumerate}
We now prove that the amplitude is typically real by recasting the dispersion relation in Robertson form and proceed to the bounds.
 \subsection{Amplitude and Typical Real-ness}
We define the variables  $z,~a$ with $s_k = a- a\frac{(z-z_k)^2}{(z+z_k)2}$ where $z_k = \pm 1$ are the square roots of unity , to make the 2-channel crossing symmetry manifest and also to connect with GFT. We can further relate $z,~a$ variables to $x,y$ using the relations $x=(s_1+s_2)= \frac{-16 a z^2}{(-1+z^2)^2}$ , $y=  s_1 s_2= \frac{-16 a^2 z^2}{(-1+z^2)^2}$ and $a=\frac{y}{x}$. Similarly, the crossing symmetric kernel can be expanded in ${\tilde z}=z^2$ as:
\bea
 H(s_1',{\tilde z}) =\frac{16 a \left(2 a-s_1'\right)}{\left(s_1'\right)^2} \frac{\tilde z}{1-\left(16 \frac{a}{s_1'} \left(1-\frac{a}{s_1'} \right)-2\right)\tilde z +\tilde z^2}= \sum_{n=0}^{\infty} \beta_n(a,s_1') 
\eea
which is a M\"{o}bius transformation of the Koebe function and hence univalent in the disk $|{\tilde z}|<1$ provided $\bigg|16 \frac{a}{s_1'} \left(1-\frac{a}{s_1'} \right)-2\bigg|<2$ to ensure there are no singularities inside the disk $|{\tilde z}|<1$ and this gives us $0<a<\frac{\mu}{3}$ since $s_1' >\frac{2 \mu}{3}$. Furthermore, since the coefficients are all real this kernel is also a {\it typically-real} function. We can see this as we did for the fully crossing symmetric case  by observing the normalised version of the kernel satisfies the Rogosinski bounds \eqref{rns}:
\bea
\beta_n(a,s_1') =\frac{2 a  \left(\left(\frac{8 a^2-8 a s_1'+4
   \sqrt{a \left(a-s_1'\right) \left(s_1'-2
   a\right)^2}+\left(s_1'\right){}^2}{\left(s_1'\right)^2}\right)^
   n-\left(\frac{8 a^2-8 a s_1'-4 \sqrt{a \left(a-s_1'\right)
   \left(s_1'-2
   a\right)^2}+\left(s_1'\right)^2}{\left(s_1'\right)^2}\right)^
   n\right)}{\sqrt{a \left(a-s_1'\right)}}\,.
 \eea \\\\
 Notice that $\beta_0(a,s_1')=0 $ and $\beta_1(a,s_1')=\frac{16 a \left(2 a-s_1'\right)}{\left(s_1'\right)^2} $. The ranges of   $s_1$ and $a$ were $\frac{2 \mu}{3}\le s_1'<\infty$ and $0 < a <\frac{\mu}{3}$ and we infer that   $\beta_1(a,s_1')<0$ for $0< a\le\frac{\mu}{3}$. \\
 \begin{figure}[h!]
 \centering
  \includegraphics[width=0.8 \textwidth] {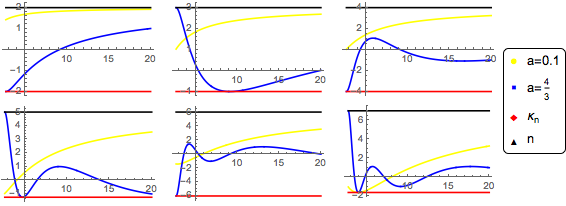}
 \caption{$\beta_n/\beta_1$ vs $s_1$ for $n=2,3,4$ and $n=5,6,7$, from left to right. The above plots show that the Rogosinksi bounds are respected the 2-channel symmetric amplitude with $\mu =4$ for the full range of $a$ namely $0< a< \frac{\mu}{3}$.}\vskip 0.5cm
 \end{figure}
 \vspace*{3 pt}
As in the three-channel case, we can show that the whole amplitude ${\mathcal M}({\tilde z},a)$, not just the kernel, is typically-real. We do this be recasting the dispersion relation in the Robertson representation as we had done for the 3-channel case. By setting  $16 \frac{a}{s_1'} \left(1-\frac{a}{s_1'} \right)-2 =2 \xi$ and making a change of variable to $\xi$ in \eqref{disp2} we get:
\bea
\frac{{\mathcal {\overline M}}_0(z,a) -\alpha_0}{\int_{-1}^1 d\xi {\bar{\mathcal A}}\left(\xi,s_2(x,~a)\right)} = \int_{-1}^1 d\mu(\xi)  \frac{\tilde z}{1-2 \xi \tilde z+\tilde z^2}\,,
\eea
where $d\mu(\xi) = \frac{{\bar{\mathcal A}}\left(\xi,s_2(\xi,~a)\right) d\xi}{ \int_{-1}^1 d\xi {\bar{\mathcal A}}\left(\xi,s_2(\xi,~a)\right)} $, is a non-decreasing probability measure.

 \vspace*{5 pt}
{\it This proves  that  2-2 scattering amplitude of massive particles with 2-channel symmetry  is typically real, when the absorptive part is positive.}
 
 We can  now expand the amplitude directly in $\tilde z,~a$ variables :
  \bea
 {\mathcal M}({\tilde z},a) =\sum_{n=0}^{\infty} \alpha_n(a) a^{n} {\tilde z}^{2n},
 \eea
 where $\alpha_n(a)$ is a polynomial in $a$.
 \bea
 \alpha_p(a) a^{p} =\sum_{n=0}^p\sum_{m=0}^n {\mathcal W}_{n-m,m} a^m (-1)^{p-n} (-16)^n a^{n} \left( {-2n \atop p-n} \right)\,.
 \eea
 \vspace*{-10 pt}
This leads to:
 \bea
 a^{ n} \alpha_n(a) =\frac{1}{\pi} \int_{\frac{2 \mu}{3}} ^{\infty} \frac{ds_1'}{s_1'} {\mathcal A}(s_1',s_2(s_1',a)) \beta_n(a,s_1')\,.
 \eea
 We note in particular that from the above expression that $\alpha_1(a)= -16 \mW_{10} ( a w_{01}+1)$ has the same sign as  $\beta_1$ as the ${\mathcal A}(s_1,s_2(s_1,a))>0$  for  $s_1 \in [\frac{2 \mu}{3},\infty) $ and  $a \in \left(0, \frac{\mu}{3} \right) $, which implies that $a ~w_{01}+1>0$ identically analogously to 3-channel case. 

  \subsection{$\mW_{pq}$ bounds} \label{bounds}
By defining the {\it schlict} function   $F({\tilde z},s_1,a)={\tilde z} +\sum_{n=2}^{\infty} \frac{\beta_n(s_1,a)}{\beta_1(s_1,a)} {\tilde z}^n$ corresponding to the kernel $H({\tilde z},s_1,a)$ we can see that
\bea
F({\tilde z},s_1,a)=\frac{{\tilde z}}{1+ \gamma{\tilde z}+z^2 }
\eea
with $\gamma= \left(16 \frac{a}{s_1'} \left(\frac{a}{s_1'}-1 \right)+2\right)$as argued earlier this a typically-real function for $\frac{2 \mu}{3} \le s_1<\infty$ and $0<a \le \frac{\mu}{3}$ and thus we can apply the {\it Bieberbach-Rogosinski} inequalities to it: 
\bea
 -\kappa_n \le \frac{\beta_n(s_1,a)}{\beta_1(s_1,a)}\le n ~{\rm for} ~n \ge 2.
\eea
However there are some crucial differences between the 3-channel case and the 2-channel case which do not allow us to bound all the $w_{pq}$'s using Bieberbach-Rogosinski inequalities alone as we shall explain in the next section. We illustrate this by looking at the $n=2$ case. We refer the reader to \ref{n235} for the analysis and bounds for higher $n$ cases. 
\bea
{\bf n=2:}~&-&2~(a w_{01}+1 )\le -16 a(a^2 w_{02}+ a w_{11}+w_{20})+2 a
   w_{01}+2\le 2~(a w_{01}+1 ) ,\\\nonumber
\eea
where since  $W_{1,0} >0$ from \eqref{posm0}  we have normalised all $\mW$'s using $w_{pq} =\frac{\mW_{pq}}{\mW_{10}}$ and $\alpha_1(a) <0 $ this gives a bound 
\bea \label{bound1}
-\frac{3}{\mu} \le w_{01}.
\eea
We can check that our bound $-0.75 \le w_{01}$ is respected by the open string amplitude $w_{01}\approx -0.59$ and 1-loop $\phi^4$ with $w_{01}\approx -0.289$ from the table in appendix (\ref{appG}). 

\subsection{Differences from the fully symmetric case}
Unlike the fully-symmetric case, the Bieberbach type inequalities do not directly lead to two-sided bounds on the Wilson coefficients in the two-channel case. This can be argued as follows.
Similar to the fully-symmetric case the Rogosinski/Bieberbach inequalities are always of the form:
\bea\label{nu}
\Bigg|\frac{P_{2n-1}(\{ w_{p,q}\} ,a)}{(a w_{01}+1)}\Bigg| \leq n
\eea  
where $P_{2n-1}(w_{p,q},a)$ are degree $2n-1$ polynomials in $a$ with coefficients that are $w_{p,q}$'s. However now we do not have  an upper bound on $|a w_{01}+1|$. Thus the polynomial $P_{2n-1}(\{w_{p,q}\},a)$ is no longer bounded and the Markov brothers' inequality \cite{Markov1,Markov2} is not applicable here. 

It is obvious that  $w_{pq}$'s can take arbitrarily large values still satisfying the above inequalities. For example $w_{01}$ can take arbitrarily large values for sufficiently small $a$ close to zero. Thus, it is obvious that the $w_{pq}$'s are unconstrained by the Bieberbach-Rogosinski inequalities! \footnote{Using considerations of low spin dominance could give an improved range of $a_{min}<0<a<\frac{16}{9}$ which would lead to two sided bounds. We do not consider this in this work (see \cite{CGHRS} for related discussion in the fully crossing symmetric case.). }However some of the $w_{pq}$'s do get one-sided bounds when we apply positivity conditions as we shall show next.
\subsection{Numerical bounds}
As we explained in the previous section we do not two sided bounds  for all $w_{pq}$'s. However, we do get a few bounds in this case by applying the positivity bounds which we call $PB_C^{(2)}$ \eqref{ps2}, which are the direct analogs of the inequalities derived in \cite{ASAZ} for the two channel case
We review  the derivation of the above expressions in appendix (\ref{appH}).
%
The $n=2$ positivity bounds  and results are :
\bea
 && w_{11}+ \frac{9}{2} w_{20} \ge 0 ,~~ w_{02}+3 w_{11}+9w_{20} \ge0, \nonumber\\\\
 &-&1.6875\leq w_{11},~ 0\leq w_{20} \leq 0.375,~-0.75\leq w_{01}, \nonumber
\eea
The higher $n$ results are provided in \ref{n235}.
 An important difference from the 3-channel situation is that the Rogosinski-Bieberbach inequalities in this case are not the reason for the constraints, rather the $PB_C^{(2)}$ inequalities are instrumental in the bounds. Thus we conclude, that in this instance, the linear inequalities arising from Typically Real-ness are not of use.
We expect that when we use the non-linear constraints from $T_R$ considerations, to be discussed in section 4, we will further constrain the $w_{pq}$-space. We  emphasise that we have not imposed locality/null constraints in the above analysis. 
\subsection{Comparison with known results}
In \cite{yutin} it was shown that the intersection of the EFT-hedron  for a fixed derivative order $k$  with the monodromy plane leads to a small island around the open string solution which further converged to the string solution as $k$ was increased. We would now do a similar analysis using the GFT to see if we get similar results. 

We begin with a lightning review of the mondoromy conditions in \cite{yutin} and recast it into our language.  The low energy expansion of a generic amplitude in $s_1$ and $s_2$ with possible massless poles: 
\bea
{\mathcal A}(s_1,~s_2) = -\frac{1}{s_1 ~s_2}+\frac{b}{s_1}+\frac{b}{s_2}+\frac{c ~s_2}{s_1}+\frac{c ~s_1}{s_2}+\sum_{k,q \ge 0} g_{k,q} s_1^{k-q} s_2^q
\eea
where, cyclic symmetry implies $g_{k,q}=g_{k,k-q}$.
The monodromy relation which follows from the fact that the worldsheet  string integrand is permutation invariant:
\bea\label{mon}
{\mathcal A}(s_1,-s_1-s_2) + e^{i ~\pi~ s_1} {\mathcal A}(s_1,s_2)+ e^{i~ \pi~ (s_1+s_2)}{\mathcal A}(s_2,-s_1-s_2)=0
\eea
and constrains $g_{p,q}$ strongly. Since the coefficients of $\frac{s_2^{2n}}{s_1}$ this implies that $b=0$. Furthermore $c=0$ due to unitarity.  The sub-leading orders  also force $g_{2n,0} = \zeta(2n+2)$  and gives the following relations up to $k=4$:
\bea 
g_{2,1}= \frac{\pi ^4}{360},~g_{3,1}=2 g_{3,0}-\frac{1}{6} \pi
   ^2 g_{1,0},~g_{4,2} = 2 g_{4,1}-\frac{\pi ^6}{15120}\,.
\eea
At $k=4$ we have  $g_{1,0}$, $g_{3,0}$ and $g_{4,1}$ as undetermined parameters that need to be constrained.
We can recast the mondoromy conditions in our language by rewriting  $\mW_{p,q}$ in terms of $g_{p,q}$ by comparing \eqref{mon} with
 \bea
{\mathcal A}(x,~y) = -\frac{1}{y}+\sum_{p,q \ge 0} \mW_{p,q} x^{p} y^q
\eea
where $x=s_1+s_2$ and $y=s_1~s_2$.
\bea
\mW_{0,0}&=& \frac{\pi
   ^2}{6},\mW_{0,1}= -\frac{7
   \pi ^4}{360},\mW_{0,2}=
   \frac{31 \pi
   ^6}{15120},\mW_{0,3}= 2
   g_{6,1}-2
   g_{6,2}+g_{6,3}-\frac{\pi
   ^8}{4725}, \nonumber \\
   \mW_{1,0}&=&
   g_{1,0},\mW_{1,1}=
   -\frac{1}{6} \pi ^2
   g_{1,0}-g_{3,0},\mW_{1,2}=
   5 g_{5,0}-3
   g_{5,1}+g_{5,2},
   \mW_{2,0}=
   \frac{\pi
   ^4}{90},\nonumber \\ \mW_{2,1}&=&
   g_{4,1}-\frac{4 \pi
   ^6}{945},\mW_{2,2}= -4
   g_{6,1}+g_{6,2}+\frac{\pi
   ^8}{1050},\mW_{3,0}=
   g_{3,0},\mW_{3,1}=
   g_{5,1}-5
   g_{5,0},\nonumber \\ \mW_{4,0}&=&
   \frac{\pi
   ^6}{945},\mW_{4,1}=
   g_{6,1}-\frac{\pi
   ^8}{1575},\mW_{5,0}=
   g_{5,0},\mW_{6,0}=
   \frac{\pi ^8}{9450} 
\eea
We  impose  the following conditions after noting our convention for the open string  $\frac{2 \mu}{3}=1$ along with manually imposing $g_{5,0} =\zeta(7)$. We do this to see what two dimensional region is carved out for the space of $g_{1,0}$ and $g_{3,0}$. Note that we impose both linear conditions $TR_U$, $PB_C^{(2)}$ and also nonlinear conditions  $D_k$ which we will introduce in \eqref{det} in the next section.

\bea\label{yt2}
{\bf D_{3}:}&& \mW_{1,0}
 \mW_{3,0}-\mW_{2,0}^2>0,\nonumber\\
 {\bf \alpha_1 <0:}  && -2<\frac{\mW_{1,0}}{\mW_{0,1 }},\nonumber\\
 {\bf D_5:}&& -\mW_{3,0}^3+2 \mW_{2,0} \mW_{4,0} \mW_{3,0}+\mW_{1,0} \mW_{5,0}
   \mW_{3,0}-\mW_{1,0} \mW_{4,0}^2-\mW_{2,0}^2 \mW_{5,0}>0,\nonumber \\
 {\bf TR_U:} && -2\leq
   -\frac{16 a^3 \mW_{0,2}}{a
   \mW_{0,1}+\mW_{1,0}}-\frac{16
   a^2 \mW_{1,1}}{a
   \mW_{0,1}+\mW_{1,0}}-\frac{16 a
   \mW_{2,0}}{a
   \mW_{0,1}+\mW_{1,0}}+\frac{2 a
   \mW_{0,1}}{a
   \mW_{0,1}+\mW_{1,0}}+\frac{2
   \mW_{1,0}}{a
   \mW_{0,1}+\mW_{1,0}}\leq
   2 \nonumber  \\
  {\bf PB_C:} &&\mW_{1,1}+\frac{9}{2} \mW_{2,0} \ge 0,~\mW_{0,2}+3\mW_{1,1}+9\mW_{2,0}\ge 0,~ \mW_{1,0}>0,~0\le \mW_{2,0}\le \mW_{1,0},~0\le \mW_{2,0}\le \mW_{3,0} \nonumber \\
\eea

which gives us the following plot:
\begin{figure}[H]
  \centering
  \begin{minipage}[b]{0.45\textwidth}
    \includegraphics[width=\textwidth]{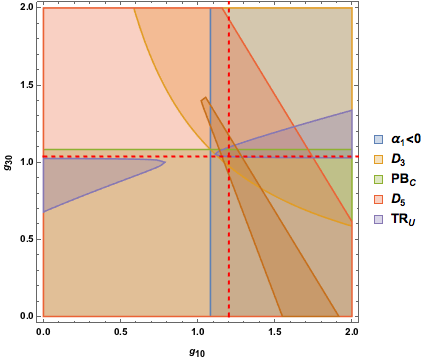}
    \label{(a)}
  \end{minipage}
  \hfill
  \begin{minipage}[b]{0.38\textwidth}
    \includegraphics[width=\textwidth]{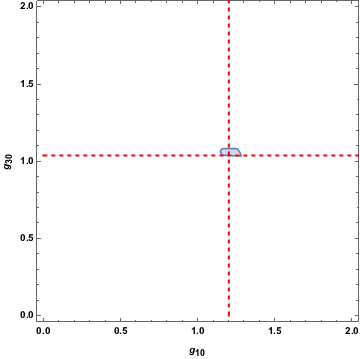}
    \label{(b)}
  \end{minipage}
 \vspace*{-30 pt}  \caption{The intersection of the red lines indicates the open string solution.}
\end{figure} 
 \

 We note again that the $TR_U$ conditions have to hold for {\it all $a$} in $(0,\frac{1}{2})$ and have been implemented above by discretising the $a$ with different step sizes. Reduction in the step size leads to smaller regions and the above plot is for step size $\frac{1}{100}$. The plots are analogous to the region obtained for $k=4$ in \cite{yutin}. We note that the range of parameters is $1.143 \le g_{1,0}\le 1.286$ and $1.033 \le g_{3,0}\le 1.082$. Since the string values are $g_{1,0} =\zeta(3)$ and $g_{3,0}=\zeta(5)$, the results are within $12\%$ and $4\%$ of the exact answer respectively.

   \subsection{Bounds in the presence of poles} 
   We now consider an amplitude with a massless pole at the origin
   \bea
{\mathcal M}_{sym} &=&- \frac{g_{ym}^2}{s_1 s_2 } + g_0+ g_2 (s_1+s_2) + g_3 s_1 s_2 +\cdots\nonumber\\
&=& -\frac{g_{ym}^2}{y} + g_0 +2 g_2 x+ g_3 y+\cdots \nonumber\\
-\frac{{16 a^2\mathcal M}_{sym}}{g_{ym}^2}&=& -\frac{1}{ z} +\left(2-16 a^2 \hat g_0\right) +\left(512 a^3 \hat g_2 -1+256 a^4 \hat g_3 \right) z+\cdots, \nonumber
\eea
where $\hat g_i =\frac{g_i}{g_{ym}^2}$.
This is in the Goodman class $TM^{*}$ with $\beta_0=\left(2-16 a^2 \hat g_0\right)$ and $\beta_1=(512 a^3 \hat g_2 -1 + 256 a^4 \hat g_3)$ but as before we cannot bound $\beta_0$ directly in this form so we again look at its dual in $TM$ namely $-\frac{g_{ym}^2}{{16 a^2\mathcal M}_{sym}}$and apply the Rogosinksi bounds \cite{rogo}, which give:
\bea
-2&\le&\left(16 a^2 \hat g_0-2\right)\le 2, \nonumber\\
-1&\le& \left(256 a^4  \hat g_0^2-256 a^4
    \hat g_3+512 a^3  \hat g_2-64 a^2
    \hat g_0+3\right)\le 3 \,.
\eea
We will just use the first equation here. This gives
 \bea
 0&\le& \hat g_0 \le \frac{1}{4 a^2} \,.
 \eea
This would bound $0\le \hat g_0\le 1$ since $0<a<\frac{1}{2}$. For the open string case, the range of $a$ is further reduced to $0<a<\frac{1}{4}$ if we demand that $-\frac{1}{{\mathcal M}}$ have no poles inside the unit disk as we shall argue now. We start with
\bea
-\frac{1}{{\mathcal M(s_1,s_2)}}=-\frac{\Gamma(1-s_1-s_2)}{\Gamma(1-s_1)\Gamma(1-s_2)}\,.
\eea
Rewriting the above in $\tilde z , a$ variables we note that the numerator has poles corresponding to $1+2(-1+8a) \tilde z +\tilde z^2=0$ and if $z_{\pm}$ are the roots then since $z_{+}z_{-} =1$ we need the roots to be complex to prevent having a real root inside the disk $|\tilde z|<1$ which gives $0<a<\frac{1}{4}$ as alluded to earlier. This gives 
\bea
 0\le \hat g_0 \le 4 \,,
\eea
which is respected by the open string which has $\hat g_0 =\pi^2/6\approx 1.644$.

   \section{Non-linear constraints and the EFT-hedron}
   So far, we have been focusing on the conditions that are linear in the Wilson coefficients. However, there are also interesting non-linear conditions, analogous to the Grunsky inequalities\footnote{The Grunsky inequalities feature in other fascinating areas of mathematical physics as recently discussed in \cite{McKay:2021ejv}. Positivity in mathematics via moment problems, as we will discuss in this section, appears in several areas; for a recent survey, see \cite{khare}.} in the univalent case, discussed in \cite{HSZ}. Since we have established that the amplitude is $T_R$, we will move to using the appropriate conditions for $T_R$ functions and not the Grunsky inequalities. In this section, we will examine some of these conditions. We begin by quoting the theorem we need which can be found in \cite{rogo}.
   
 \subsection{Toeplitz determinant conditions} \label{toe}
 In ${\mathbb R}^{n-1}$ if the point $P_n = (a_2, a_3, ...,a_n)$ consists of the coefficients of  a function $f (z)$ in $T_R$, then $P_n$ is in the closed convex hull $K_n$ of the $(n-1)$- dimensional curve
\bea
 \frac{\sin 2 \vartheta}{\sin \vartheta},~\frac{\sin 3 \vartheta}{\sin \vartheta} ,\cdots,\frac{\sin n \vartheta}{\sin \vartheta} ;~~~ 0\le \vartheta\le \pi \nonumber
\eea
 and conversely, every point of $K_n$ corresponds to a function in $ T$.
The points of $K_n$ are characterized by the fact that for them the $n - 1$ Toeplitz determinants
\bea \label{det}
D_k =
\begin{vmatrix}
2 & a_{2} & a_{3}-a_1 & \dots & a_{n}-a_{n-2} \\ 
a_2 & 2 & a_2 & \dots & a_{n-1}-a_{n-3} \\
a_3-a_1 & a_2 & 2 & \dots \\
\hdotsfor{5} \\
a_n-a_{n-2} & a_{n-1}-a_{n-3}  & \dots &\dots & 2
\end{vmatrix} 
\eea
for $2\leq k\leq n$, $a_0=0, a_1=1$ are all are nonnegative.  We will give a short derivation of this property using the trigonometric moment problem \cite{pinkus,shohat,schmudgen}. 

To see this, we begin with the Robertson representation  for a typically-real the function $f(z)=z+\sum_{n=2}^{\infty} a_n z^n$ and by observing that the kernel is the generating function of Chebyshev polynomials :
\bea
f(z)& =&\int_{0}^{\pi}d \mu(x)  \frac{z}{1- 2 z \cos x+ z^2} \nonumber\\
&=& \int_{0}^{\pi} d \mu(x) \sum_{n=1}^{\infty} \frac{\sin n x}{\sin x} z^n
\eea
By assuming uniform convergence and interchanging the sum and the integral we can read off the coefficients $a_n$ as:
\bea
a_n = \int_{0}^{\pi} d \mu(x)  \frac{\sin n x}{\sin x}\,.
\eea
Since $\mu(x)$ is a probability measure by organising $\vec{a}=(a_2,a_3,\cdots,a_n)$, we see that $\vec{a}$ has to belong to the convex hull of the curve $K_n$ as claimed above.

To see why the second part is true we need to use \eqref{142}, the one-one correspondence between typically-real functions and Caratheodory functions, $\phi(z)$, also called functions with positive real part.
\bea\label{142}
\phi(z)=\frac{(1-z^2)}{z} f(z) = 1+a_2 z+\sum_{n=2}^{\infty} (a_{n+1}-a_{n-1}) z^n\,.
\eea
We shall argue that every Caratheodory function provides a solution to the {\it Trigonometric moment} problem and thereby directly gives us  \eqref{det}. 
Every Caratheodory function $\phi(z)$ has a unique Herglotz representation similar to the Robertson representation of the typically-real functions 
\bea \label{143}
\phi(z) &=& \int_{0}^{2\pi}d \nu(x) \frac{e^{i x} +z}{e^{i x}-z}  \nonumber\\
&=& 1+2 \sum_{n=1}^{\infty} \int_{0}^{2\pi}d \nu(x)  e^{-i n x} z^n,
\eea
where, $\nu(x)$ is a probability measure.
Comparing \eqref{142} with \eqref{143} we get:
\bea
a_2 &=& 2 \int_{0}^{2 \pi}d \nu(x)  e^{-i x}, ~~\nonumber\\
a_{n+1}-a_{n-1}&=& 2  \int_{0}^{2 \pi}d \nu(x)  e^{-i n x} \quad \forall n\ge 2\,.
\eea 
In other words  the sequence $\{1,\frac{a_2}{2},\cdots,\frac{a_{n+1}-a_{n-1}}{2},\cdots\}$ provides a solution to trigonometric moment problem. A necessary condition for a sequence $\{\alpha_n\}$ to be a solution to the trigonometric-moment problem with respect to $\mu(x)$ and non-finite support  is that the infinite Toeplitz matrix $\{\alpha_{i-j}\}_{i,j=0}^{\infty}$ with $\alpha_{-i} =\bar{\alpha_i}$ is positive definite. If we only want a solution to the truncated trigonometric moment problem then the matrix is allowed to be positive semi-definite\cite{pinkus,shohat,schmudgen}.

In our case this leads to \eqref{det} since all $a_n$'s are real and after multiplying each row by $2$. Note that a positive definite matrix  has its leading principal minors non-negative and these are precisely the $D_k$'s listed in \eqref{det}.
\subsubsection{ Connection with the crossing symmetric EFT-hedron}
In this section,  we will try to connect with the EFT-hedron \cite{nima}.
 There is a crucial difference in what we will discuss below as compared to \cite{nima}, since in our approach crossing symmetric is inbuilt with the constraints coming from the locality/null constraints. Thus unlike \cite{nima} where either a 3-channel EFT-hedron or a two channel symmetric EFT-hedron is discussed by intersection with the suitable crossing plane, our discussion is applicable using the crossing symmetric formulae for the Wilson coefficients in eq.\ref{Belldef1} and \ref{Belldef} for the fully-symmetric and 2-channel symmetric cases respectively.  We shall discuss only the fully symmetric case here for simplicity. \\
First note that since $\mW_{pq}$ mutliplies $x^p y^q$, the mass dimension of $\mW_{n-m,m}$ is $2(2n+m)$. As in \cite{nima}, we define $\Delta\equiv 2n+m$. Now consider as an example $\Delta=10$ which corresponds to the possibilities $(n,m)=(1,8),(2,6),(3,4),(4,2),(5,0)$. This enables us to write for $\Delta=10$
\be
\begin{pmatrix} \mW_{-7,8}\\ \mW_{-4,6} \\ \mW_{-1,4} \\ \mW_{2,2} \\ \mW_{5,0} \end{pmatrix}=\sum_{\ell \in {\rm even}} P_{\ell} \begin{pmatrix} \mB_{1,8}\\ \mB_{2,6} \\ \mB_{3,4} \\ \mB_{4,2} \\ \mB_{5,0}\end{pmatrix}\,, \quad P_\ell=\int_{\Lambda^2}^\infty \frac{ds_1}{2\pi s_1^{\Delta+1}}\Phi(s_1)(2\ell+2\alpha)a_\ell(s_1)>0\,.
\ee
Now the first 3 rows above should vanish using the null constraints. So we focus only on the last 2 rows. Let us focus on $\alpha=1/2$ for definiteness. It is easy to verify that for $\ell\geq 4$ any $2\times 2$ matrix formed out of $\mB_{4,2}, \mB_{5,0}$ have positive minors! 
Specifically we mean that the matrix
\be
\begin{pmatrix} \mathcal{B}^{(\ell_1)}_{5,0} &  \mathcal{B}^{(\ell_1)}_{4,2}\\
\mathcal{B}^{(\ell_2)}_{5,0} &  \mathcal{B}^{(\ell_2)}_{4,2}\end{pmatrix} {\rm ~has~positive~minors~} \forall \ell_2>\ell_1\geq 4.
\ee
Explicitly, we find
\be
\mB_{5,0}=4\,, \quad \mB_{4,2}=30+(\ell-4)\ell(\ell+1)(\ell+5)\,,
\ee
from which it is easy to see that for $\ell\geq 4$ these entries are positive for any spin while for $\ell=2$ we have $\mB_{4,2}<0$--using the explicit formulas it is possible to check the positivity of the minors for $\ell\geq 4$ quoted above.
Thus in this simple example, there appears to be a critical spin above which the cyclic polytope picture in \cite{nima} emerges. \\

{\it 
More accurately, we should say that after subtracting out the spin-2 contribution (we can retain spin-0) from $\mW_{2,2}, \mW_{5,0}$ we will find that they lie inside a cyclic polytope.}
\\

This story persists for $\Delta=8,6$ as well and is trivial for $\Delta=2,4$ where the only possibilities are $(n,m)=(1,0)$ or $(n,m)=(2,0)$ which are positive definite for any spin. For $\Delta=11$ we have $(n,m)=(5,1), (4,3)$ after imposing the null constraints and here we find that the critical spin is 6. 
If we had not imposed the null constraints, even then a similar finding would have emerged for the $\Delta=10$ case, using the bigger $5\times 5$ matrix for $\ell\geq 8$. 

For $\Delta=12,14$ we have a $3\times 3$ matrix after imposing the null constraints. Example, for $\Delta=12$, we find
\be
\begin{pmatrix} \mathcal{B}^{(\ell_1)}_{6,0} &  \mathcal{B}^{(\ell_1)}_{5,2} & \mathcal{B}^{(\ell_1)}_{4,4}\\
\mathcal{B}^{(\ell_2)}_{6,0} &  \mathcal{B}^{(\ell_2)}_{5,2} & \mathcal{B}^{(\ell_2)}_{4,4}\\
\mathcal{B}^{(\ell_3)}_{6,0} &  \mathcal{B}^{(\ell_3)}_{5,2} & \mathcal{B}^{(\ell_3)}_{4,4}\\
\end{pmatrix} {\rm ~has~positive~minors~} \forall \ell_3>\ell_2>\ell_1\geq 8.
\ee
For $\Delta=16$ we again have a $3\times 3$ matrix but the condition is $\ell_3>\ell_2>\ell_1\geq 10$. For $\Delta=18$ we have a $4\times 4$ matrix with the critical spin now being 12. Therefore the general statement that we seem to be making is\\

{\it 
After subtracting out the contribution of a finite number of spins  we will find that the $\mW_{n-m,m}$ for $n>m$ lie inside a cyclic polytope.}

\vskip 0.5cm

The emergence of a cyclic polytope in this manner is profound,  and reminiscent of the results in \cite{nima}.
We also have the Toeplitz-determinant conditions which are the analogues of the Hankel determinant conditions for the EFT-hedron in \cite{nima}. There are a couple of key differences however between the two EFT-hedron and our case:
\begin{itemize}
\item As we mentioned earlier in our approach crossing is inbuilt and does not require intersection with the appropriate crossing plane as in the EFT-hedron.
\item Unlike the Hankel-determinant positivity conditions in \cite{nima} which implied {\it all} minors of the Hankel matrix were positive, the Toeplitz determinant conditions only translate to positivity of the {\it principal} minors of the Toeplitz matrix. However our non-linear conditions similar to the linear $TR_U$ conditions come with the $a$-parameter and the positivity of the Toeplitz matrix has to hold for the entire range of $a$ this leads to {\it infinitely many conditions} as we shall see in the next section. 
\end{itemize}
We do not attempt to implement these constraints in this work. It would be interesting to understand how one could distill non-linear constraints independent of $a$ from the Toeplitz conditions and we make a few comments regarding this now.
\subsection{Analysis of nonlinear constraints}
To start analysing these conditions, the strategy we will employ is to expand these conditions around $a\sim 0$. The simplest condition is $k=2$. For the 3-channel (2-channel works exactly analogously):
\be
a_n=\frac{a^{2n} \alpha_n(a)}{a^2 \alpha_1(a)}\,,
\ee
where $\alpha_n$ is given in eq.(\ref{alphadef}). Expanding around $a\sim 0$ and assuming $\mW_{10}>0$, we find
\be
D_2=108 w_{20} a^2\geq 0\implies w_{20}\geq 0\,.
\ee
Thus, remarkably, we have recovered a condition that we proved using the positivity of Gegenbauer polynomials! The surprise does not stop here. Expanding $D_3$ around $a=0$ gives
\be\label{w30nl}
D_3=2^3 27^3 a^6 (w_{20}w_{30}-w_{20}^3)\geq 0\implies w_{30}\geq w_{20}^2\geq 0\,.
\ee
This is a condition that was derived in \cite{tolley} using Cauchy-Schwarz inequality and also follows from the leading order Grunsky inequalities \cite{HSZ}. The next condition is equally simple.
\be\label{w40nl}
D_4=2^4 27^6 a^{12} (w_{30}-w_{20}^2)(w_{40}w_{20}-w_{30}^2)\geq 0\implies w_{40}w_{20}\geq w_{30}^2\,,
\ee
which is again a condition that follows from Cauchy-Schwarz inequality considerations in \cite{tolley} and independently leads to $w_{40}\geq 0$ since we have already shown that $w_{20}\geq 0$. This is a new derivation of this condition and does not automatically follow from the Grunsky inequalities in an obvious manner.  $D_5>0$ leads to\footnote{We get the more complicated looking condition listed in eq.(\ref{yt2}) which can be verified to be eq.(\ref{w50nl}) in Mathematica by incorporating the reverse inequality which yields false.}
\be\label{w50nl}
w_{50}w_{30}\geq w_{40}^2\,,
\ee
which again follows from the Cauchy-Schwarz inequalities in \cite{tolley}. 
Note that $w_{50}\geq 0$ again follows from $D_5\geq 0$. Thus we seem to have recovered the conditions $w_{n0}\geq 0$ as well as the Cauchy-Schwarz conditions, only with the assumption that $\mW_{10}>0$ and using $D_k\geq 0$ expanded around $a\sim 0$. Note that since $a$ lies in a range, we have an infinite set of nonlinear conditions and we expect to get further constraints examining the $D_k\geq 0$ conditions away from $a\sim 0$.

Now let us examine the effect of eq.(\ref{w30nl}) on the allowed $w_{30}, w_{20}$ domains which followed from the linear conditions.  The linear allowed region found was $0\leq w_{20}\leq 0.140625$ and $0\leq w_{30}\leq 0.01977$. To remind the readers, these conditions used the properties of the partial wave expansion. If we now restrict the rectangular allowed region using eq.(\ref{w30nl}), we find the figure below. As can be seen, the 3 benchmarking theories, 1-loop $\phi^4$, closed string as well as the pion S-matrix all now lie close to the boundary! The situation is similar in the 2-channel case although the string values do not lie at the corner of the allowed region. 
   \begin{figure}[ht]
 \begin{tabular}{cc}
  \includegraphics[width=0.4 \textwidth] {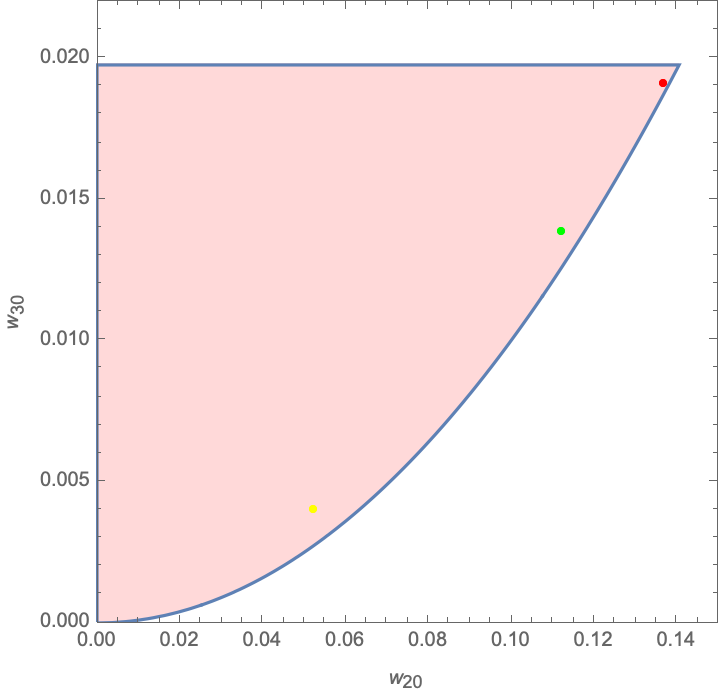}& \includegraphics[width=0.4 \textwidth] {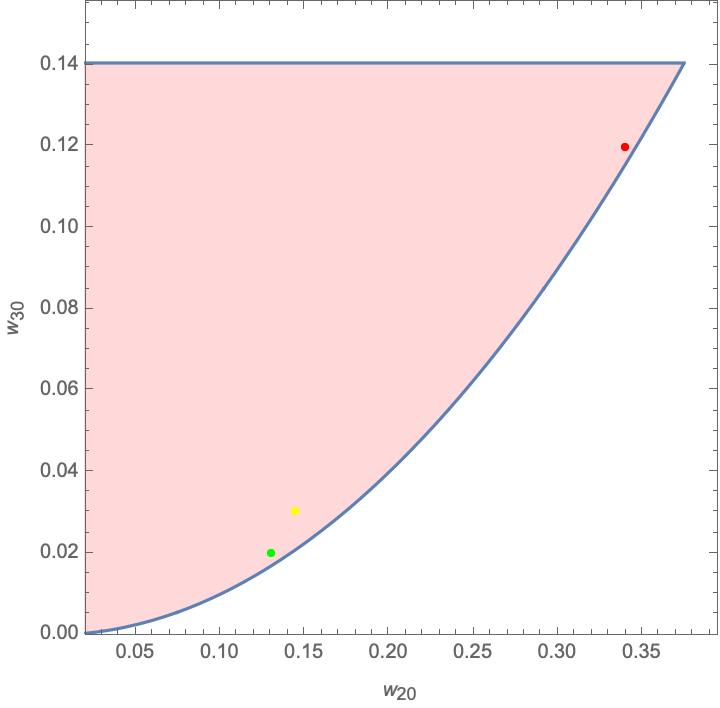}\\
  (a) & (b)
  \end{tabular}
 \caption{Nonlinear constraints in 3-channel case. Red is the closed string, yellow is 1-loop $\phi^4$ and green is the pion S-matrix at $s_0=0.35$ from the bootstrap. (a) 3-channel (b) 2-channel}\label{cluster}
 \end{figure} 
 \vskip 0.5cm
 Next let us now use eq.(\ref{w40nl}) again on the allowed  $w_{30}, w_{20}$ domains  using the range $0\leq w_{40}\leq 0.00278$ bound worked out earlier. The strategy is to use $w_{20}\geq w_{30}^2/w_{40}$ for various values of $w_{40}$ with the maximum allowed value being $w_{40}^{max}=0.00278$. The result is plotted in fig.\ref{rose}(a), with a similar plot for the 2-channel case in fig.\ref{rose}(b). This shows how constrained the space of theories gets on using the $D_k>0$ constraints. Further, notice that the clustering of the benchmarking theories happens near the lower boundary of the allowed region. 
 
 These results were obtained by considering the $D_k\geq 0$ constraints in the neighbourhood of $a\sim 0$. However, these constraints are supposed to be valid for a continuous range of $a$ parameter values. A general analysis of the strongest bounds arising out of such consideration is left for future work. For now, we will content ourselves by examining what happens to the plot in fig.\ref{rose}(a) on using the full $D_3\geq 0$ condition for the 3-channel case, if we set all the $w_{pq}$ values, except $w_{30},w_{20}$ to their minimum or maximum allowed values (see tables (\ref{mint}), (\ref{maxt})). This leads to fig.(\ref{rosecut}) which demonstrates that stronger results are indeed possible using the full allowed $a$-range and $D_k\geq 0$. Note that inequalities such as $-3/2\sqrt{w_{20}}\leq w_{01}\leq 8\sqrt{w_{20}}$ \cite{tolley} which form the parabolic bounding lines in fig.(\ref{schcomp}) do not follow from $D_k\geq 0$ in an obvious manner; it is not inconceivable that such inequalities will require null/locality constraints input.
 
   \begin{figure}[H]
 \begin{tabular}{cc}
  \includegraphics[width=0.45 \textwidth] {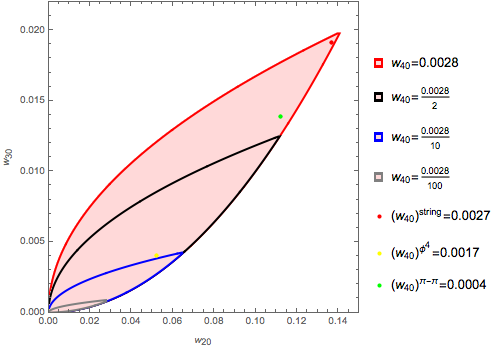}& \includegraphics[width=0.45 \textwidth] {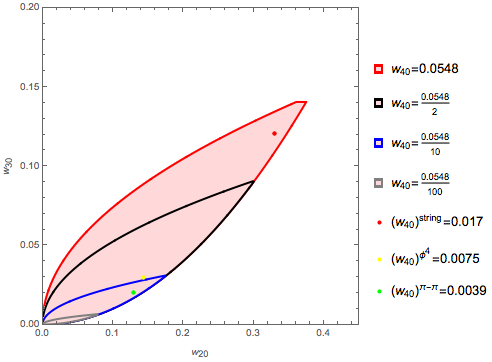}\\
  (a) & (b)
  \end{tabular}
 \caption{The rose. Allowed region shrinks on using $D_4$. The big outside petal is for $w_{40}^{max}$, with the inside petals being $w_{40}^{max}/2, w_{40}^{max}/10, w_{40}^{max}/100$. (a) 3-channel (b) 2-channel}  \label{rose}
 \end{figure}
 \vskip 0.5cm

 \begin{figure}[htb]
 \begin{tabular}{cc}
  \includegraphics[width=0.4 \textwidth] {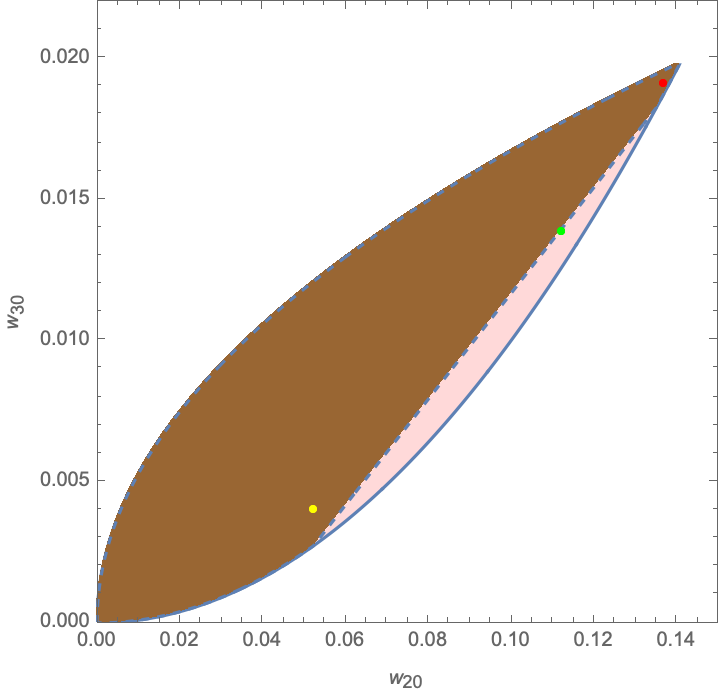}& \includegraphics[width=0.4 \textwidth] {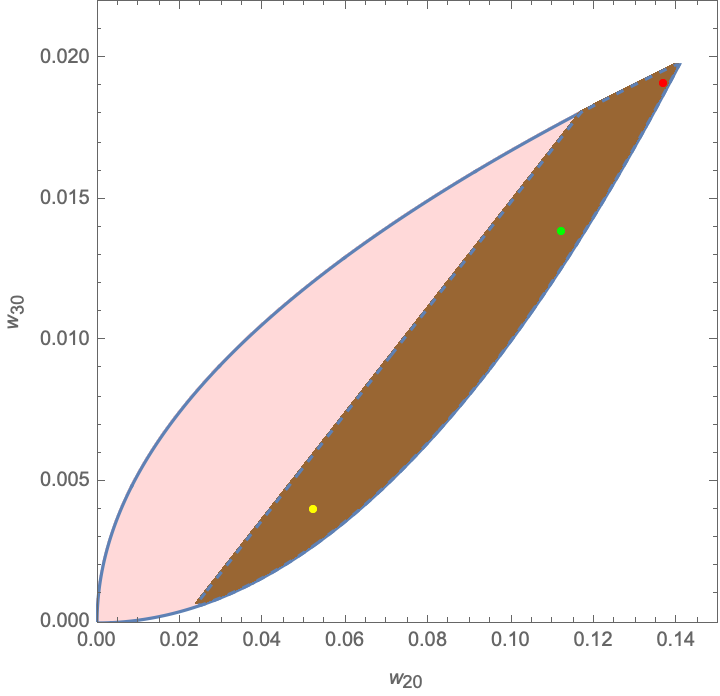}\\
  (a) & (b)
  \end{tabular}
 \caption{Using the full $D_3\geq 0$ on fig.\ref{rose}(a). (a) When all $w_{pq}$ except $w_{30}, w_{20}$ set to maximum with $w_{40}=w_{40}^{max}$. (b) When all $w_{pq}$ except $w_{30}, w_{20}$ set to minimum with $w_{40}=w_{40}^{max}$. Brown indicates the restricted allowed range.} \label{rosecut}
 \end{figure}
 \vskip 0.5cm
 

{\bf Future:} Note that each $D_k$ is a complicated polynomial in $a$ where we have only looked at the leading term as $a\rightarrow 0$ in the considerations above. We outline some future directions here. 
 By looking at each $D_k$ condition as positive polynomial in $a$ over the interval range for $a$, we can analyse the coefficients using certain {\it positivstellensatz} results\footnote{Positive locus theorems.} from real algebraic geometry such as the many  generalisations of Poly\'{a}-Sz\'{e}go theorem\cite{polyaszego}:
 
 {\it Any polynomial $p(x)$ on $[-1,1]$ that is positive on the entire interval can be expressed as follows:
 \bea
 p(x) =\sum_{i=0}^m d_i (1+x)^i (1-x)^{m-i},
 \eea
 where $d_i\ge 0$ for sufficiently large $m$.
 }
 
 We could start from $D_k$ and by changing variables map  the range of $a$ to $[-1,1]$. Then we can apply the above theorem to get certain combinations of $w_{pq}$'s to be positive. The main issue in the practical implementation of the above is an estimate the lowest value of $m$ above which all $d_k$'s are guaranteed to be positive \cite{powers, putinar}. We will mention briefly here that though we do not, in general, have a good estimate for $m$, we can consider {\it large} $m$ such as $m=50$ and look at the corresponding $d_i$. {\it Rather remarkably, we find that some of the positivity conditions $PB_C$ are obtained this way, without even referring to partial waves or Gegenbauer positivity!} This further emphasises the close connection between typical-realness and positivity. We leave further exploration of this fascinating observation to future forays.  
   \section{Discussion}
   In this paper, we applied techniques in Geometric Function Theory, pertaining to Typically-Real functions, to 2-2 scattering amplitudes in QFTs. We used the 3-channel crossing symmetric dispersion relation, originally in \cite{AK}, and studied recently in \cite{ASAZ}.  We also established a new 2-channel crossing symmetric dispersion relation, which makes use of the square roots of unity unlike the cube roots of unity in \cite{AK, ASAZ}. 
   The main findings of this paper are as follows:
   \begin{itemize}
   
   \item Crossing symmetric amplitudes, in a suitable {\it real} range of the parameter $a$ which is the ratio of the crossing symmetric Mandelstam invariants, are Typically-Real functions. This range of $a$ translates into bounds on the Wilson coefficients.
   \item There are 3 kinds of constraints. i) Constraints linear in the Wilson coefficients which follow from the Bieberbach-Rogosinski inequalities for Typically-Real Functions. ii) The Toeplitz determinant conditions for $T_R$, as discussed in \cite{rogo} and summarized in section 4. iii) Positivity conditions arising from the positivity of the Gegenbauer polynomials in the crossing symmetric partial wave expansion, which enters the dispersion relation. In the 3-channel case, we found that (i) was more constraining than the Bieberbach inequalities considered in \cite{HSZ}, which is expected since Typically-Real functions are more constrained than univalent functions. In the 2-channel situation, the linear conditions were not constraining. In both cases the nonlinear constraints were constraining. Finally, we supplemented our analysis with (iii) which depend on unitarity ($a_\ell(s)\geq 0$) and the properties of the Gegenbauer polynomials. 
    \item  We were able to argue why we will always get bounded regions in $w_{pq}$-space and that all $w_{pq}$'s will be $o(1)$ numbers directly from the Bieberbach-Rogosinski inequalities before numerically implementing them for the fully crossing symmetric case. Eq.(\ref{markov2}) in fact readily suggests the schematic relation $$|\Delta w|_{max} \Delta a\sim c$$ where $\Delta w$ is the spread in the Wilson coefficient values, $\Delta a$ is the spread in the allowed $a$ parameter and $c$ is proportional to the bound on $a w_{01}+1$.
   \item We compared our findings with \cite{SCH} and \cite{yutin}. In both cases, we found remarkable agreement. In particular, compared to \cite{SCH}, all our bounds were dimension independent and in a few cases seemingly stronger than the SDPB methods employed there. We explained why this was the case by showing agreement of our results and ones in \cite{SCH} with the bounds obtained by solving the locality constraints \cite{ASAZ} which were the analogues of the null constraints \cite{SCH,tolley} in the fixed-$t$ dispersion relations for the massive and massless (EFT) cases respectively. In our method, we did not impose these conditions. Our primary tool was positivity and the constraints we applied did not depend on the spacetime dimensions. It remains an important open problem to show what role Typically Real-ness plays in the fixed-$t$ dispersion analysis and if it is possible to show this property of the amplitude directly using that dispersion relation.
   \item We were also able to use extremal functions well known in GFT and the Krein-Milman theorem (\ref{A22}) to explain an empirical observation in \cite{SCH} about the regions obtained being a convex hull of extremal amplitudes.
   
    \item Importantly, we also established that positivity and typically-realness considerations will not be enough to give us all interesting bounds. For instance the $g^2$ bound in \cite{smat3} follows from nonlinear unitarity and the techniques in \cite{Goodman}, while appearing to superficially suggest bounds on such couplings, were inadequate to bound them\footnote{While we were careful, it is plausible that we did not use the results of \cite{Goodman} cleverly enough to find such bounds.}. Nonlinear unitarity plays an important role in the studies of \cite{massless, r4} and it will be interesting to connect our techniques to the results of these papers.
    
   \item We were able to consider a massless pole using the techniques in \cite{Goodman} and derived  bounds for maximal supergravity as considered in \cite{SCH2}. 
  
   \item A combination of typical-realness and positivity of Gegenbauer polynomials led us to low-spin dominance discussed in \cite{nima, bern}. For scalar EFTs where the cutoff $\Lambda^2\gg m^2$, so that the scalars are effectively massless, we found that the singularity free condition inside the disk led to a range of $a$ that was $O(\Lambda^2)$. This coupled with the Gegenbauer positivity led to the conclusion that only a few spins could contribute; see sec.(\ref{low}). The fact that our bounds when translated to the EFT case agreed well with \cite{SCH} suggests that it should be possible to establish low spin dominance for the theories living near the boundaries as in fig.(\ref{cluster}), of the allowed region, by examining the solutions to the null/locality constraints. We will elaborate more on this in an upcoming work \cite{CGHRS}. 
   \end{itemize}
 
 Now for some thoughts on immediate future directions.
The positivity conditions arising out of the Toeplitz deteriminants should be examined in more detail in the future. We restricted our attention to the leading order in $a\sim 0$ conditions and found that they coincide with the Cauchy-Schwarz conditions in \cite{tolley}. A preliminary study of the more general conditions did suggest that stronger results are possible. In \cite{AZ}, the techniques used in this paper will be applied for the full pion scattering amplitude and useful and interesting inequalities for all physical pion scattering (e.g. $\pi^+\pi^-\rightarrow \pi^0\pi^0$) will be derived. It should be possible to extend our methods to external particles carrying spin \cite{nimayutin, nima}. This turns out to be indeed the case and we shall say more about this in the upcoming work \cite{CGHRS}. This will be useful for the Standard Model Effective Field Theory. A natural question to ask is if imposing nonlinear unitarity conditions (not the nonlinear Toeplitz conditions) will sharpen the bounds. The spread of values that we have found using the S-matrix bootstrap examples (see tables \ref{mint}, \ref{maxt}) seem to suggest that the bounds will not tighten substantially.  One reason for this optimism, is that tightening of the bounds is related to {\it increasing} the allowed range of $a$ values arising from typically-real considerations. Our arguments for the allowed range of $a$, suggest that such an increase is not possible. 

More general classes of functions that appear in the context amplitudes, also seem to be naturally typically-real. For instance, logarithms: $-\log(1- z) =\int_{0}^{z} d\zeta \frac{1}{1-\zeta}$ and classical-polylogarithms: $Li_{s+1}(z)=\int_{0}^{z} d\zeta \frac{Li_{s}(\zeta)}{\zeta},~ s \geq -1$ are typically-real  by applying \eqref{rogoint} repeatedly to the Koebe function $Li_{-1}(z)=\frac{z}{(1-z)^2}$ which suggests our techniques are applicable more generally. This merits further study and we leave this for future work.

Another area worth exploring is CFT Mellin amplitudes in light of Typical Real-ness. Many implicit assumptions made in QFTs are on firmer footing in CFTs (eg. \cite{Kravchuk:2021kwe}) and as such lessons from CFTs will be useful guiding beacons for QFTs. Since the crossing symmetric dispersion relation was set up for CFT Mellin amplitudes in \cite{GSZ}, a reasonably straightforward extension of the methods set up in this paper should be possible. It will be fascinating to connect CFT-Typically-Realness with swampland conditions discussed recently in \cite{kundu, rastelli2021} as well as the intriguing correlator bounds found in \cite{miguel}.

   \section*{Acknowledgments} We thank Parthiv Haldar and Ahmadullah Zahed for discussions, Apoorva Khare and Vasudeva Rao for  correspondence and, Simon Caron-huot, Rajesh Gopakumar and Yu-tin Huang for comments on the draft. We also thank Shaswat Tiwari and Vinay Vikramaditya for providing us the numbers arising from the S-matrix bootstrap.
   AS thanks the theory groups at CERN, IHES and ENS, Weizmann Institute, ICTS Bengaluru and NTU Taiwan for virtual hospitality during virtual seminars and the ensuing virtual discussions.
   \appendix
   \section{Some consequences of Null constraints}
   
   \subsection{Fully crossing Symmetric case} \label{nullapp}
   This expression \eqref{Belldef1} is valid for both $n\geq m, n\geq 1$ as well as $n<m, n\geq 1$. The $n<m$ Wilson coefficients should vanish for a local theory i.e., $\mW_{n-m,m}=0$ for $n<m$ giving us the locality/null constraints. 
   
 We first note that in this case, unless $j\geq n$  the $\Gamma(n-j)/\Gamma(n-m+1)$ factor is zero, as
 $n<m$ the denominator goes to infinity and the whole term vanishes. If we also have $j\ge n$ then using the Euler-reflection identity for gamma functions we can rewrite the above as $\Gamma(m-n)/\Gamma(n+1-j)$ (up-to an overall sign). Furthermore since $p_{\ell}^{(j)}(\xi_0)=\partial^j C_\ell^{(\a)}(\sqrt{\xi})/\partial\xi^j|_{\xi=\xi_0}$ we have  $j\le \frac{\ell}{2}$, since the $C_\ell^{(\a)}(\sqrt{\xi})$ is a polynomial of degree $\ell/2$ (argued below) and differentiating it more than $\ell/2$ times  gives zero. 
 Thus combining these two statements gives us the result $\ell \ge 2 n$.
 
We shall now prove that $\ell \ge 2 j$ which we used above. The square-root argument of  $C_\ell^{(\a)}(\sqrt{\xi})$ that complicates matters here. However, a convenient expression in terms of the Gauss hypergeometric function or equivalently in terms of Jacobi polynomials can be found.
\begin{eqnarray}
p_{\ell}^{(j)}(\xi_0)&=&\frac{ \sqrt{\pi}(-1)^\ell \xi_0^{\frac{\ell}{2}-j}\Gamma(\ell+\a)}{\Gamma(\frac{\ell+1}{2})\Gamma(\frac{\ell}{2}-j+1)\Gamma(\a)}\,{}_2F_1\left(-\frac{\ell-1}{2},-\frac{\ell}{2}+j,1-\ell-\alpha,\frac{1}{\xi_0}\right)\,,\\
&=&\frac{\sqrt{\pi}(-\xi_0)^{\frac{\ell}{2}-j}\Gamma(\frac{\ell}{2}+j+\a)}{\Gamma(\frac{\ell+1}{2})\Gamma(\a)}P^{(-\ell-\a,j+\a-\frac{1}{2})}_{\frac{\ell}{2}-j}\left(1-\frac{2}{\xi_0}\right)\,.
\end{eqnarray}
Note that the $\Gamma(\frac{\ell}{2}-j+1)$ factor in the denominator vanishes unless $\ell\geq 2j$. \\

It can be checked that $p_\ell^{(j)}(\xi_0)\geq 0$ for $\xi_0\geq 1$. This is an important positivity property since the sign of $\mathcal{B}^{(\ell)}_{n,m}$ is then governed by the other factors in eq.(\ref{Belldef1}). 
Let us address these signs. 

We will be interested in what happens in the null constraints for which $n<m$. After using the reflection identity, the key factor governing the signs in the null constraints is
\be
\frac{(3j-m-2n)\Gamma(m-n)}{\Gamma(j+1)\Gamma(m-j+1)\Gamma(1+j-n)}\,.
\ee
To get a non-zero answer we need $j\geq n$ as observed before. Further the Gamma factors are positive. Thus the sign is controlled entirely by $(3j-m-2n)$ and the analysis is now very simple! The first non-zero answer needs $j=n$. This gives $(3j-m-2n)=(n-m)<0$. Next for $j=n+1$ we have $(3j-m-2n)=n-m+3$ which may be positive or negative depending on the how big $n-m$ is. Thus we conclude that the first non-zero contribution is {\it always} negative. Further since $\ell\geq 2j$ we will have the first spin contributing at $\ell=2n$ and the contribution will be negative. Thus we reach the remarkably simple result: \\
\vskip 0.2cm
{\it In the null constraints, we will have the first non-zero spin contribution $\ell=2n$ to be always negative.}
\vskip 0.2cm
Subsequent spins may contribute either negatively or positively. Note that the positivity of the Jacobi polynomials is not sufficient to fix their signs since more than one $j$ contributes in this case, with the leading $j=n$ being negative and higher terms in the sum being positive or negative depending on the sign of $(3j-m-2n)$ being positive or negative. However, there always exists some $\ell>L_c$ where the total contribution after the $j$-sum is positive for all $\xi_0>1$. For $n=1, m=2$ we find that $\ell=2$ contributes negatively while all higher spins contribute positively. This immediately tells us that there has to be at least one spin-2 partial wave contributing. For $n=2,m=3$ a similar statement can be made for spin-4 since higher spins are all positive and only spin-4 is negative. This observation continues till $n=8, m=9$ where the spin-18 contribution is no longer manifestly positive for all $\xi_0$. In this particular case, we find that spin-20 contribution onwards is again manifestly positive.

\subsection*{Infinite number of higher spin partial waves have to be non-zero}
Using the above observations, we can argue that for the null constraints to be satisfied we will need infinite number of higher spins contributing. Let us go through this argument. Let us start with $n=1,m=2$. For this, the $\ell=2$ contribution is negative while higher spins are positive. This means we will need at least one higher spin, $\ell>2$, for this condition to be respected. Say this higher spin is $\ell=10$. Then we look at the $n=5,m=6$ null constraint. Here $\ell=10$ is negative while beyond $\ell=12$, higher spins are positive. So here too we will need another higher spin $\ell \geq 12$ contribution. As is obvious, repeating this argument leads to the conclusion that we will need an infinite number of higher spin partial waves to be non-zero. In a different guise, this result was already proved in \cite{gribov}.

\subsection*{Constraining EFTs}
Since we are interested in EFTs, we will be interested in the limit where the range of $s_1$ integration in the dispersion relation starts at some $\Lambda\gg \mu$ so that we can approximate $\xi_0=1$. Alternatively, this is also true for massless external particles. In this case the argument of the truncating ${}_2F_1$ is unity and we can use the Gauss summation formula. Then the sum over $j$ can be done giving
\be\label{Beft}
\mathcal{B}^{(\ell)}_{n,m}=2{C_\ell^{(\alpha)}(1)}\frac{(2n+m)(-n)_m \, _4F_3\left(-m,-\frac{m}{3}-\frac{2 n}{3}+1,-\frac{\ell }{2},\frac{\ell }{2}+\alpha ;1-n,-\frac{m}{3}-\frac{2 n}{3},\alpha +\frac{1}{2};4\right)}{ n~  m!}\,.
\ee
This relation holds when $s_1\gg \mu$. Using this expression we will now correlate the $n\geq m$ Wilson coefficient relations to the analysis in \cite{yutin2}.
Using this explicit expression we can redo the positivity checks. In this case we find that for $m=n+1$, until $n=14$, the pattern is always that $\ell=2n$ is negative while $\ell=2n+2$ onwards are positive. For $n=15$ we find that $\ell=30,32$ are both negative and higher spins are positive. This first pair of spins having same sign trend continues till $n=32$ when the first 3 spins, namely $\ell=64,66,68$ are negative and higher spins are positive. 
 \vskip 0.2cm
{\it  Thus all spins up-to $\ell=28$ have to be present and infinitely many higher spins are non-zero}\\
 \vskip 0.2cm
 This is reminiscent of the findings of \cite{yutin2}.

\subsection{Two-channel symmetric case} \label{nullapp2}

The expression \eqref{Belldef} is valid for both $n\geq m, n\geq 1$ as well as $n<m, n\geq 1$. The $n<m$ Wilson coefficients should vanish for a local theory i.e., $\mW_{n-m,m}=0$ for $n<m$ giving us the locality/null constraints. 

Note that as before we have, $j\geq n$ for the $\Gamma(n-j)/\Gamma(n-m+1)$ factor to be non-vanishing which gives $\ell\geq n$ (here odd spins also contribute unlike the fully symmetry case) since $p_{\ell}^{(j)}(\xi_0)=\partial^j C_\ell^{(\a)}(\xi)/\partial\xi^j|_{\xi=\xi_0}$ we have $\ell \ge j$ as we now argue. A convenient expression for $p_\ell$'s can be found, which is a lot simpler than the 3-channel expression:
\begin{eqnarray}
p_{\ell}^{(j)}(\xi_0)&=& \frac{\Gamma(\alpha+1/2)~\Gamma(\ell+2 \alpha+j)}{2^{j} ~\Gamma(2 \alpha)~\Gamma(\ell+\alpha+1/2)} P_{\ell-j}^{(\alpha+j-1/2,\alpha+j-1/2)}(\xi_0)\nonumber\\
&=& 2^j ~(\alpha)_j  ~ C_{\ell-j}^{(\a+j)}(\xi_0)\,.
\end{eqnarray}
It is obvious that $p_\ell^{(j)}(\xi_0)\geq 0$ for $\xi_0\geq 1$, since $C_{\ell-j}^{(\a+j)}(\xi_0)>0$ for $\xi_0 = \frac{s_1}{s_1-\frac{2\mu}{3}}>1$. Since the Gegenbauers are polynomials of degree $\ell$ differentiating them more than $\ell$ times would yield zero so we need $\ell \ge j$ for the term to be nonzero. Combining these two conditions gives us $\ell \ge n$.
\subsection*{Constraining EFT's}
For examining the null constraints with $n<m$ we use  the reflection identity and obtain
\be
\frac{(2j-m-n)\Gamma(m-n)}{\Gamma(j+1)\Gamma(m-j+1)\Gamma(1+j-n)}\,,
\ee
as the factor governing the signs.
 Furthermore, since the Gamma factors are positive,  the sign is controlled  by $(2j-m-n)$ and the analysis is similar to the 3-channel case. To get a non-zero answer we need $j\geq n$. The crucial difference in this case is that {\it all spins contribute not just the even spins}.  The first non-zero answer needs $j=n$. This gives $(2j-m-n)=(n-m)<0$. Next for $j=n+1$ we have $(2j-m-n)=n-m+2$ which may be positive or negative depending on the how big $n-m$ is. Thus we conclude that the first non-zero contribution is {\it always} negative.  Since $\ell \ge j$ we have :
 \vskip 0.2cm
 {\it In the null constraints, we will have the first non-zero spin contribution $\ell=n$ to be always negative.}
 \vskip 0.2cm


As before, the positivity of the Gegenbauer polynomials is not sufficient to fix the signs since more than one $j$ contributes with the leading $j=n$ being negative and higher terms in the sum being positive or negative depending on the sign of $(2j-m-n)$ being positive or negative. However, there exists some $\ell>L_c$ where the total contribution after the $j$-sum is positive for all $\xi_0>1$. For $n=1, m=2$ we find that $\ell=1$ contributes negatively, $\ell=2$ contribution is zero and all higher spins contribute positively. This immediately tells us that there has to be at least one spin-1 partial wave contributing. For $n=k,~m=k+1$  with $k\ge 2$ we see that spins $k,k+1,\cdots, k+\lfloor\frac{k}{2}\rfloor$ are negative while all higher spins are positive. Thus, we can only conclude that at least one of the partial waves corresponding to these spins must contribute. It is clear, however, that an infinite number of higher spin partial waves have to be non-zero for satisfying all the null-constraints that arise. 
\subsection{Implementation of $N_c$}
We will now present a few details of how $N_c|_{\xi_0>1}$ and $N_c|_{\xi_0=1}$ were implemented to get bounds in tables \ref{mint} and \ref{maxt}  in section \ref{schtables}. The dual optimization problem we need to solve to get bounds for $\mW_{p,q}$'s is as follows. We have the following data:\\
$\bullet$ A set of functions $\mW_{n-m,m}$ \eqref{Wdef} of that we need to bound for $n\ge m\ge 1$.\\
$\bullet$ A set of null constraints $N_c$ that vanish for $\mW_{n-m,m}$ for $m>n$.\\
Recall that we bound $w_{p,q}= \frac{\mW_{p,q}}{\mW_{1,0}}$. Thus to bound $w_{p,q}$ we do the following we consider the vector $v(\ell,s_1)= \left(\mW_{10},\mW_{p,q}, \mW_{n_1-m_1,m_1},\mW_{n_2-m_2,m_2},\cdots \mW_{n_k-m_k,m_k}\right)$ for all  $n_i\le m_i$ such that $2n_i+m_i \le N_{max}$.
We then solve the following set of dual problems to get a bound for $w_{p,q}$.
\[
    \left\{
                \begin{array}{ll}
                  
                  \text{Maximize :}  A\\
\text{subject~ to} ~ 0\le (-A,1,c_1,c_2,\cdots,c_k).v(\ell,s_1) ~~~ \forall s_1 \ge \Lambda^2,~~ \ell =0,2,4,\cdots
                \end{array}
              \right.
  \]
  and 
\[
    \left\{
                \begin{array}{ll}
                  
                  \text{Minimize :}  B\\
\text{subject~ to} ~ 0\le (-B,1,c_1,c_2,\cdots,c_k).v(\ell,s_1)~~~ \forall s_1 \ge \Lambda^2,~~ \ell =0,2,4,\cdots

                \end{array}
              \right.
  \]
  which lead to respectively
  \bea
  0&\le& -A \mW_{10}+ \Lambda^2 \mW_{p,q} +0 \,, \nonumber\\
  {\text and} \\
  0&\le& B \mW_{10}- \Lambda^2 \mW_{p,q} +0 \,,
  \eea
which together give $A\le w_{p,q}\le B$. 
We can implement the above by truncating in spin to $\ell=\ell_{max}$ and discretising in $\Lambda^2\le s_1\le \Lambda_{max}^2$. We get the bounds given in last two columns of the tables \ref{mint}, \ref{maxt} for $\ell_{max}=8$, $\Lambda_{max}^2=50$ and $N_{max}=8$ . 

 \section{Typically-Real Functions: A compendium of useful results} \label{appA}
   Here, we will collect in one place several useful results for typically real functions---see \cite{Goluzin, Wigner, anna}.
    A {\it typically-real \footnote {The ubiquitousness of typically real functions in physics was first emphasised to our knowledge by Wigner in \cite{Wigner} who called them $R$-functions. Related functions focusing on the upper half-plane are also called Nevanlinna functions, Herglotz functions, Nevanlinna-pick functions and many others  in the mathematics literature. We will keep using the nomenclature of Typically Real, which is more appropriate for our context of the unit disk.}} $f(z)$ in class $T$ is a function that has positive imaginary part in the upper half plane and negative imaginary part  in the lower half plane i.e., satisfy:
   \bea
   \Im f(z)~~ \Im z > 0 ,
   \eea
 whenever $\Im z \neq 0$. 
  An important subclass of typically real functions is:
 \bea
 f(z) = z+\sum_{n=2}^{\infty} a_n z^n,
 \eea
which are regular and typically real inside the unit disk $|z|<1$. We will refer to this as $T_R$ in the disk. Similar to univalence, typical real-ness of a function also constrains the coefficients $a_n$ quite strongly.   Note however,  an example of a typically real function that is not univalent:
\bea \label{counterex}
 f(z)=\frac{(1+z^2)z}{(1-z^2)^2} =z+3 z^3+5 z^5 +\cdots ~~~ .
\eea
To see that this function is not univalent, note that it violates the Nehari necessary condition for univalence, which can be found for example, in \cite{HSZ}. However we note that

{\emph {A {\it schlicht} function  $f(z)$ that is univalent inside the unit disk $|z|<1$ with all real coefficients $a_n$ is also typically real.}} 

The proof is easy. As $\Im f(z) =\frac{f(z)-{\bar f(z)}}{2 i}$, for $\Im f(z)$ to change sign we would need $f(z)=f(\bar{z})$ (Note ${\bar f(z)} =f(\bar{z})$ since $f(z)$ has only real coefficients) somewhere, but since $f(z)$ is univalent this implies $z={\bar z}$. Thus $f(z)$ is real only on the real axis and $\Im f$ can be positive is $\Im z >0$ and negative in $\Im z <0$. The last statement is true as $f$ is schlicht thus$f(z)=z+\cdots$ and in the neighbourhood of the origin $\Im f \Im z >0$.

We shall refer to this subclass of $T_R$ as $TR_U$. And this is particularly relevant class of functions for us as the kernel $H(s'_1,z,a)$ in \eqref{kernel} belongs to $TR_U$ as $s'_1$ and $a$ are both real and $H(s'_1,z,a)$ is univalent. We shall argue several analytic  properties for $T$ which shall hold for both $T_R$ and $TR_U$ as well and will be useful:
\begin{enumerate}
\item It is obvious from the definition that $f(z)\in T$ is real only on the real axis. 
\item For $f(z)\in T$, $f'(x)>0$ at all regular points on the real axis.

 Since at any regular point $z_0$, $f(z)$ has a power series expansion of the form $f(z)= a_0+a_1(z-z_0)^n+\cdots$  with $a_0=f(z_0)$ real and $a_1\neq 0$, now in a tiny neighbourhood of $z_0$ namely $(z-z_0)= r^n e^{i n \phi}$ with $r\ll 1$ and $0 < \phi \leq 2 \pi$, the first two terms of the series dominate and we have
\bea
\Im f(z) \sim a_1 r^n \sin{n \phi} >0 {\rm ~for} ~ \Im z>0
\eea
and vice versa. This can only happen if $a_1>0$ and $n=1$ since $\sin{n \phi}$ changes sign in $0\leq \phi\leq \pi$ for $n>1$. since $f'(z_0) =a_1$ we conclude $f'(z_0)>0$.
\item $f(z)\in T $ has no essential singularity.

Due to the previous result $f(z)$ is a monotonically increasing function on the real interval  except at possible singularities. So it cannot assume any of the real values it assumes inside the unit disk $|z|<1$ outside of it. Now by Picard's big theorem if $f(z)$ has an essential singularity at $z=z_0$ then in a neighbourhood of $z_0$, $f(z)$ assumes all possible values in the  complex plane except at most one value. This is in contradiction with the previous statement. 

\item All poles of $f(z)\in T $, if any, can lie only on the real axis.

Since in the neighbourhood of a pole $z=z_0$, $f(z) \sim a(z-z_0)^{-n} = a r^{-n} e^{- i n \phi}$ changes sign if $z_0$ is not on the real axis this would contradict $\Im f \Im z >0$.  

\item All poles of $f(z)\in T$ are simple and their residues are negative.

Let us consider a pole at $z=z_0$ and we have $f(z) \sim a(z-z_0)^{-n} = a r^{-n} e^{- i n \phi}$ since $f(z)$ is real on the real axis $\phi=0$ this implies $a$ is real. We also note that if $n>1$ then as before $f(z)$ changes sign in the upper half plane $0<\phi<\pi$. Furthermore, since its derivative is positive on the real axis we have $- \frac{a}{r} >0$ which implies $a<0$. 

\item The only  entire typically real functions are linear $f(z)= a z+b$ with $a,b \in {\mathbb R}$ and $a>0$.
since $f(z)$ is entire it has power series of the form for $|z|<\infty$
\bea \label{herglotzlinear}
f(z)&=& \sum_{n=0}^{\infty} a_n z^{n}=\sum_{n=0}^{\infty} a_n r^{n} e^{i n \phi} \nonumber \\
\implies \Im f &=&  \sum_{n=0}^{\infty} a_n r^{n} \sin{n \phi} \,.\nonumber
\eea
Since, $n \sin{\phi} \geq \sin{n \phi}$  we have
\bea
Sgn(n \sin{\phi} \pm \sin{n \phi}) &=&Sgn(\sin{\phi}) \nonumber \\
\implies \int_{0}^{\pi} d\phi (n \sin{\phi} \pm \sin{n \phi}) \Im f &=& \frac{\pi}{2} (n a_1 r \pm a_n r^n) \nonumber \\
\implies Sgn(n a_1 r \pm a_n r^n) &=&Sgn(\sin \phi ) \,. \nonumber
\eea
But since we can make $r$ arbitrarily large the above statement cannot be true unless $a_n=0$ for $n\ge 2$. Since, $f(z)$ is also real on the real axis with positive derivative we have $f(z) =a_0+ a_1 z$ with $a_0,a_1 \in {\mathbb R}$ and $a_1>0$. 
\end{enumerate}

Thus we can consider three subclasses of $T$ namely $T_R$ which are regular inside the unit disk, $TM$ which have simple poles inside the disk except at the origin and $TM^*$ which have simple poles inside the disk with a pole at the origin.
We state the results for each $T_R$, $TM$ and $TM^*$ relevant for our purposes and refer the reader to the seminal works by Rogosinski\cite{rogo}, Goodman\cite{Goodman} and Komatu, Nehari-Schwarz \cite{komatu} respectively. See also \cite{anna} for a more recent review of the subject.
We begin with the class $T_R$.\\\\\\

\subsection{Regular typically-real functions inside the disk}
 \begin{itemize}
 \item {\bf Characterisation theorem:} If the function $f(z)=z+a_2 z^2+\cdots$ is analytic in the unit disk $|z|<1$ then the following are equivalent:\\
 \noindent 1) $f(z)$ is typically real.\\
 \noindent 2) The function 
 \bea
 (1-z^2)\frac{f(z)}{z} =1+a_2 z+\sum_{n=2}^{\infty} (a_{n+1}-a_{n-1}) z^n 
 \eea
 has positive real part in $|z|<1$ called a Caratheodory function.\\
 3) There exists a nondecreasing probability measure $\mu(t)$ with $-\pi\le t\le \pi$ such that:
 \bea\label{robo}
 f(z)=\int_{-1}^{1} \frac{z}{1-2 z \cos t+z^2} d\mu(t)\,.
 \eea
 We note that  the final condition is the Robertson representation up to a change of variable.
 
 We briefly state the following results for the Carath\'{e}odory class:
 
 An analytic function in $|z|<1$,  $\phi(z) =1+c_1 z+c_2 z^2+\cdots$  has $\Re \phi(z)>0$ in $|z|<1$ then:
 
 1) $|c_n|\le 2$ \\
 2) $\phi(z)$ has the Herglotz representation:
 \bea\label{cara}
 \phi(z) = \int_{0}^{2 \pi} \frac{ e^{i t} +z }{ e^{i t}-z } d\mu(t)~ for~ |z|<1 \,
 \eea
 and $\mu(t)$ is a positive measure.
  Both the integral representations \eqref{robo} and \eqref{cara} are consequences of the Krein-Milman theorem (see (\ref{A22}) for more details). The condition (1) allows us to do a preliminary check if a given function can be  typically-real. For example $f(z) = \frac{z}{(1-z^2)^3}$ is not typically real as $\phi(z) = \frac{(1-z^2)}{z} f(z) = 1+2 z^2+3 z^4+\cdots$ is not in the Carath\'{e}odory class. In fact we have the following \cite{rogo}:

 \item {\bf Bieberbach-Rogosinski bounds:} For $f(z) \in T_R$ in the disk we have the following:
  \bea 
 \begin{cases}
-n~~~~~~~~~~~~~~~~~~~~~~~~~~~~~~~~ & , n~even \\
\\  ~~~~~~~~~~~~~~~~~~~~~~~~~~~~~~~~  \le a_n \le n  \\
-n<k_n = \frac{\sin n~\vartheta_n}{\sin \vartheta_n} & ,n ~odd
\end{cases} \,,
\eea
where $\vartheta_n$ is the smallest solution of $\tan n \vartheta = n \tan \vartheta$ located in $( \frac{\pi}{n},~\frac{3 \pi}{2n} )$. 

\item  For any $n\ge 0$  we have the following combinations of the coefficients being positive:
\bea\label{rogo2}
a_1+a_3+\cdots+ a_{2n+1} \ge0 \,,\nonumber \\
1+\frac{a_2}{2}+\dots+\frac{a_n}{n}\ge 0.
\eea
\item {\bf Szeg\"{o}:}  All sections 
\bea
s_m(z)= \sum_{n=1}^m a_n z^n \,,
\eea
 of a typically real power series in  $|z| <1$ are typically real in $| z| < \frac{1}{4}$. This enables us to find a natural connection with EFTs as discussed in \cite{HSZ}.
 
 \item{\bf Distortion theorem:} The distortion theorems that holds for schlict functions also hold for typically real functions on the real line:
 \bea\label{disto}
 \frac{x}{(1+x)^2}&\le& f(x)\le \frac{x}{(1-x)^2} \,,\nonumber \\
\frac{1-x}{(1+x)^2} &\le& f'(x)\le \frac{1+x}{(1-x)^2} ~{\rm for}~ 0\le x<1 \,.
\eea 
For $TR_U$ these bounds are valid inside the unit disk.

\item {\bf Convexity:} If $f_1 (z$) and $f_2  (z)$ belong to $T_R$ then  
\bea\label{convexity}
t f_1 (z) + (1-t)f_2(z),~0\leq t\leq 1\,
\eea
 also belongs to $T_R$. This is obvious since $\Im f(z)$ has the same sign as $\Im f_i(z)$ and is more generally true for any positive linear combination but if we want to restrict to functions inside the disk $|z|<1$ such as $T_R$, $TM$ and $TM^{*}$ then we need to take a convex sum to ensure the image lies inside the disk.
\item If  $f (z)$ belongs to $T_R$ then the following functions also belong to $T_R$:
\bea\label{rogoint}
{\rm {\bf Reflection}}&:&-f (-z) \,, \nonumber\\
{\rm {\bf Integral~ transform~ 1}}&:&\int_0^z\frac{f(\zeta)}{\zeta}d\zeta\,,\\
{\rm {\bf Integral~ transform~ 2}}&:&\frac{2}{z}\int_0^zf(\zeta)d\zeta\,,\\
 {\rm {\bf ~n^{th} ~root ~transform}}&:&~ \sqrt[n]{f(z^n)}  \,.\nonumber
\eea
\item Furthermore, If $f _i(z)$ with $i=1,\cdots,n$ belongs to $T_R$ then so does 
\bea\label{improgo}
\sqrt[n]{f_1(z)f_2(z)\dots f_n(z)}\,.
\eea
 \end{itemize} 
We will now consider the cases where the function is meromorphic inside the unit disk. There a a couple of complementary approaches by Goodman and Komatu for this class of functiuons.
 \subsection{Meromorphic typically-real functions inside the disk} \label{boundswithpoles}
 In this section, we will first consider the case where all such poles are within a disk of radius $\rho$ so that the function is regular in the annulus $\mathfrak A$. This corresponds to the situation in fig.4(a). The other situation is when the poles start beyond a certain radius, with possibly a puncture at the origin as indicated in fig.4(b). This situation will be considered in the next section.
  \begin{figure}[h!]
 \centering
  \includegraphics[width=0.6 \textwidth] {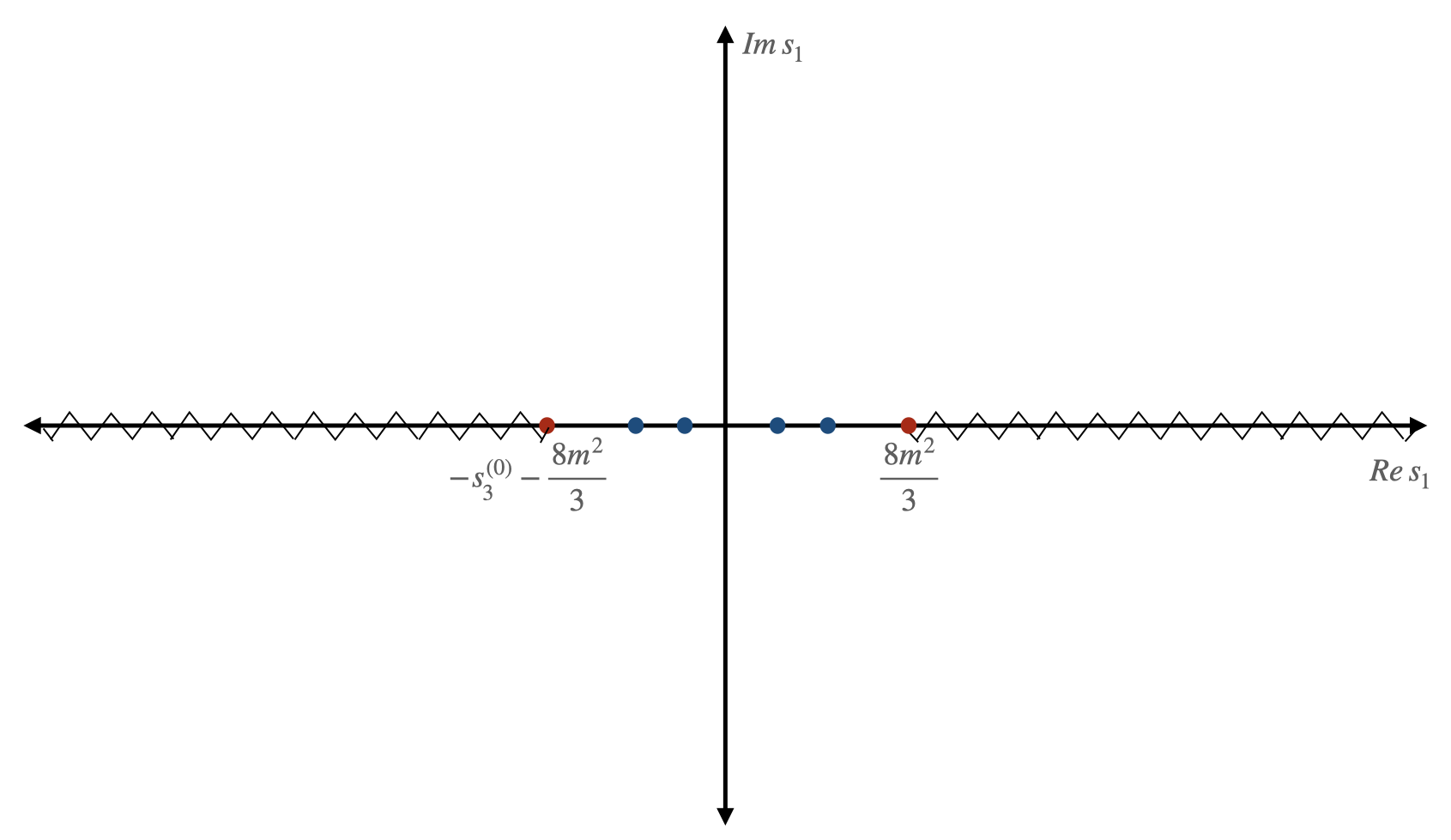}
 \caption{The location of the poles and branch cuts in the complex $s_1$ plane.}\vskip 0.5cm
 \end{figure}
\begin{figure}[h!]
  \centering
  \begin{minipage}[b]{0.49\textwidth}
    \includegraphics[width=\textwidth]{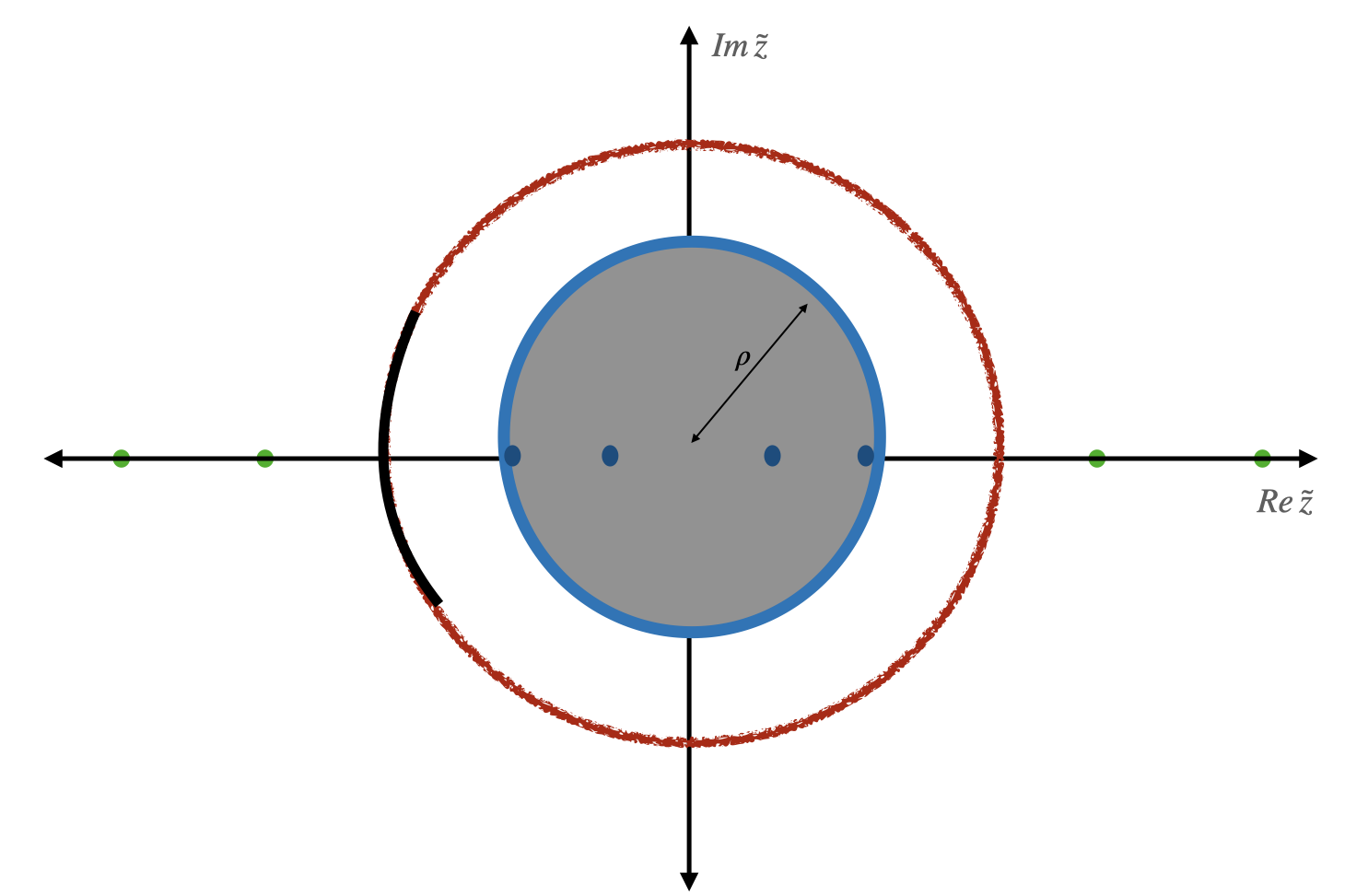} 
    \caption*{}
    \label{4(a)}
  \end{minipage}
  \hfill
  \begin{minipage}[b]{0.49\textwidth}
    \includegraphics[width=\textwidth]{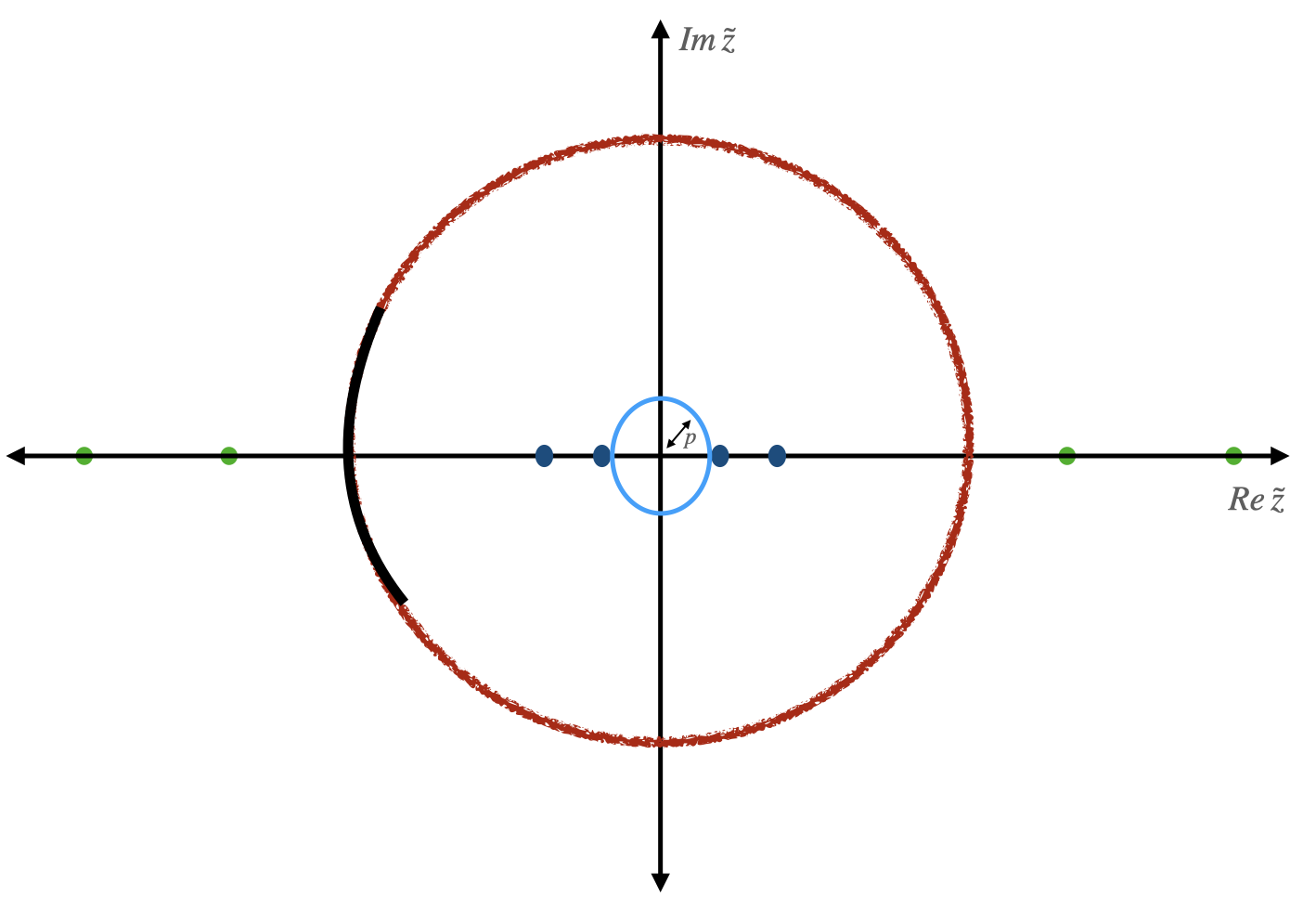} 
    \caption*{}
    \label{4(b)}
  \end{minipage}
  \vspace*{-15 pt}
 \caption{The image in the $z$-plane with branch cuts being mapped arcs on the boundary and poles being mapped onto $[-1,1]$ with image poles being mapped outside the disk.The amplitude can now be treated as (a) An analytic function inside the annulus leading to the RNS bounds. Here $\rho$ is the radius within which all poles are located. (b) A meromorphic function inside the disk leading to the Goodman bounds.  Here $p$ is the radius beyond which the massive poles are located. In the $TM$ case there is no puncture at the origin while for the $TM^*$ case there is.}\vskip 0.5cm
\end{figure} 

Proving coefficient bounds of the Bieberbach kind is easier for such functions compared to the general case proved by de Branges. This was first done by Rogosinski \cite{rogo} in 1932 for the disc and later in 1954 by Nehari and Schwarz \cite{NS} for the univalent Laurent series defined in the annulus. A generalization was found by Komatu \cite{komatu}, where essentially the lower side of the bound was a suitable generalization of the Rogosinski result. We will refer to these bounds as the RNS-bounds after the authors\footnote{Not to be confused with the nomenclature RNS used in string theory!}. 

\subsubsection{RNS bounds}

We will closely follow  \cite{komatu}. In what follows, we will only be using $\Im z \Im f(z)>0$ for functions defined on the annulus with $\Im z \Im f(z)=0$ on the real axis. 

We start by defining
\be
\Phi(r,\theta)\equiv {\Im f(r e^{i\theta})}{\Im z}=\sum_{m=1}^\infty(a_m r^{m-1}-a_{-m}r^{-m-1}){\sin m\theta}\,{\sin\theta}\equiv \sum_{m=1}^\infty \alpha_m(r){\sin m\theta}\,{\sin\theta}\geq 0\,.
\ee
The last inequality is simply a consequence of $\Im z \Im f(z)\geq 0$ and without loss of generality we have set $a_0=0$. Now consider
\be
\frac{2}{\pi}\int_0^\pi d\theta \, \Phi(r,\theta) \left(\frac{\sin n \theta}{\sin\theta}\right)=\alpha_n(r)\,.
\ee
For $n=1$ since $\phi(r,\theta)\geq 0$ we have
\be
\alpha_1(r)\geq 0\,.
\ee
Next we see that for $n=2,3,\cdots$ we have for $0\leq \theta \leq \pi$
\be \label{rog}
-\kappa_n\equiv {\rm Min}\left( \frac{\sin n\theta}{\sin\theta}\right){\leq}\frac{\a_n(r)}{\a_1(r)}\leq {\rm Max} \left(\frac{\sin n\theta}{\sin\theta}\right)=n\,.
\ee
$\kappa_n$ is obtained by solving $\partial_\theta \frac{\sin n\theta}{\sin\theta}=0$ in $0\leq \theta \leq \pi$ which is given by:
\be\label{rog1}
\kappa_n=\frac{\sin n\varphi_n}{\sin \varphi_n}\,,\quad n \csc\varphi_n \cos (n\varphi_n)=\sin(n \varphi_n) \cot(\varphi_n)\,. \footnote{Equivalently we can state this by saying its a solution of $\tan n \varphi_n = n \tan \varphi_n$ located in $( \frac{\pi}{n},~\frac{3 \pi}{2n} )$ for $n>3$ with $\kappa_3=1$.}
\ee
For $n$ even, one can show \cite{rogo, komatu} that $\kappa_n=n$ while for $n$ odd, we have $\kappa_n$ given by solutions to the transcendental equation above. For instance
\be
\kappa_3=1,\quad \kappa_5\approx 1.24,\quad \kappa_7\approx 1.60\,, \quad {\rm with} \quad \kappa_{2n+1}\xrightarrow{n\rightarrow \infty}{} 0.217 (2n+1)\,.
\ee
This will lead to the conclusion below that for the odd $n$, we have a stronger result than what follows from the general Bieberbach considerations, eg. used in \cite{HSZ}. This also explains the findings in fig.(\ref{fig22}).
 Using eq.(\ref{rog1}) we have
\begin{eqnarray}\label{inn1}
-\kappa_n \alpha_1(1) && \leq a_n-a_{-n} \leq n \alpha_1(1)\,,\\ \label{inn2}
-\kappa_n \alpha_1(\rho) &&\leq a_n \rho^{n-1}-a_{-n} \rho^{-n-1}\leq n \alpha_1(\rho)\,.
\end{eqnarray}
These lead to
\begin{eqnarray}
a_n &\leq& n \alpha_1(1)+a_{-n}\leq n \alpha_1(1)+(a_n\rho^{n-1}+\kappa_n \alpha_1(\rho))\rho^{n+1}\,,\\
a_n &\geq& -\kappa_n\alpha_1(1)+a_{-n}\geq -\kappa_n \a_1(1)+(a_n \rho^{n-1}-n\a_1(\rho))\rho^{n+1}\,,\\
a_{-n}\rho^{-n-1} &\leq& \kappa_n \a_1(\rho)+a_n\rho^{n-1}\leq \kappa_n\a_1(\rho)+(a_{-n}+n \a_1(1))\rho^{-n-1}\,,\\
a_{-n}\rho^{-n-1}&\geq & -n\a_1(\rho)+a_n\rho^{n-1}\geq -n\a_1(\rho)+(a_{-n}-\kappa_n \a_1(1))\rho^{n-1}\,.
\end{eqnarray}
The first two sets above use eq.(\ref{inn1}) followed by eq.(\ref{inn2}) while the order is reversed in the next two sets.
From here we get the final inequalities for $n=2,3,\cdots$:
\begin{eqnarray}
-\frac{n}{1-\rho^{2n}}\left(\frac{\kappa_n}{n}\a_1(1)+\a_1(\rho)\rho^{n+1}\right)&\leq& a_n \leq \frac{n}{1-\rho^{2n}}\left(\a_1(1)+\frac{\kappa_n}{n}\a_1(\rho)\rho^{n+1}\right)\,,\\
-\frac{n\rho^{2n}}{1-\rho^{2n}}\left(\frac{\kappa_n}{n}\a_1(1)+\a_1(\rho)\rho^{-n+1}\right)&\leq& a_{-n} \leq \frac{n\rho^{2n}}{1-\rho^{2n}}\left(\a_1(1)+\frac{\kappa_n}{n}\a_1(\rho)\rho^{-n+1}\right)\,,
\end{eqnarray}
For the case where $\rho=0$ and $a_{-n}=0$ for $n\geq 1$, we recover the {\bf Bieberbach-Rogosinski bounds} \cite{rogo} for analytic functions in the disk, which we call $T_R$ in the disk. One typically asks at this point if these bounds are sharp, i.e., if there is a function which saturates these inequalities. Due to the existence of the $n$-dependent $\kappa_n$, addressing this question in general is hard \cite{komatu}.  However, if we set $\kappa_n=n$, a simpler answer can be found in \cite{NS} and is given by the famous Weierstrass-elliptic function\footnote{ This is the Nehari Schwarz extremal function for the univalent case. Komatu gives the extremal function for the general case using Weierstrass-Elliptic functions $\zeta(z; i \pi,\log \rho)$ as  
$f_0(z,\phi)= \frac{\zeta(\log{z} - i\phi)- \zeta(\log{z} + i \phi)}{2 i \sin \phi}= \sum \limits_{{n=-\infty}\atop{n\neq 0}}^{\infty} \frac{\sin n\phi}{\sin \phi} \frac{z^n}{1-\rho^{2n}} + \frac{i \eta_1 \phi}{\pi \sin \phi} $ which saturates lower bound of \eqref{rns} for appropriate $\phi$ (see \cite{komatu}).}
\bea\label{wp}
f_0(z)&=&{\wp}(\log z; i\pi, \log \rho) \nonumber\\
&=& c_0 +\sum_{k=1}^{\infty} \frac{k }{1-\rho^{2k}} z^{k} +\sum_{k=1}^{\infty} \frac{k \rho^{2k}}{1-\rho^{2k}} z^{-k}\,.
\eea
A proof of the above is given in the appendix \ref{appendixA}.
We will mostly be interested in cases where $a_{-n}=0$ for all $n\geq 1$ or $n \geq 2$. 
\subsubsection{The Goodman bounds}
In this section, we will generalize to typically real functions which are meromorphic in the unit disc following Goodman \cite{Goodman}. The analytic structure in the complex $\tilde z$-plane is indicated in fig.4(b). Goodman introduces two classes of such functions. The first denoted by $TM$ has a Taylor series around the origin of the form
\be\label{gmcl1}
f(z)=z+\sum_{n=2}^\infty b_n z^n\,,
\ee
while the second denoted by $TM^*$ has a Laurent expansion about the origin of the form
\be\label{gmcl2}
\phi(z)=-\frac{1}{z}+\sum_{n=0}^\infty \beta_n z^n\,.
\ee

In a manner that is similar to the bounds derived in the previous section, we can derive bounds for the coefficients $b_n$ and $\beta_n$'s. First let us order the poles $p_j$ such that
\be
0<|p_1|\leq |p_2|\leq \cdots <1\,.
\ee
Then denoting the residue at pole $p_j$ by $-m_j$ with $m_j>0$, we have the following useful theorem \cite{Goodman}. 
For each $r$ in $|p_k|<r<|p_{k+1}|$, we have
\bea\label{polres}
\sum_{j=1}^k m_j\left(\frac{r^2}{p_j^2}-1\right)&\leq& r^2\,,\\
\sum_{j=1}^\infty m_j \left(\frac{1}{p_j^2}-1\right)&\leq& 1\,,\\
0<m_j &\leq& \frac{p_j^2}{1-p_j^2}\,, \quad j=1,2,\cdots \label{good3}
\eea
If we further define $p$ such that $|p_j|\geq p>0, j=1,2,3\cdots$, and
\be
B(n,p)=\frac{p(p^{2n}-1)}{p^n(p^2-1)}\,,
\ee
we have for $TM$
\bea 
-B(n,p)&\leq& b_n \leq B(n,p) \,, \qquad n~{\rm even}\,,\\
-\kappa_n &\leq& b_n \leq B(n,p)\,, \qquad n~{\rm odd}\,.
\eea
Finally for $TM^*$ we have $\beta_1\geq -1$ and for $n\geq 2$,
\bea
-(1+\beta_1)B(n,p)&\leq& \beta_n \leq (1+\beta_1)B(n,p)\qquad n~{\rm even}\,,\\
-(1+\beta_1)\kappa_n&\leq& \beta_n \leq (1+\beta_1)B(n,p)\qquad n~{\rm odd}\,.
\eea
Notice that since $|p_j|>p$ these bounds are complementary to the RNS-bounds in the previous section.
A crucial result about typically-real functions is the Schiffer-Bargmann representation theorem \eqref{SB1}\eqref{SB2} which we shall now argue following \cite{Goodman,Wigner}.
 \subsection{Proof of Schiffer-Bargmann representation} \label{SBproof}
 If $f(z)$ has a pole $p\neq 0$ with residue $-m<0$, then $t(z) = m \frac{(1-p^2)}{p^2} \frac{z}{1-2s z+z^2}$ with $2 s= p+1/p$ has the same pole and same residue thus $g(z)=f(z)-t(z)$ has the same pole structure as $f(z)$ except being regular at $z=p$ and except for the pole $g(z)$ is real on the real axis (since $t(z)$ is an univalent function with real coefficients it is also typically real).

We shall now argue that $\Im g(z) \Im z \ge 0$ which implies either $g(z) =0 $ or $\frac{g(z)}{g'(0)}$ is typically real inside the disk.
 \begin{figure}[h!]
 \centering
  \includegraphics[width=0.55 \textwidth] {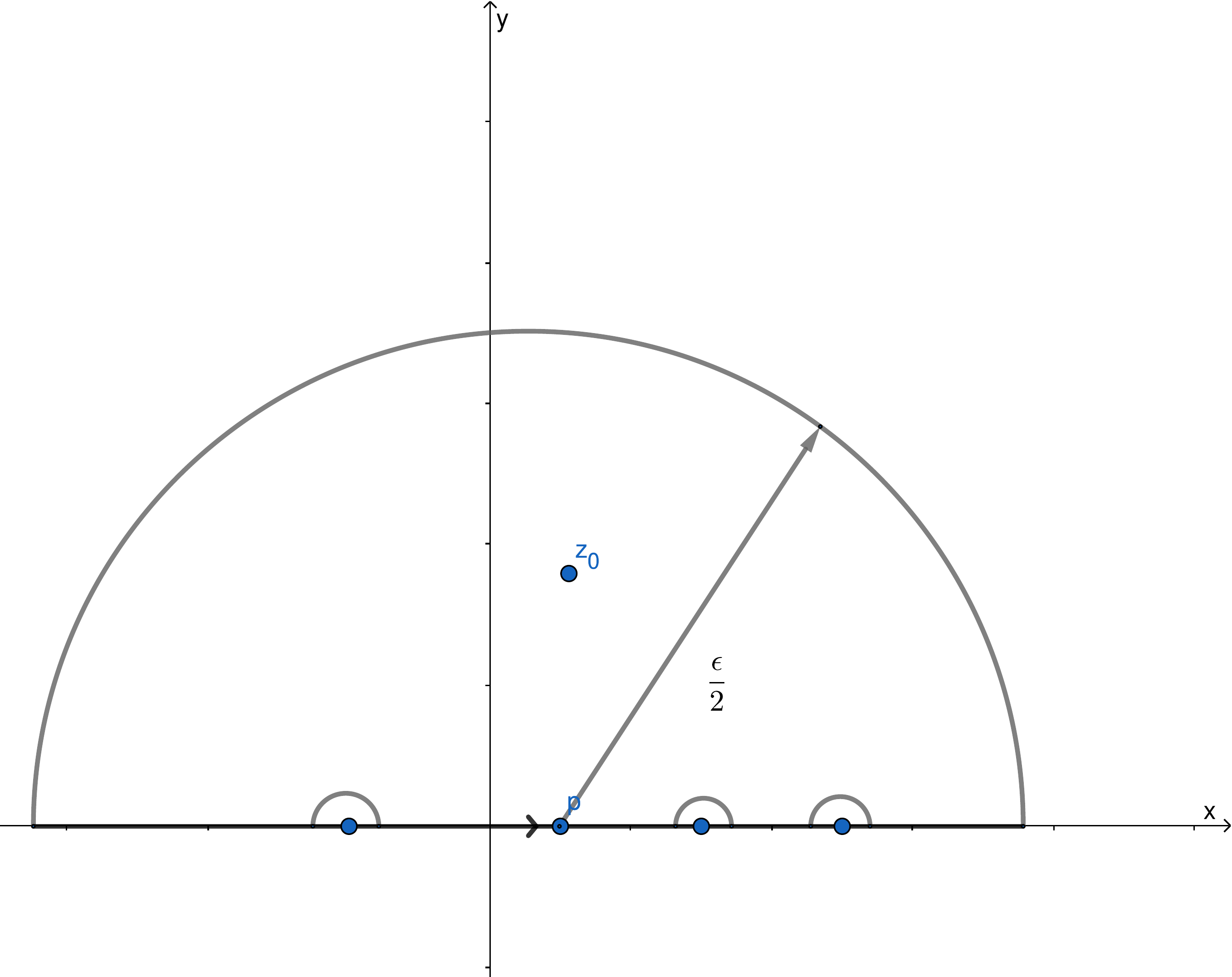}
 \end{figure}
Suppose for some $z_0$ with $\Im z_0 >0$ we have $\Im g(z_0) = -\epsilon$ with $\epsilon>0$. For any $\epsilon >0$ since $t(z)$ is regular everywhere except $z=p<1$, it is continuous in a sufficiently narrow domain containing $|z|=1$ i.e there exists an $r\le |z|<1$ s.t $|\Im t(z)|<\frac{\epsilon}{2}$ . We can consider the contour shown in above figure which omits all poles of $f(z)$ except $z=p$. Since in the neighbourhood of any other pole $p_i \neq p$, $t(z)$ is regular we can take radius of any of semicircles encircling the poles to be as small as we want so that $|\Im t(z)| < \frac{\epsilon}{2}$ while making the big semicircle enclose $z_0$. This implies that for the full region enclosed by the contour, we have
\bea
\Im g(z) = \Im f(z) -\Im t(z)  \le  -\frac{\epsilon}{2} \,. \nonumber
\eea
Since $\Im f >0$, this contradicts our assumption that there exists $z_0$ with $\Im z_0>0$ for which $\Im g(z_0) > -\epsilon$ . Thus, $\Im g \Im z >0$ but since $g(0)=0$ if its also true that $g'(0)=0$ then $g(z)=0$. Otherwise $g'(0) = 1-m \frac{(1-p^2)}{p^2}>0$ and we have $\frac{g(z)}{g'(z)}$ is typically real inside the disk.

We can iteratively repeat this process for and finite set of poles to get:
\bea
g_k(z) =f(z) -\sum_{i=1}^k m_i  \frac{(1-p_i^2)}{p_i^2} F_{s_i}(z)
\eea
with $F_{s_i}=  \frac{z}{1-(p_i+p_i^{-1})z+z^2}$ , ~~$s_i = p_i+\frac{1}{p_i}$.
If $f(z)$ has only a finite number of poles we are done, otherwise since 
\bea
g'_k(0)=1- \sum_{i=1}^k m_i  \frac{(1-p_i^2)}{p_i^2} >0 
\eea
for every $k$, the sequence of functions $g_k(z)$ converges uniformly on any closed subset of $|z|<1$, not containing any poles of $f(z)$. Since the limit function $g(z)$ would have no poles in $|z|<1$ either $g(z)=0$ identically if $g'(0)=0$ or  if $g'(0)=\mu >0 $ we have $g(z) = \mu a(z) $ where $\mu>0$ and $a(z)$ is a regular typically real function in $|z|<1$ and $f(z)$ satisfies \eqref{SB1}. Furthermore $\mu$ satisfies \eqref{mu1}.
Notice that \eqref{SB2} follows as if $f(z)=\frac{-1}{z}+\sum_{n=0}^{\infty} \beta_n z^n $ has a pole at $z=0$ then we can apply the above result to $g(z)= \frac{1}{z}-\beta_0 +z +f(z)  $ and conclude $g(z)$ has a representation \eqref{SB2} and noticing that $g'(0) = (1+\beta_1)$ the results \eqref{SB2} and \eqref{mu2} readily follow\footnote{We also note here for completeness that if we were considering $f(z)$ which are typically real on the whole complex plane then $a(z)$ would have been linear due to the fact that the only entire typically real functions are linear and the proof for this case is virtually identical and was the one first considered by Schiffer and Bargmann and also reproduced by Wigner in \cite{Wigner}. We restricted to the unit disk as its best suited for our purposes following Goodman \cite{Goodman}.  }.

\subsection{Proof of relation  \eqref{wp} }\label{appendixA}
We begin by given the standard definition of the Weierstrass $\wp$ function:
\bea
\wp(z;\omega_1,\omega_2) = \frac{1}{z^2} +\sum_{{(m,n)\neq (0,0)}} \frac{1}{(z-2 ~m ~\omega_1-2 ~n~\omega_2)^2} - \frac{1}{(2~ m~ \omega_1+2 ~n~\omega_2)^2} \,,
\eea
where $\omega_1$,$\omega_2$ are called the half periods. 
Its more conventional to scale to $z \rightarrow \frac{z}{2 \omega_1}$ and $\omega_2 \rightarrow \tau = \frac{\omega_2}{\omega_1}$, which is usually denoted as $\wp(z,\tau)$. We shall continue to denote the variable by z with the understanding that its actually $\frac{z}{2 \omega_1}$.
\bea
\wp(z;\tau) = \frac{1}{4 \omega_1^{2} z^2} +\frac{1}{4 \omega_1^{2}}\sum_{{(m,n)\neq (0,0)}} \frac{1}{(z- m- ~n~\tau)^2} - \frac{1}{(m+~n~\tau)^2} \,.
\eea
We note that in \eqref{wp} $\wp(\log[z], i \pi, \log[\rho])$ has argument $z \rightarrow \frac{\log[z]}{2 \pi i}$ and $\tau= \frac{\log[\rho^2]}{2 \pi i}$. Thus $\rho^2$ plays the role of the "Nome" $q=e^{2 \pi i\tau}$. We shall now derive some relations for $\wp(z,\tau)$ before finally substituting for $z,\tau$ to get the result \eqref{wp}.
We begin with the following well known relation for $\wp(z,\tau)$:
\bea\label{wp1}
\wp(z,\tau)= \frac{1}{4 \omega_1^{2} z^2} +\frac{1}{4 \omega_1^{2}}\sum_{n=1}^{\infty} (2n+1) G_{2n+2} z^{2n}\,,
\eea
where $G_{2n+2}$ is the Eisenstien series defined as:
\bea
G_{2n+2} = 2 \zeta(2n+2) \left[ 1+\frac{2}{\zeta(1-(2n+2))} \sum_{k=1}^{\infty} \frac{k^{2n+1} q^{k}}{1-q^{k}}\right]\,.
\eea
Substituting the above definition into \eqref{wp1} gives:
\bea\label{wp3}
\wp(z,\tau)= \frac{1}{4 \omega_1^{2} z^2} +\frac{1}{4 \omega_1^{2}}\sum_{n=1}^{\infty} 2 (2n+1) \zeta(2n+2) z^{2n}+ \frac{1}{4 \omega_1^{2}}\sum_{n=1}^{\infty} \frac{4(2n+1)\zeta(2n+2)}{\zeta(1-(2n+2))}  \sum_{k=1}^{\infty} \frac{k^{2n+1} q^{k}}{1-q^{k}}\,.
\eea
The second term can be simplified as:
\bea
\sum_{n=1}^{\infty} 2 (2n+1) \zeta(2n+2) z^{2n}=-\frac{\pi^{2}}{3} +\frac{\pi^2}{(\sin[\pi z])^{2}} -\frac{1}{z^{2}}
\,.\eea
We can simplify the last term using the reflection identity for zeta functions:
\bea
\frac{\zeta(z)}{\zeta(1-z)} = \frac{(2 \pi)^z}{2 \Gamma[z] \cos\left(\frac{\pi z}{2}\right)}\,,
\eea
using these relations \eqref{wp3} becomes:
\bea
\wp(z,\tau)&=& \frac{1}{4 \omega_1^{2} z^2} +\frac{1}{4 \omega_1^{2}} \left(-\frac{\pi^{2}}{3} + \frac{\pi^2}{(\sin[\pi z])^{2}} -\frac{1}{z^{2}}\right) +\frac{2}{\omega_1^{2}}  \sum_{k=1}^{\infty} \frac{k q^{k}}{1-q^{k}} \sum_{n=1}^{\infty} \frac{-(2 \pi)^{2n+2}}{(2n)!} (i k z)^{2n}\,, \nonumber\\
\wp(z,\tau)&=& \frac{1}{4 \omega_1^{2}} \left(-\frac{\pi^{2}}{3} + \frac{\pi^2}{(\sin[\pi z])^{2}}\right) -\frac{16 \pi^2}{4 \omega_1^{2}}  \sum_{k=1}^{\infty} \frac{k q^{k}}{1-q^{k}} \sin[k\pi z]^2\,,
\eea
where in the last term we performed the sum over $n$.
We can now  apply the above result for the case of interest:

\bea
\wp(\log[z], i \pi, \log[\rho])&= &\frac{-1}{4 \pi^{2}} \left(-\frac{\pi^{2}}{3} + \frac{\pi^2}{\left(\sin\left[\frac{\log[z]}{2 i }\right]\right)^{2}}\right) + 4   \sum_{k=1}^{\infty} \frac{k \rho^{2k}}{1-\rho^{2k}} \sin\left[\frac{k \log[z]}{2 i}\right]^2 \,,\nonumber \\
&=& \frac{1}{12} +\frac{1}{z +\frac{1}{z} -2} +  \sum_{k=1}^{\infty} \frac{k \rho^{2k}}{1-\rho^{2k}} \left(z^{k} +z^{-k} -2\right)\,, \nonumber \\
&=& c_0 +\sum_{k=1}^{\infty} \frac{k }{1-\rho^{2k}} z^{k} +\sum_{k=1}^{\infty} \frac{k \rho^{2k}}{1-\rho^{2k}} z^{-k}\,.
\eea
This proves the result  \eqref{wp}.
\subsection{Extremal functions in GFT}\label{A22}
For several interesting classes of functions in GFT extremal functions are well known and have integral representations due to the Krein-Milman theorem\cite{brickman}. The Krein-Milman theorem is a generalisation to arbitrary compact convex subsets (possibly infinite dimensional) of the fact that finite dimensional convex sets such as polytopes can be expressed as a convex hull of their vertices (extreme points).

Let $X$ be a complex vector space.  For any elements $x,y \in X$ by $[x,y]:= \{t x+(1-t) y:0\le t\le 1 \}$ is called the closed interval between them. If $K$ is a subset of $X$ and $p\in K$, then $p$ is called an {\it extreme point} if $\nexists ~distinct~ x, y \in K$ such that $p=t x+(1-t) y$ for $0<t<1$. The set of all extreme point of $K$ is denoted $extreme(K)$
A set $S$ is called {\it convex} if for any $x,y \in S$ then $[x,y]\in S$. The smallest convex set $co(S)$ containing $S$ is called the convex-hull of $S$. The smallest closed convex set containing $S$ is called the closed convex hull of $\bar{co}(S)$.

{\it {\bf Krein-Milman theorem:} Suppose $X$ is a Hausdorff locally convex topological vector space and  $K$ is a compact and convex subset of $X$.Then $K$ is equal to the closed convex hull of its extreme points:
$K=\bar{co}(extreme(K))$}

\begin{table}[ht]
\centering 
\begin{tabular}{|c| c| c| c|} 
\hline 
Class & definition & Extremal function & Integral Representation \\ [0.5ex] 
\hline 
Univalent  & $f(z_1)\neq f(z_2)$ & $\frac{z}{(1-x z)^2}$ &\\ 
$U$ & & &  $\int_{|x|=1}\frac{z}{(1-x z)^2}$\\
& if $z_1\neq z_2$& with $|x|=1$& \\
\hline
Typically-real & $\Im f(z) \Im z >0$   & $\frac{z}{(1-2 x z +z^2)}$ & \\
$TR_U$ & & & $\int_{-1}^{1} \frac{z}{1-2 z \cos t+z^2} d\mu(t)$\\
&for $\Im z \neq 0$& with $|x|=1$& \\
\hline
Carth\'{e}odory  & $\Re f(z)>0$ & $\frac{ e^{i t} +z }{ e^{i t}-z }$ &  \\
&&&$\int_{0}^{2 \pi} \frac{ e^{i t} +z }{ e^{i t}-z } d\mu(t)$\\
${\mathcal P}$& for $|z|<1$  &$0\le t\le 2 \pi$ &  \\
\hline
Convex & $\Re \left(1+\frac{z f''(z)}{f'(z)} \right)>0$ & $\frac{z}{(1-x z)}$&  \\ 
 $C$ &&&$\int_{|x|=1}\frac{z}{(1-x z)}$\\
 &&with $|x|=1$ &\\
 [1ex] 
\hline 
\end{tabular}
\caption{Extremal functions for various classes inside the unit disk $|z|<1$}\vskip 0.5cm
\label{table:nonlin} 
\end{table}
\vspace*{20 pt}
We already knew extremal functions such as as the Koebe function \cite{HSZ} as those satisfying the coefficients bounds or distortion bounds and above theorem justifies our expectation as well as giving a nessecary and sufficient conditions for functions to belong to these classes. Note that the Robertson representation for $T_R$ did not say anything about the extremal function. Since from the Krein-Milman theorem, we find that the extremal function has an integral representation whose kernel coincides with what enters in the Robertson representation, we can say that the Krein-Milman theorem implies the Robertson representation.
\section{Typical-Realness in physics} \label{trp}
As alluded to in the introduction, Typical-realness/ Herglotz property  has been used in the physics in the context of scattering both in quantum mechanics and relativistic quantum field theory since the seminal work of Wigner in 1950 \cite{Wigner}. We shall briefly review a few of these.
\subsection{Typical-realness in quantum mechanics}
\begin{enumerate}
\item{\bf R-matrix theory of scattering:} In \cite{WignerEisenbud}, the R-matrix method for describing nuclear reactions with a compound nucleus point of view was introduced. The idea was to remain agnostic about the physics inside the nuclear sphere of radius $a$ and describe the observed reactions purely in terms of the known wave functions outside the nuclear sphere $r> a$ and the derivative of the logarithm of the wave function on the boundary of the nuclear sphere $r=a$ called the $R$-function or $R$-matrix. It was later generalised to any spherically symmetric scattering of particles in the field of some scattering center \cite{Wigner} by dividing the problem into ``short range interaction"  $r<a$ and ``long range interaction"  $r\ge a$ regions. It was shown that the knowledge of this $R$-function along with the radius $a$ was enough to determine the cross-section. It was argued that the $R$-function was Herglotz. We shall give a short proof of this fact following \cite{Wuomura} for the one-channel case and refer the interested reader to \cite{Wigner,WignerEisenbud} for the general theory.

Consider the simple case of the one-channel collision of a slow neutron with a nucleus. The interaction $V(r)$ between the neutron and the nucleus is:
\bea
V(r)=\begin{cases}
V(r),& 0<r\le a  \\
0,&a<r 
\end{cases}  \,, \nonumber
\eea
where $a$ is the radius of the nuclear sphere (known from experimental data). The radial wave function for the $s$-wave inside the nucleus $(0\le r \le a)$ satisfies
\bea \label{R1}
\frac{d^2 \mathcal{F}}{d r^2} +[k^2-U(r)] \mathcal{F}(r)=0 \, ,
\eea
with $U(r) =\frac{2m}{\hslash^2} V(r)$ and $\mathcal{F}(0)=0$.
We do not have knowledge of $V(r)$ in general so the above equation cannot be solved to obtain the wave function inside the nuclear sphere $r<a$. However, it was argued that using just the information about the value of the derivative of the wave function on the boundary of the nuclear sphere is sufficient to obtain the cross section.
Let $w_{\lambda}(r)$ be a complete set of eigenfunctions in $0\le r \le a$ of 
\bea\label{R2}
\left[ \frac{d^2 }{d r^2} +\epsilon_{\lambda}-U(r) \right] w_{\lambda}(r) =0, ~\epsilon_{\lambda}=\frac{2m}{\hslash^2} E_{\lambda} \, ,
\eea 
subject to the boundary conditions $w_{\lambda}=0$, $\frac{d w_{\lambda}}{dr}\big|_{r=a}=0$ and normalised as $\int_{0}^{a}w^2_{\lambda}(r) dr =1$.
Combining the equations  \eqref{R1} and \eqref{R2} for $\mathcal{F}(r)$ and $w_{\lambda}(r)$ gives:
\bea
\int_{0}^a \mathcal{F} w_{\lambda}(r) dr = \frac{w_{\lambda}(a)}{\epsilon_{\lambda}-\epsilon} \left(\frac{d\mathcal{F}}{dr}  \right) \bigg|_{r=a}\,, \nonumber 
\eea
where $\epsilon = k^2= \frac{2m}{\hslash^2}E$. 
Thus the wavefunction $\mathcal{F}$ inside the nuclear sphere  is completely determined interms of $w_{\lambda}(r)$  once we specify the value of its derivative at the boundary $ \left(\frac{d\mathcal{F}}{dr}  \right) \bigg|_{r=a}$:
\bea
\mathcal{F}(k r) =  \left(\frac{d\mathcal{F}}{dr}  \right) \bigg|_{r=a} \sum_{\lambda}\frac{w_{\lambda}(a)}{\epsilon_{\lambda}-\epsilon} w_{\lambda}(r) 
\eea
The above can be used to define the $R$-function or the ``derivative-function":
\bea
R (\epsilon)= \frac{\mathcal{F}(a)}{\mathcal{F}^{\prime}(a)} = \sum_{\lambda}\frac{w^2_{\lambda}(a)}{\epsilon_{\lambda}-\epsilon} \,,
\eea
which purely depends on parameters $w^2_{\lambda}(a)$ and $\epsilon_{\lambda}$ which are properties of the nucleus.\\

{\it It is clear from the above expression that $R(\epsilon)$ is a typically-real/ Herglotz function since $w^2_{\lambda}(a)$ is positive.}\\\\
In the free region $a\le r < \infty$, for $\ell\neq 0$, we get wave functions $G_{\ell}(r)$ and $H_{\ell}(r)$ that satisfy:
\bea
G_{\ell}^{\prime}H_{\ell} &-& H_{\ell}^{\prime}G_{\ell}= k \,, \nonumber \\
G_{\ell}(k r) &\rightarrow& \sin\left( k r - \frac{\ell \pi}{2} \right), \nonumber \\
H_{\ell}(k r) &\rightarrow& \cos\left( k r - \frac{\ell \pi}{2} \right), ~ as ~ r\rightarrow \infty \,.\nonumber
\eea
By requiring the continuity of the wave functions at $r=a$ one gets the relations:
\bea
\mathcal{F}_{\ell}(k r) &=& \left(\frac{d\mathcal{F}}{dr}  \right) \bigg|_{r=a}( G_{\ell}(r) + S_{\ell} H_{\ell}(r))\,, \nonumber \\
\tan{\delta_{\ell}} &=&  \frac{k R G'_{\ell}(a)-G_{\ell}(a)}{H_{\ell}(a)- k R H'_{\ell}(a) } \nonumber 
\eea
where $S_{\ell} = \exp{(2 i \delta_{\ell})}$.
The total cross section $\sigma= \frac{4 \pi}{k^2} \sum_{\ell=0}^{\infty} (2l+1) \sin^2{\delta_{\ell}}$ is thus determined entirely in terms of $R$ and the wave functions $G_{\ell},~H_{\ell}$ and their derivatives on the boundary of  the nuclear sphere.
\item {\bf Perturbative expansion of energy levels:} In \cite{BenderWu1,BenderWu2,Simon1,Simon2} , typical-realness/ Herglotz property was used in the context of Rayleigh-Schr\"{o}dinger perturbation theory for the expansion of the Energy levels $E(\epsilon)$ of anharmonic oscillator in terms of the perturbation parameter $\epsilon$. It was argued that this function was Herglotz and this was further used in showing several features such as the global three sheeted structure and branch points for the quartic oscillator. We give a brief argument for why \emph{ the energy levels of any regular perturbation problem with the perturbation term having a fixed sign is Herglotz}  following \cite{Benderorzag} and refer the interested reader to the series of works \cite{BenderWu1,BenderWu2,Simon1,Simon2} for further details.
We begin by considering the Schr\"{o}dinger equation:
\bea
\left[-\frac{d^2}{d x^2} +V(x)+ \epsilon W(x) -E(\epsilon) \right] \Psi(x) =0\, , \nonumber
\eea
Since the above is a regular perturbation problem for all $\epsilon$, $\epsilon W(x)$ is negligible compared to $V(x)$ as $|x| \rightarrow \infty$. Thus the asymptotic behaviour of $\Psi(x)$ for large $|x|$ is independent of $\epsilon$ i.e. $\Psi(x) \rightarrow 0$ as $|x|\rightarrow \infty$. We can multiply by  $\Psi(x)^{*}$ and integrate to get:
\bea
\underbrace{\int_{-\infty}^{\infty} \Psi^{'}(x)^{*} \Psi^{'}(x) dx}_{Real,>0} & + &\underbrace{\int_{-\infty}^{\infty} V(x)\Psi(x)^{*} \Psi(x) dx}_{Real} + \epsilon \underbrace{\int_{-\infty}^{\infty} W(x) \Psi(x)^{*} \Psi(x) dx}_{Real} \nonumber \\ &=& E \underbrace{\int_{-\infty}^{\infty} \Psi(x)^{*} \Psi(x) dx}_{Real,~ >0}  \, ,\nonumber  \\
& \implies& \Im E(\epsilon) =\frac{\int_{-\infty}^{\infty} W(x) \Psi(x)^{*} \Psi(x) dx }{\int_{-\infty}^{\infty} \Psi(x)^{*} \Psi(x) dx }  \Im (\epsilon) \, .\nonumber
\eea
Thus assuming that $W$ is one-signed, either $E(\epsilon)$ or $-E(\epsilon)$ is Herglotz. The Herglotz property also explains why most perturbation problems have branch points, since the only entire Herglotz function is linear as we proved in \eqref{herglotzlinear}. We can see that any $E(\epsilon)$ which has a nonzero $\epsilon^n$ term for $n\ge 2$ and satisfying the above hypothesis about fixed sign of $W(x)$ will nessecarily have branch points \cite{Benderorzag}.
\end{enumerate}
\subsection{Typical-Realness in QFT}
\begin{enumerate}
\item {\bf Fixed-$t$ dispersion relations:} In \cite{Jinmartin1,Jinmartin2} typical-realness was encountered in the fixed-t dispersion relations, the authors showed that:\\
$\bullet$ Assuming that the Froissart bound is satisfied for the cross-section, the number of subtraction required for fixed-$t$ dispersion relations in the region $0< t< 4\mu^2$ is at most $2$.\\
$\bullet$ In the 2-channel symmetric case ($s,u$ symmetry) be defining $z= \left(s- 2 \mu^2+\frac{t}{2}\right)^2$ and rewriting the dispersion in $z,t$ variables:
\bea\label{zdisp}
F(z,t)= f(t) + \frac{z}{\pi} \int_{(2 \mu^2+t/2)^2}^{\infty} \frac{\Im F(z',t) dz'}{z'(z'-z)}\,. 
\eea
Since the absorptive part is positive for $z \ge (2 \mu^2+t/2)^2$ this leads to $\Im F(z,t) \Im z >0$ or equivalently $F(z,t)$ is a Herglotz function.
 
Using the above property several facts about the asymptotic properties of the amplitude and the cross-section were  derived. One such result was that for $t\ge0$,  $F(s,t)$ cannot decrease faster than $\frac{1}{s^2}$  as $|s| \rightarrow \infty$. The Herglotz property continued to hold in the case with $N$-subtractions and the number of zeros of $F(z,t)$ were constrained. In particular, it was shown that for $z< (2 \mu^2+t/2)^2$, $F(z,t)$ can have at most one zero. We refer the interested reader to \cite{Jinmartin1,Jinmartin2} for several other interesting results and details. 

In \cite{Jinmartin1,Jinmartin2} however the Bieberbach-Rogosinski inequalities were not applied to constrain the coefficients. We shall briefly comment on how to do this since its closely related to the current work. Since $F(z,t)$ is analytic in $z\le (2 \mu^2+t/2)^2$ we can apply the Bieberbach-Rogosinski bounds on $|\tilde z| =\frac{|z|}{(2+t/2)^2} \le 1$ by considering $F(z,t)= F((2 \mu^2+t/2)^2 \tilde z, t )$. Converting to {\it schlicht} form we get:
\bea
 F(\tilde z,t)\rightarrow \frac{F(\tilde z,t)-(f(t)+F(0,t))}{F'(0,t)}= \tilde z+\sum_{k=2}^{\infty} \frac{(t+4)^{2k-2}}{k! 4^{k-1}} \frac{F^{(k)}(0,t)}{F^{\prime}(0,t)} \tilde z^k \,. \nonumber
 \eea
We can get inequalities of the type:
\bea
-\kappa_n  \frac{ n! ~4^{n-1}}{(t+4)^{2n-2}}\le \frac{F^{(n)}(0,t)}{F^{\prime}(0,t)} \le \frac{n~ n!~ 4^{n-1}}{(t+4)^{2n-2}}\,, ~{\rm for}~0\le t\le 4
\eea
By noting that $x=4-t$ and $y=\frac{(t+4)^2-4 z}{4}$ we could convert the above to bounds on $\mW_{p,q}$'s where $F(s,u)= \sum_{p,q=0}^{\infty} \mW_{pq} x^p y^q$. These do not however lead to simple looking constraints on $\mW_{pq}$'s as can be seen from the expansion below

{\scriptsize \[ F(z,t)= z-z^2 \left(\frac{2 \left((t-4) \left((t-4) \mW_{2,2}- \mW_{1,2}\right)+ \mW_{0,2}\right)}{(t-4) \left((t-4)
    \mW_{0,2}+(t-4) \left((t-4)^2  \mW_{2,2}-(t-4)  \mW_{1,2}+2  \mW_{2,1}\right)-2  \mW_{1,1}\right)+2
    \mW_{0,1}}\right)+ \cdots \, . \]}
A systematic examination of these constraints is warranted, which we leave for future work.    
\item {\bf Vacuum polarisation in QED:} In \cite{Mickens} the vacuum polarisation function $\Sigma(s)$ was shown to be  Herglotz and several properties we deduced using this fact. We briefly review the argument here. The vacuum polarisation function can be represented as :
\bea
\Sigma(s) = \frac{s}{\pi} \int_{4 \mu^2}^{\infty} \frac{dw}{w} \frac{\rho(s)}{w-s} \,,\nonumber
\eea 
where the above is the K\"{a}llen-Lehmann spectral representation of the vacuum polarisation two point function $ i \int d^{4} x e^{i q x} \langle 0 |T j^{\mu}(x) j^{\nu}(0) |0 \rangle = -( q^2 g^{\mu \nu} -q^{\mu} q^{\nu}) \Sigma(q^2) $ without the tensor structure and $\rho(s)= \Im \Sigma(s)$ is a positive function for $s\ge 4 \mu^2$.

Let, $s=x+i y$ and $h(w)=\frac{\rho(w)}{\pi w}$ which is non-negative for $w\ge 4 \mu^2$. We can straightforwardly see that \
\bea
\Sigma(x+i y) =  \int_{4 \mu^2}^{\infty} \left[\frac{x(w-x)-y^2}{(w-x)^2+y^2} \right]h(w) dw + i y \int_{4 \mu^2}^{\infty} \frac{x h(w) dw}{(w-x)^2+y^2}  \,,\nonumber
\eea 
 which immediately  implies that $Sign(\Im  \Sigma(s)) =Sign(s)$ or that $\Sigma(s)$ is typically real. 
 
 This allowed the author to conclude that  $\Sigma(s)$ has no complex zeros, and can only be real on the  real axis. Furthermore, since $\Sigma(s)$ is analytic for $|s|\le 4 \mu^2$ by redefining $z=\frac{s}{4 \mu^2}$ it was shown by applying Bieberbach bounds (the first such application of these bounds in physics to our knowledge) to $\Sigma(z)$ in Schlicht form that:
 \bea 
\bigg| \frac{\Sigma^{(n)}}{\Sigma^{(1)}}\bigg| \le \frac{n (n!)}{(4 \mu^2)^{n-1}} \,, \nonumber
 \eea 
 where  $\Sigma^{(n)} = \frac{d^n \Sigma(s)}{d s^n}\big|_{s=0}$.
\end{enumerate}
\section{Results for higher $n$} 
\subsection{Fully-crossing symmetric case}\label{n35}
We shall provide the details of the $n=3,4,5$ cases here. 
\subsection*{$n=3$ results:}
We impose the following conditions for $n=3$ as alluded to:\\ 
1) There is a single  constraint at $n=3$ in $TR_U$ which follows from eq.(\ref{rns}) with $\rho=0$ which reads: 
\bea \label{1}
&&-({a w_1+1})\leq \nonumber \\ &&729 a^4(a^3 w_{03} + a^2 w_{12}+a w_{21}+w_{30})-108 a^2(a^2 w_{02}+a w_{11}+w_{20})+ 3 ({a w_1+1})\nonumber \\ && \quad \quad \qquad\quad \leq 3 ({a w_1+1})\,.
\eea
To emphasise again, compared to the Bieberbach inequalities in \cite{HSZ}, the lower bound which follows from \cite{rogo} for typically real functions is stronger.

\noindent 2) There are the following constraints at $n=3$ in $PB_C$:
\bea\label{2}
w_{21}+\frac{21 w_{30}}{16}\geq 0,\quad w_{12}+\frac{21
   w_{21}}{16}+\frac{297 w_{30}}{256}\geq 0,\\ w_{03}+\frac{9
   w_{12}}{16}+\frac{81 w_{21}}{256}+\frac{729 w_{30}}{4096}\geq 0,\quad 0\leq
   w_{30}\leq \frac{9 w_{20}}{64}\nonumber \,.
\eea
\noindent  Imposing  conditions \eqref{1},\eqref{2}  gives us finite regions as shown in the figures below.\\

 \begin{figure}[h!]
  \centering
   \includegraphics[width=0.7\textwidth]{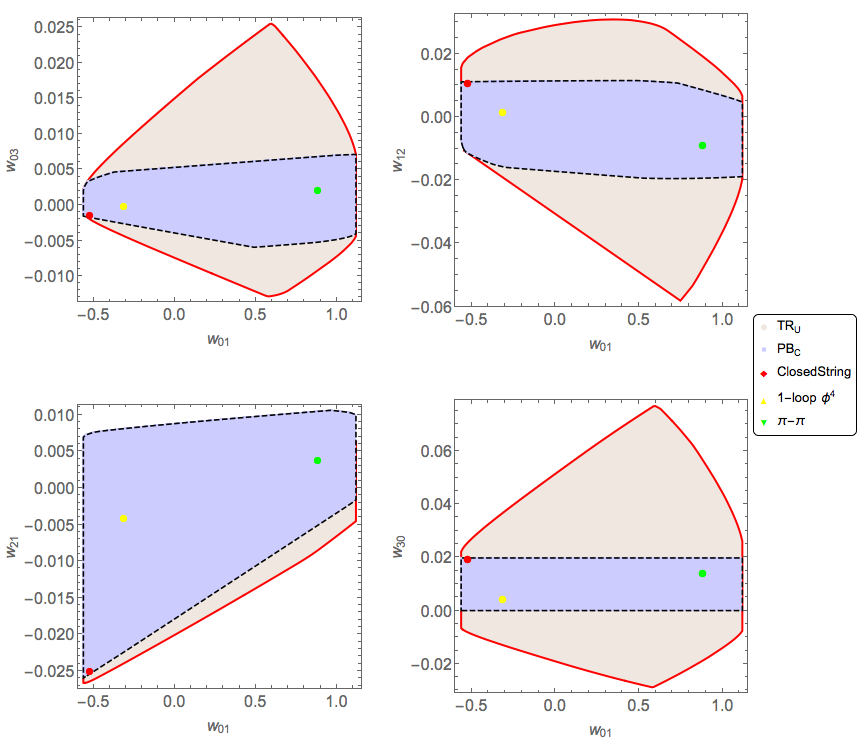}
\caption{The comparison of the regions obtained for $w_{03}$, $w_{21}$ and $w_{12}$ with $w_{30}$.}\vskip 0.5cm
\end{figure} 
\vspace{10 pt}
We have indicated the closed string, 1-loop $\phi^4$ and pion results in the plot above as special points. Note that the string solution is again very close to boundary of the allowed regions, similar to the findings in \cite{nima}.

The results we get at $n=3$ \footnote{The results have been obtained by discretising $a$ between $-\frac{8}{9}$ to $\frac{16}{9}$ with step size $\frac{1}{211}$. For smaller step sizes the results don't change for the first 2 significant digits (Except $w_{01}$ and $w_{n0}$ which were bounded a priori and the full range is realized for them)  so we quote these values. We have attached the Mathematica notebooks along with the submission. }are:
\bea
-0.089&\leq& w_{02}\leq 0.039,~~ -0.1318\leq w_{11}\leq
   0.1054,~0 \leq
   w_{20}\leq 0.140625,\nonumber \\
   -0.5625&\leq& w_{01}\leq 1.125,\quad -0.019\leq
   w_{12}\leq 0.011,~~-0.026\leq w_{21}\leq
   0.0105,\nonumber \\ -0.0059&\leq& w_{03}\leq 0.0071,\quad 0\leq w_{30}\leq
   0.01977\,.
\eea
 As a comparison, if we used $-3$ as the lower bound, rather than $-1$, which follows from the Bieberbach conjecture, we would find the following weaker bounds for $w_{12}, w_{21}, w_{03}$: $-0.028\leq w_{12}\leq 0.022$, $-0.026\leq w_{21}\leq 0.022$, $-0.011\leq w_{03}\leq 0.0089$. 
\subsection*{$n=4$ results:}
We impose the following conditions for $n=4$ as alluded to:\\ 
1) There is a single  constraint at $n=4$ in $TR_U$ and this reads:
\bea 
&&-4 (a w_{01}+1)\leq\nonumber\\ &-&19683 a^{6} (a^4 w_{04}+a^3 w_{13}+ a^2 w_{22}+a w_{31} +w_{40})+4374 a^4( a^3 w_{03} + a^2 w_{12}+a w_{21}+w_{30})\nonumber\\ &-&270 a^2(a^2 w_{02}+ a w_{11}+ w_{20})+4 (a w_{01}+1) \nonumber \\ &&~~~~~~~~~~~~~~~~~\leq 4 (a w_{01}+1) \,.
\eea

\begin{figure}[H]
  \centering
    \includegraphics[width=\textwidth]{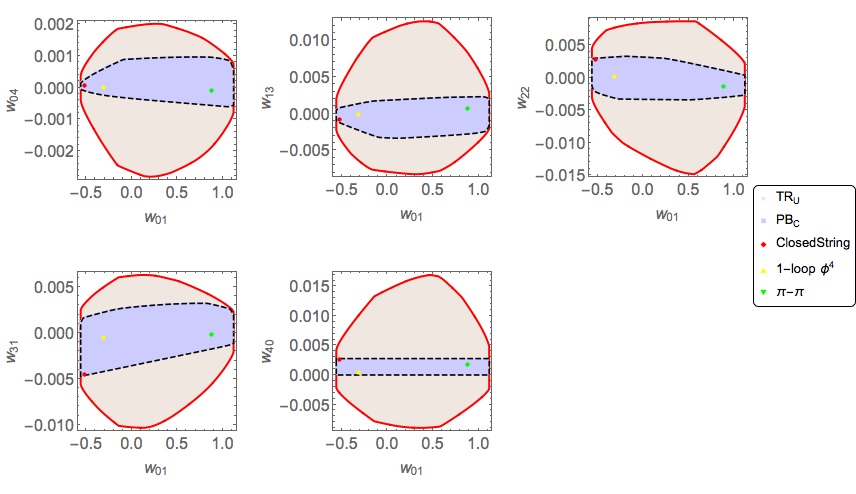}
  \caption{The comparison of the regions obtained for $w_{04}$, $w_{13}$, $w_{22}$  $w_{31}$ and $w_{40}$ with $w_{01}$.}\label{n4} \vskip 0.5cm
\end{figure}

\noindent 2) There are the following constraints at $n=4$ in $PB_C$:
\bea
w_{31}&+&\frac{27 w_{40}}{16}\geq 0,~~ w_{22}+\frac{27}{256} \left(16 w_{31}+17 w_{40}\right)\geq 0, ~w_{13}+\frac{9 \left(768 w_{22}+752 w_{31}+591
   w_{40}\right)}{4096}\geq 0,\nonumber \\   w_{04}&+&\frac{9 \left(4096 w_{13}+9 \left(256 w_{22}+144 w_{31}+81 w_{40}\right)\right)}{65536}\geq 0 \,.
\eea
\vspace*{30 pt}
\noindent Imposing these conditions gives us fig.(\ref{n4}) and bounds below.


%
\bea
-0.5625&\leq& w_{01}\leq 1.125,~0.\leq w_{20}\leq 0.140625,~-0.1318\leq w_{11}\leq 0.1055,~0\leq w_{30}\leq 0.01977,\nonumber\\-0.089&\leq& w_{02}\leq 0.039,~-0.026\leq
   w_{21}\leq 0.0105,~0.\leq w_{40}\leq 0.00278,\nonumber \\~-0.0194&\leq& w_{12}\leq 0.0114,~-0.0047\leq w_{31}\leq 0.0032,~-0.0059\leq w_{03}\leq
   0.0066,\nonumber \\-0.0034&\leq& w_{22}\leq 0.0033,~-0.0033\leq w_{13}\leq 0.0023,~-0.00059\leq w_{04}\leq 0.00095 \,.
\eea

\subsection*{$n=5$ results:}
We impose the following conditions for $n=5$:\\ 
1) There is a single  constraint at $n=5$ in $TR_U$ and this reads:
\bea 
&-&1.24995 \left(a w_{01}+1\right)\leq \nonumber \\ &&531441 a^{8}(a^5 w_{05}+a^4 w_{14}+a^3 w_{23}+a^2 w_{32}+a w_{41}+w_{50})\nonumber\\&-& 157464 a^{6}(a^4 w_{04}+a^3 w_{13}+a^2 w_{22}+a w_{31}+w_{40})\nonumber\\ &+& 15309 a^4(a^3 w_{03}+a^2 w_{12}+ a w_{21}+w_{30})-540 a^2(a^2 w_{02}+a w_{11}+ w_{20})+5 (a w_{01}+1)\nonumber\\ &&~~~~~~~~~~~~~~~~~~~~~~~\leq 5 \left(a w_{01}+1\right) \,.
\eea
where we have obtained the lower bound by using \eqref{rog}.

\noindent 2) There are the following constraints at $n=5$ in $PB_C$:
\bea
w_{41}&+&\frac{33 w_{50}}{16}\geq 0,~ w_{32}+\frac{33 w_{41}}{16}+\frac{657 w_{50}}{256}\geq 0,~w_{23}+\frac{3 \left(2816 w_{32}+3312 w_{41}+3015
   w_{50}\right)}{4096}\geq 0\nonumber \\ w_{14}&+&\frac{3 \left(45056 w_{23}+49920 w_{32}+43056 w_{41}+32427 w_{50}\right)}{65536}\geq 0,\nonumber\\ w_{05}&+&\frac{9 w_{14}}{16}+\frac{81
   w_{23}}{256}+\frac{729 w_{32}}{4096}+\frac{6561 w_{41}}{65536}+\frac{59049 w_{50}}{1048576}\geq 0 \,. 
\eea
\noindent Imposing these conditions gives us finite regions as shown in the figures below. Curiously, in all our plots, the closed string answer is close to a corner of the allowed regions. It will be very interesting to try to construct the amplitudes which live at the boundaries (especially corners) of the allowed regions.

 \begin{figure}[H]
  \centering
    \includegraphics[width=\textwidth]{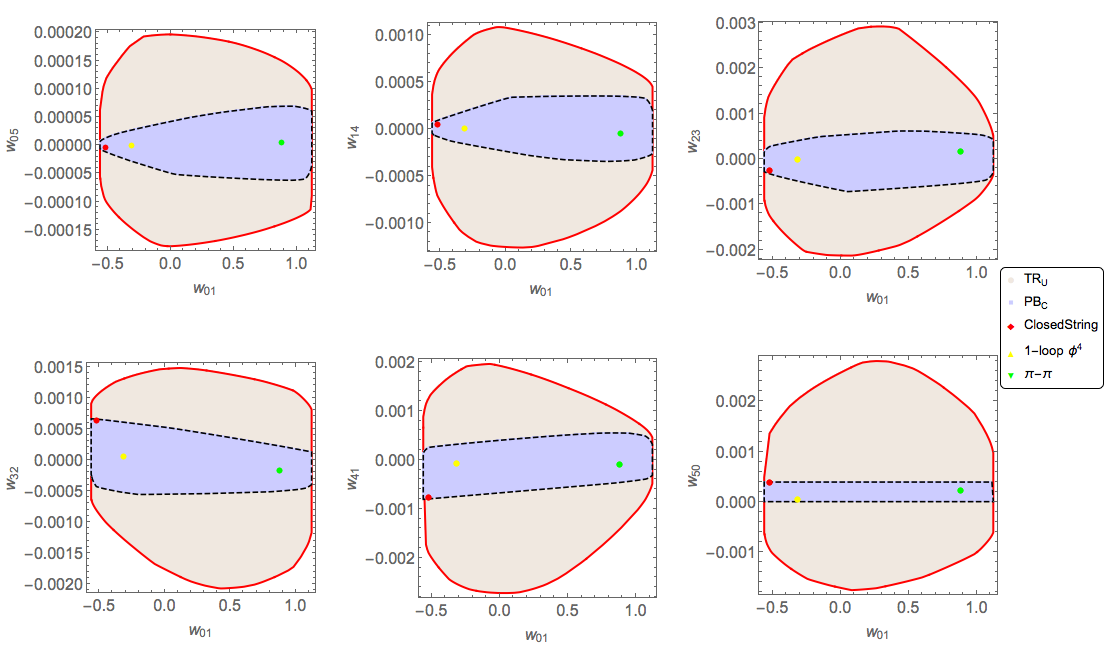}
  \caption{The comparison of the regions obtained for $w_{05}$, $w_{14}$, $w_{23}$,  $w_{32}$, $w_{41}$ and $w_{50}$ with $w_{01}$.}\vskip 0.5cm
\end{figure} 


The results we get  at $n=5$ are:
\bea
-0.5625&\leq& w_{01}\leq 1.125,~0.\leq w_{20}\leq 0.140625,~-0.1318\leq w_{11}\leq 0.1055,~0\leq w_{30}\leq 0.01977,\nonumber\\ -0.089&\leq& w_{02}\leq 0.039,~-0.026\leq
   w_{21}\leq 0.0105,~0\leq w_{40}\leq 0.00278,\nonumber \\ -0.019&\leq& w_{12}\leq 0.011,-0.0047\leq w_{31}\leq
   0.0022, ~0\leq w_{50}\leq 0.00039,\nonumber\\-0.0049&\leq& w_{03}\leq 0.0065,~ -0.0033\leq w_{22}\leq 0.0031,~-0.00081\leq w_{41}\leq
   0.00055,\nonumber\\ -0.0025&\leq& w_{13}\leq 0.0022,~-0.00056\leq w_{32}\leq 0.00066,~-0.00051\leq w_{04}\leq 0.00064,\nonumber\\ -0.00072&\leq& w_{23}\leq
   0.00062,~-0.00034\leq w_{14}\leq 0.00035,~-0.000062\leq w_{05}\leq 0.000069  \,.
\eea
\subsection{Two-channel symmetric case}\label{n235}
We will now provide the details of $n=3$ to $n=5$ cases.The Bieerbach-Rogosinski inequalities give:
\bea
{\bf n=3:}~&-&(a w_{01}+1 )\le 256 a^2( a^3 w_{03}+ a^2 w_{12}+a w_{21}+w_{30})-64 a(a^2 w_{02}+a w_{11}+w_{20})
  \nonumber\\&+&3 a
   w_{01} +3\le 3 ~(a w_{01}+1 ), \\\nonumber\\ 
{\bf n=4:}~&-&4(a w_{01}+1 )\le -4096 a^3(a^4 w_{04}+ a^3
   w_{13}+a^2 w_{22}+aw_{31}+w_{40})\nonumber\\&+&1536 a^2(a^3 w_{03}+a^2 w_{12}+a w_{21}+w_{30}) -160 a(a^2 w_{02}+a w_{11}+w_{20})\nonumber\\&+&4 (a w_{01}+1)  \le 4 (a w_{01}+1 ) ,\\
{\bf n=5 :}~&-&1.25 ~(aw_{01}+1) \le 65536 a^4(a^5 w_{05}+a^4 w_{14}+a^3 w_{23}+a^2 w_{32}+a w_{41}+w_{50})\nonumber\\&-&32768 a^3(a^4 w_{04}+a^3 w_{13}+a^2 w_{22}+a w_{31}+w_{40})+5376 a^2(a^3 w_{03}+ a^2 w_{12}+a w_{21}+w_{30})\nonumber\\&-&320 a(a^2 w_{02}+a w_{11}+w_{20})+5(a w_{01}+1) \le 5(a w_{01}+1),
\eea
The positivity bounds $PB_c^{(2)}$ \eqref{ps2} are:
\bea
{\bf n=3:}~ w_{21}&+&6 w_{30} \ge 0, ~~w_{12}+\frac{9}{2} w_{21}+\frac{63}{4} w_{30} \ge 0,\nonumber\\ w_{03}&+&3 w_{12}+9w_{21}+27w_{30}\ge0, \nonumber\\
{\bf n=4:}~ w_{31}&+&\frac{15}{2}w_{40}\ge 0,~w_{22}+6 w_{31}+\frac{99}{4}w_{40},\nonumber\\  w_{13}&+&\frac{9}{2}w_{22}+\frac{63}{4}w_{31}+\frac{405}{8}w_{40}\ge0,\nonumber\\ w_{04}&+& 3 w_{13}+9w_{22}+27w_{31}+81w_{40}\ge0, \nonumber\\
{\bf n=5:}~ w_{41}&+&\frac{9}{2}w_{50}\ge 0,~w_{32}+\frac{15}{2} w_{41}+36w_{50},\nonumber\\  w_{23}&+&6w_{32}+\frac{99}{4}w_{41}+\frac{351}{4}w_{50}\ge0,\nonumber\\ w_{14}&+& \frac{9}{2} w_{23}+\frac{63}{4}w_{32}+\frac{405}{8}w_{41}+\frac{2511}{16}w_{50}\ge0\nonumber\\ w_{05}&+& 3 w_{14}+9w_{23}+27w_{32}+81w_{41}+243 w_{50}\ge0,
\eea

The bounds we get are the following :
\bea
{\bf n=3:}~ &-&1.6875\leq w_{11},~ 0\leq w_{20} \leq 0.375,~-0.75\leq w_{01}, \nonumber\\~&-&0.84375\leq w_{21},~ 0\leq w_{30}\leq 0.140625, \nonumber\\
{\bf n=4:}~ &-&1.6875\leq w_{11},~ 0\leq w_{20} \leq 0.375,~-0.75\leq w_{01}, \nonumber\\~&-&0.84375\leq w_{21},~ 0\leq w_{30}\leq 0.140625,~\nonumber\\ &-&0.4113\leq w_{31},~0\leq w_{40} \leq 0.0548, \nonumber \\
{\bf n=5:}~ &-&1.6875\leq w_{11},~ 0\leq w_{20} \leq 0.375,~-0.75\leq w_{01}, \nonumber\\~&-&0.84375\leq w_{21},~ 0\leq w_{30}\leq 0.140625,~ \nonumber\\ &-&0.4113\leq w_{31},~0\leq w_{40} \leq 0.0548 \nonumber \\&-&0.0924\leq w_{41},~0\leq w_{50} \leq 0.02055,
\eea
The open string values from table in appendix (\ref{appG}) when converted to our units are 
\bea
w_{11}\approx -0.132\,,\quad w_{20}\approx 0.127\,,\quad w_{21}\approx -0.025\,,\quad w_{30}\approx 0.045 \nonumber \\
w_{31}\approx -0.031\,,\quad w_{40}\approx 0.016\,,\quad w_{41}\approx -0.014\,,\quad w_{50}\approx 0.006.
\eea
All values are in the allowed range.

\section{Two channel-parametric dispersion} \label{appB}
We would like to find parametric dispersion relations for 2-channel crossing symmetry following \cite{AK, ASAZ}. We begin by considering conics (quadratic hyper-surfaces) in the variables $s_1 = s- \frac{\mu}{3},s_2= t- \frac{\mu}{3}$ for a fixed $u=u_0<  \mu $ (so that we can ignore the u-channel cut)  and $\mu=4 m^2$ which satisfy the equation
${\mathcal C}: (s_1(z)-a_1)(s_2(z)-a_2)=K$ with $a_i,K$ being a real parameters. 
We would need to choose $a_i$ and $K$ such that:\\
(1) ${\mathcal C} \subset D$.\\
(2) There is a complex coordinate $z$ on ${\mathcal C}$ such that the mapping $z\rightarrow (s,t)$ is rational.\\
where, $D$ is the Martin domains  i.e., regions where the amplitude is analytic in $s$ and $t$ in $D$ minus the physical cuts $s ,t  \ge \mu $.\\
Point \noindent(2) can be  easily satisfied since we know that all irreducible conics have rational parametrizations and choosing $a_1,a_2$ such that $a_1 a_2= K$ makes ${\mathcal C}$ irreducible.
For (1) note that there are domains $F \subset D$ of  of the form:
\bea
F= \{ s,t | |(s-a_1)(t-a_2)|< A_1\} \cup \{ s,t | |(t-a_2)(u-a_3)|< A_2\} \cup \{ s,t | |(u-a_3)(s-a_1)|< A_3\} \nonumber
\eea
For a fixed $u=u_0 < \mu$  by choosing $A = Min\left (A_1,\frac{A_2 A_3}{(u_0-a_3)^2}\right ) $ and $K<A$ guarantees (1). Since we are interested mainly in the symmetric case we choose $a_1=a_2=a, K=a^2$  to get the family of conics:
\bea
{\mathcal C}: (s_1(z)-a)(s_2(z)-a)=a^2 
\eea
with $a$ being a real parameter. \\The $s_i$'s can then be parametrised as:
\bea
s_k = a - a \frac{ (z_k+ z)^2}{(z_k- z)^2}, ~~k=1,2
\eea
where $z_k =\pm 1$ are the square roots of unity. 

The exact range of $a$ for analyticity of the amplitude is actually much bigger $-\infty< a \le \frac{2 \mu}{3}$ and can be seen by repeating the analysis in \cite{AK}. The domain of analyticity of the amplitude is in the interior of the Martin-Lehmann ellipse:

\[
  s_2(s_1,a) +\frac{4}{3} < 
  \begin{cases}
  16+ \frac{64}{s-4}, & \text{for } 4< s <16 \\
      \frac{256}{s}, & \text{for } 16<s< 32 \\
      4+ \frac{64}{s-16}, & \text{for } s \ge 32 \\
  \end{cases}
\]
By using 
\bea
a= \frac{s_1 s_2}{s_1+s_2} 
\eea
we get $s_2(s_1,a) = \frac{a s}{s-a}$. We can check that the above conditions are all satisfied for a in $(-\infty ,\frac{2 \mu}{3} ]$. As we shall see later we can restrict further for our purposes to just $0 \le a <\frac{\mu}{3}$.

Note that $a$  and $z^2$ are manifestly crossing symmetric  (as $s_1\rightarrow s_2 $ is taking $z \rightarrow -z$ ) in $s_1,s_2$. The amplitude can be written as a function of $z,a$ as $ {\bar {\mathcal M}} (z,a) = {\mathcal M}(s_1(z),s_2(z))$ and has physical cuts for $s_1 \ge \frac{2 \mu}{3}$ and $s_2 \le -u_0-\frac{\mu}{3}  $ and vice versa. These physical cuts in $s_k$ plane  get mapped to arcs on the unit-circle in the complex $z$ plane. We can show this as follows:\\
\begin{figure}[h!]
 \centering
  \includegraphics[width=0.8\textwidth] {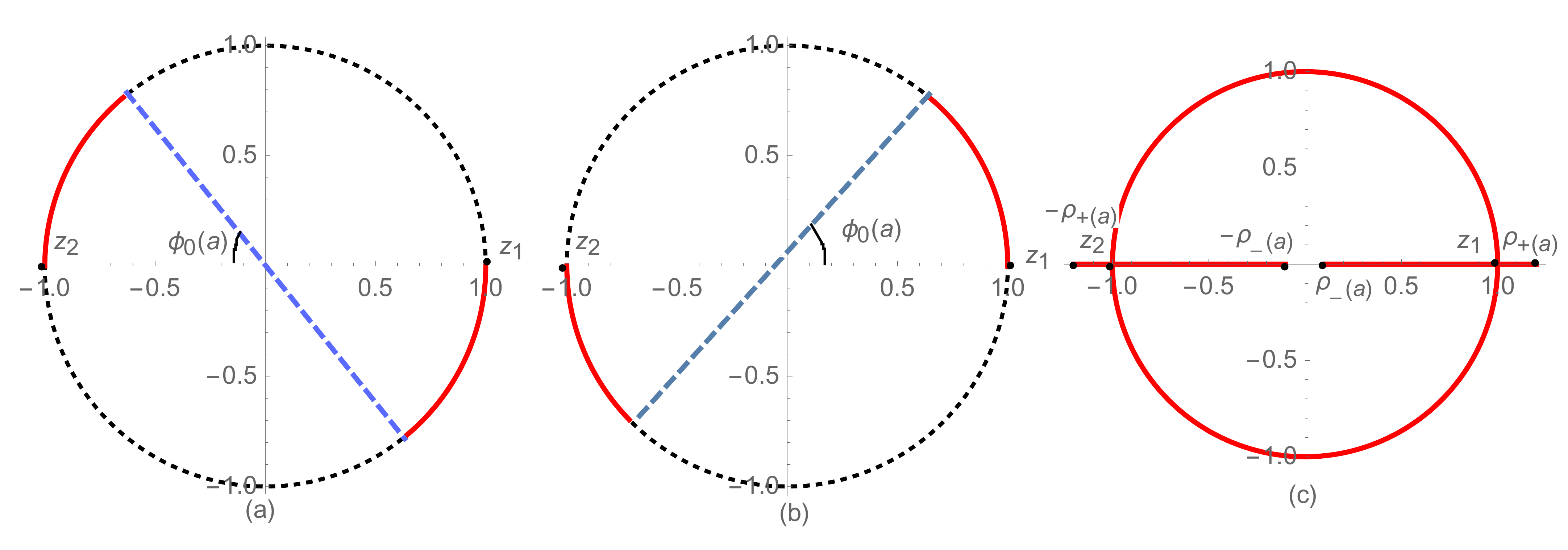}
 \caption{ The three image of the physical cuts corresponding to the the three cases (a) $0<a< \frac{\mu}{3}$(b)$\frac{\mu}{3} <a< \frac{2\mu}{3}$ and (c) $a \le 0$}
 \end{figure} 
 
\noindent Inverting $z$ for $s_1$ gives:
\bea
z_\pm = \frac{(s_1-2a)\pm 2 ~ i~  \sqrt{a (s_1- a)}  } {s_1}\,.
\eea
Note that $z_{+} z_{-} = 1 $ which would mean $|z_{+}|=|z_{-}|=1$ provided $a(s_1-a) >0$ or equivalently either $0<a <s_1$ or $s_1<a<0$. If we choose  $a$ in $[-\frac{\mu}{3} ,0) \cup (0, \frac{2 \mu}{3}]$ we can see that the whole branch cut  gets mapped to the boundary of the unit circle.

We can determine the exact location of the branch cuts $V(a)=V_1(a) \cup V_2(a)$ as follows $s_1 \ge \frac{2 \mu}{3}$ is possible only if $s_1$ is real and this corresponds to either $a=0$ or $|z|=1$ or $Arg(z)= 0$. Considering these cases we get:
\[
  V_1(a) = 
  \begin{cases}
  \{ z~ |~ |z|=1, 0 \le Arg(z) \le\phi_ 0(a)\} , & \text{for } 0<a< \frac{\mu}{3} \\
     \{ z ~|~ |z|=1,  \phi_0(a) \le Arg(z) \le \pi \} , & \text{for } \frac{\mu}{3} <a< \frac{2\mu}{3} \\
    \{ z~ |~ |z|=1, \frac{-\pi}{2} \le Arg(z) \le 0\} \cup \{ z~ |~ \rho_{-}(a)\leq |z| \leq \rho_{+}(a) , & \text{for } a \le 0
  \end{cases}
  \]

where~$\phi_0(a)= \tan^{-1} \Bigg\{ \frac{(a\left(\frac{2 \mu}{3}-a\right))^{1/2}}{(\frac{ \mu}{3}-a)}\Bigg\}, 0< \phi_0(a) \le \frac{-\pi}{2}$ and $\rho_{\pm} =\frac{(\mu -3 a)\pm \sqrt{3 a(3a- 2 \mu)} }{\mu}$ with $V_2(a)= \exp {({\rm i} \pi)} V_1(a)$.
Inside each Martin domain $D$ for fixed $s_3$ we assume 
\bea
{\mathcal M}(s_1,s_2) =o(s_1)  ~ {\rm for} ~|s_1|\rightarrow \infty,~ s_2={\rm fixed}, ~(s_1,s_2) \in D
\eea
which translates to 
\bea
{\mathcal {\bar M}}(z,a) =o\left(\frac{1}{(z-z_k)^2} \right), ~ {\rm}~ z \rightarrow z_k, ~ a~{\rm fixed}\,.
\eea
Since $z \rightarrow z_k$ are the values of $z$ that keeps $s_j$ fixed and take $s_i$ to $\infty$.
Crossing symmetric dictates that ${\mathcal {\bar M}}(z,a)$ is a function of $z^2$:
\bea
{\mathcal {\bar M}}_0(z,a) = \sum_{n=0}^{\infty} \alpha_n(a) z^{2n}
\eea
and we can write a dispersion relation:
\bea
{\mathcal {\bar M}}_0(z,a) =\alpha_0 + \frac{z^2}{\pi(1-z^2)}  \int_{V(a) } dz'
\frac{z'^{2}-1}{z'^{2} (z'-z)} {\bar{\mathcal A}}(z',a)\,.
\eea
Since only powers $z^{2n}$ appear we can rewrite the above as follows:
\bea
{\mathcal {\bar M}}_0(z,a) =\alpha_0 + \frac{z^2}{\pi(1-z^2)}  \int_{V(a) } dz'
\frac{z'^{2}-1}{z' (z'^2-z^2)} {\bar{\mathcal A}}(z',a)\,.
\eea
Using $z'=z'(s^{+}_{1})$ we can evaluate the above to get:
\bea \label{dispersion}
{\mathcal {\bar M}}_0(z,a) =\alpha_0 + \frac{1}{\pi}  \int_{\frac{2 \mu}{3}}^{\infty} \frac{ds'}{s'}
{\bar{\mathcal A}}\left(s_1',s_2(s_1^{'},a)\right) H(s_1',s_1,s_2)\,,
\eea
\vspace*{-20 pt}
where,
\bea
H(s_1',s_1,s_2)&=&\left[\frac{s_1}{(s_1'-s_1)}+\frac{s_2}{(s_1'-s_2)}\right]\\
s_2(s_1',a) &=& \frac{ a~s_1'}{s_1'-a} \,.
\eea
In the $z$-variable the kernel looks like:
\bea 
H(s_1',s_1(z),s_2(z))&=& \frac{16 a(2a-s_1')z}{s_1'^2 \left(1-\left(-2+4\frac{(2a-s_1')^2}{s_1'^2}\right)z+z^2\right)}
\eea
since, $0\le \frac{(2a-s_1')^2}{(s_1')^2} \le1$ for $\frac{2 \mu}{3} \le s_1'<\infty$ and $0 <a\le \frac{\mu}{3}$ for which the kernel has no poles inside the disk $|z|<1$. Note that we also need $2a-s_1' <0$, as for large $s_1'$ we have  $2a-s_1' <0$ we demand this maintained throughout to prevent  the kernel from vanishing anywhere inside the integration domain, the  above is of the form 
\bea \label{kernel}
H(s_1',s_1(y),s_2(y))&=& \frac{16 a(2a-s_1')}{s_1'^2} \frac{y}{1-\gamma y+y^2}
\eea
with $y=z^2$  and $|\gamma| \le 2$ which is just a M\"{o}bius transform of the Koebe function and so the kernel is univalent everywhere within the integration domain for $0<a\le \frac{\mu}{3}$.\\\\
\vspace*{-30pt}
\subsection*{Sanity checks}
We now compare the crossing symmetric dispersion relation with exact result for the pole subtracted open string amplitude by computing the absorptive part, which is sum of delta functions and  substituting it into the dispersion relations, which reduces the integral into an infinite sum. We then truncate the series upto $k_{max} =100$ and compare with the exact result. We find a reasonable agreement between the two. \footnote{The open string amplitude has a Regge behaviour  ${\mathcal M}(s_1,s_2) ~\sim o(s_1^{s_2})$ so its reasonable to expect a good agreement for $s_2<1$ here, eventhough for $s_2>1$ there is continues to be a decent agreement between the two.}

The open string amplitude is
\bea
-s_1 s_2 {\mathcal M}(s_1,s_2) = \frac{\Gamma(1-s_1) \Gamma(1-s_2)}{\Gamma(1-s_1-s_2)}
\eea

From the crossing symmetric dispersion we get,
\bea
-s_1 s_2 {\mathcal M}^{crossing}(s_1,s_2)=1-\sum_{k=0}^{\infty} \frac{(-1)^{k+1} \left(\frac{1}{k-s_1+1}+\frac{1}{k-s_2+1}-\frac{2}{k+1}\right) \Gamma \left(\frac{(k+1)^2}{-k+\frac{s_1 s_2}{s_1+s_2}-1}+k+2\right)}{k!~ \Gamma
   \left(\frac{k (k+1)+\frac{s_1 s_2}{s_1+s_2}}{-k+\frac{s_1 s_2}{s_1+s_2}-1}\right)}
\eea
\begin{figure}[h!]
 \centering
  \includegraphics[width=1.1 \textwidth] {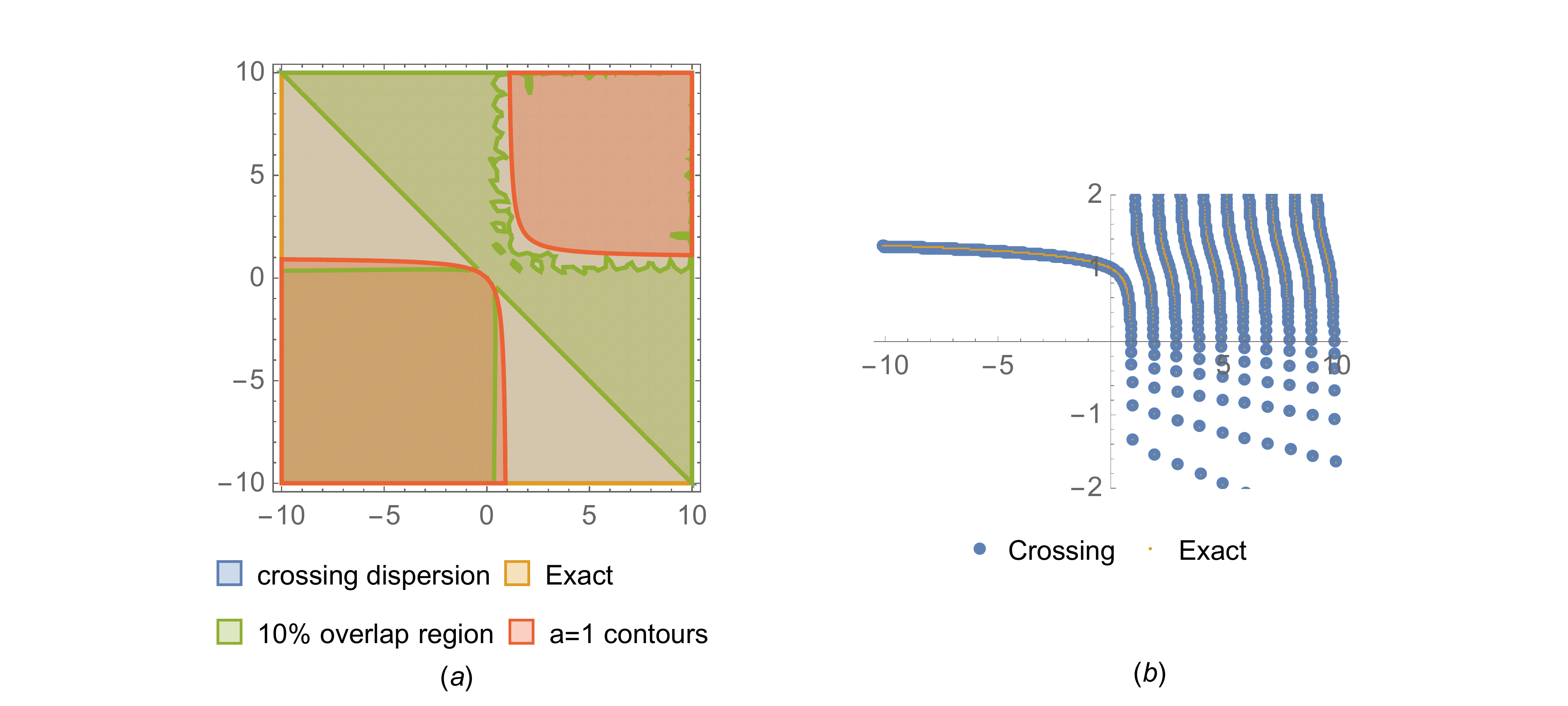}
 \caption{The comparison between the result of the dispersion relation and exact result of the pole subtracted open string amplitude (a)Region in $s_1,s_2$ space where there is a 10 \% overlap for $k_{max}=100$ is the green region(b)The curves of both functions  for $s_2=\frac{1}{11}$.}
 \end{figure} 
 \vspace*{20 pt}
 The plot confirms the validity of the dispersion relation.
 
 \section{Some details about \eqref{ps2}}\label{appH}
 We give some algebraic details of how \eqref{ps2} was obtained here.
 We begin by noting that the sign of $ \mW_{n-m,m}$ depends purely on $\mathcal{B}_{n,m}^{(\ell)}$'s.
 \bea
 \mW_{n-m,m}=\int_{\frac{2\m}{3}}^{\infty}\frac{d s_1}{2\pi s_1^{n+m+1}}\Phi(s_1)\sum_{\ell=0}^{\infty}\left(2\ell+2\a\right)a_\ell(s_1)\mathcal{B}_{n,m}^{(\ell)}(s_1)
 \eea
We consider arbitrary linear combinations of $\mathcal{B}_{n,m}^{(\ell)}$ and demand that the result is a positive sum of Gegenbauer polynomials $C_{\ell}^{(\alpha)}$'s. Since the $C_{\ell}^{(\alpha)}$ are positive for argument bigger than $\xi \ge1$ this guarantees the positivity of the combination of $\mW_{n-m,m}$ thereby giving us \eqref{ps2}. We illustrate this for a few cases:
\bea
\mathcal{B}_{n,1}^{(\ell)}\left(\delta+\frac{2\mu}{3}\right)+\chi_n^{(0,1)}\mathcal{B}_{n,0}^{(\ell)}\left(\delta+\frac{2\mu}{3}\right)&=&\frac{(n+1)}{\left(\delta+\frac{2 \mu}{3}\right)^{n+2}}  C_{\ell}^{(\alpha)}\left(\delta+\frac{2\mu}{3}\right)\nonumber \\&+&\frac{4 \alpha \xi_0}{  \left(\delta+\frac{2 \mu}{3}\right)^{n+2}} C_{\ell-1}^{(\alpha+1)}\left(\delta+\frac{2\mu}{3}\right) \ge 0 \,,
\eea
\bea
\mathcal{B}_{n,2}^{(\ell)}\left(\delta+\frac{2\mu}{3}\right)+\chi_n^{(1,2)}\mathcal{B}_{n,1}^{(\ell)}\left(\delta+\frac{2\mu}{3}\right)+\chi_n^{(0,2)}\mathcal{B}_{n,0}^{(\ell)}\left(\delta+\frac{2\mu}{3}\right)=  \frac{(n-1)(n+2)}{2 \left(\delta+\frac{2 \mu}{3}\right)^{n+3}}C_{\ell}^{(\alpha)}\left(\delta+\frac{2\mu}{3}\right)\nonumber \\ +\frac{4 n \xi_0 \alpha}{ \left(\delta+\frac{2 \mu}{3}\right)^{n+3}}C_{\ell-1}^{(\alpha+1)}\left(\delta+\frac{2\mu}{3}\right)+\frac{8 \xi_0^2 \alpha (\alpha+1)}{2 \left(\delta+\frac{2 \mu}{3}\right)^{n+3}}C_{\ell-2}^{(\alpha+2)}\left(\delta+\frac{2\mu}{3}\right) \ge 0  \,,\nonumber \\
\eea
 where the coefficients $\chi_n^{(r,m)} (\mu,\delta) $ satisfy the following recursion relation:
\bea
\chi_n^{(m,m)} (\mu,\delta) &=&1\nonumber\\
\chi_n^{(r,m)} (\mu,\delta) &=& \sum_{j=r+1}^m (-1)^{j+r+1} \chi_n^{(j,m)} \frac{{\mathscr U^{\alpha}}_{n,j,r} (\delta + \frac{2 \mu}{3})}{{\mathscr U^{\alpha}}_{n,r,r}(\delta + \frac{2 \mu}{3})} \,, 
\eea
with  ${\mathscr U}^{\alpha}_{n,m,k}=  \frac{2^j (2 \xi_0)^j (\alpha)_k (m+n-2j)\Gamma(n-j)}{(\frac{2}{3})^{m+n+1}j!(m-j)!(n-m)!}$ and $\alpha=\frac{d-3}{2}$.

 \section{Twice subtracted dispersion and typical-realness} \label{appF}
 We note that if we attempted to write down a twice-subtarcted dispersion relation for the two channel symmetric case we would get a two-channel symmetric kernel of the form:
\bea
H(s_1',s_1,s_2) = \frac{(s_1 s_2- s_1' (s_1+s_2))}{(s_1')^2} \left(\frac{s_1}{s_1-s_1'}+\frac{s_2}{s_2-s_1'}\right)\,,
\eea
which we shall argue now is also typically real. 

In the $\tilde z$-variable the kernel looks like:
\bea 
H(s_1',\tilde z,a)&=& \frac{16 a(2a-s_1')\tilde z}{s_1'^2 \left(1-\left(-2+4\frac{(2a-s_1')^2}{s_1'^2}\right) \tilde z+\tilde z^2\right)} \times \frac{16 a (a-s_1') \tilde z}{(s_1')^2 (1-\tilde z)^2} \,,
\eea
which apart from normalisation factors is the product of two typically real functions $\frac{\tilde z}{1+\gamma \tilde z+\tilde z^2 }$ and the Koebe function $\frac{\tilde z}{(1-\tilde z)^2}$ and hence is typically real  due to the \eqref{improgo} once we choose $f_1( \tilde z) =f_2(\tilde z)=\frac{\tilde z}{1+\gamma \tilde z+\tilde z^2 }$ and $f_3(\tilde z)=f_4( \tilde z)= \frac{\tilde z}{(1-\tilde z)^2}$ as
\bea
H(s_1',\tilde z,a)= \sqrt[4]{f_1(\tilde z)f_2( \tilde z) f_3(\tilde z)f_4( \tilde z)}\,.
\eea

We choose to work with once subtracted dispersion relation for its simplicity and identical structure to that of the fully crossing-symmetric dispersion relation\footnote{Interestingly since the Regge behaviour of  the open string amplitude is $o(s_1^{s_2})$ which exactly squares to the closed string Regge behaviour $o(s_1^{2 s_2})$, so our choice in an amusing way encodes the double copy structure of the string amplitude.}.  Note that we get the same same range for $a$ namely $0< a<\frac{\mu}{3}$ if we demanded the amplitude is regular inside the unit disk $|\tilde z| <1$ however the sign of $\beta_1$ is now identically positive. We leave the analysis of this case for future explorations.
\section{Various Amplitudes}\label{appG}
Here we will summarize various amplitudes we will use to benchmark the 2-channel theories. The 3-channel benchmarking amplitudes can be found in the appendix of \cite{HSZ}.
\subsection{Pole subtracted  open string amplitude}
We consider the open string amplitude (see eg.\cite{green}) with the massless pole subtracted out and also suitable kinematic factors stripped off to be constant with $o(s_1)$ fall-off:
\bea\label{os}
{\mathcal M}(s_1,s_2) =- \frac{\Gamma(1-s_1) \Gamma(1-s_2)}{s_1 s_2\Gamma(1-s_1-s_2)} +\frac{1}{s_1s_2}
\eea
We look the low energy expansion where the coefficients  involve Zeta functions and we have listed the numerical values of these in the table below. 

\begin{table}[h!]
\centering
$\begin{array}{|c|c|c|c|c|c|c|}
   \hline
  \mW_{p,q} & q=0 & q=1 &q= 2 &q= 3 &q= 4 &q= 5 \\
   \hline
p= 0 & 1.64493 & -1.89407 & 1.9711 & -1.99247 & 1.99808 & -1.99952 \\
 \hline
 p=1 & 1.20206 & -3.01423 & 4.9908 & -6.99406 & 8.99756 & -10.9991 \\
  \hline
 p=2 & 1.08232 & -4.02884 & 9.0032 & -15.9979 & 24.9983 & -35.9991 \\
  \hline
 p=3 & 1.03693 & -5.02339 & 14.0076 & -30.0009 & 54.9994 & -90.9994 \\
  \hline
 p=4 & 1.01734 & -6.01581 & 20.0078 & -50.0024 & 105. & -196. \\
  \hline
 p=5 & 1.00835 & -7.00989 & 27.0063 & -77.0028 & 182.001 & -378. \\
  \hline
\end{array} $
\caption{The $\mW_{p,q}$ of the pole subtracted open string amplitude.}
\end{table}

We note that while we are looking at the pole subtracted amplitude with a kinematic factor stripped off, $a_{\ell}(s_1)$ still continues to be positive for spacetime dimensions, $d\leq 10$. 
We have the following formula for the $a_{\ell}$'s interms of the  absorptive part 
\bea
a_{\ell}(s_1) =\int_{-1}^1 dx {\mathcal A}(s_1,s_2) (1-x)^{\a-1/2} C_{\ell}^{\a}(x)
\eea
where $x=1+\frac{2 t}{s}$.
Since the string amplitude has poles on the positive real axis at $s_1=n$ for $n\in \mathbb{Z}_{+}$. We can compute the absorptive part as the residues at the poles and get the following formula for $a_{\ell}$:
\bea
a_{\ell}(s_1=n) =\int_{-1}^1 dx\frac{2 (-1)^n}{n^2(-1+x)(n-1)!} \left(\frac{2-n-n x}{2}\right)_n (1-x)^{\a-1/2} C_{\ell}^{\a}(x)
\eea
We need $a_{\ell} >0$ and we can check by evaluating the integral above that for $d\le 10$ and all $n\geq 1$ this is indeed true. Since this the main ingredient in our analysis, it is justified to work with this case.

\subsection{1-loop $\phi^4$ amplitude for complex scalar}
We consider the standard 1-loop result (omitting the couping in front since it will cancel in the ratio $w_{pq}$) for $\phi +\phi^{*} \rightarrow \phi+\phi^{*}$:
\bea\label{bs}
{\mathcal M}(s_1,s_2) =-\frac{2 \sqrt{s_1-\frac{8}{3}} \tanh
   ^{-1}\left(\frac{\sqrt{s_1+\frac{4}{3}}}{\sqrt{s_1-\frac{8}{3}}}\right)}{\sqrt{s_1+\frac{4}{3}}}-\frac{2 \sqrt{s_2-\frac{8}{3}} \tanh
   ^{-1}\left(\frac{\sqrt{s_2+\frac{4}{3}}}{\sqrt{s_2-\frac{8}{3}}}\right)}{\sqrt{s_2+\frac{4}{3}}}
\eea
We can look the low energy expansion\footnote{We thank D. Chowdhury and A. Das for verifying these numbers for us.}  of the above amplitude and list the numerical values of the coefficients in the table below:

\begin{table}[ht]
\centering
$\begin{array}{|c|c|c|c|c|c|c|}
   \hline
  \mW_{p,q} & q=0 & q=1 &q= 2 &q= 3 &q= 4 &q= 5 \\
   \hline
p= 0 & -1.7408& -0.06635 & 0.00344& -0.0002673& 0.0000245 &
   -2.482\times 10^{-6} \\
 \hline
 p=1 & 0.2292 & -0.02092 & 0.00233& -0.0002795 & 0.0000348 & - \\
  \hline
 p=2 &0 .03317 & -0.006892 & 0.001203 & -0.0001966 & 0.0000310 &-
   \\
  \hline
 p=3 & 0.006974 & -0.00233 & 0.000559 & -0.0001161& - & -\\
  \hline
 p=4 & 0.001723 & -0.0008022 & 0.0002458 & -0.0000620 & - & - \\
  \hline
 p=5 & 0.0004661 & -0.0002795 & 0.0001045 &  -&  -&  -\\
  \hline
\end{array} $
\caption{The $\mW_{p,q}$ corresponding to the 1-loop $\phi^4$ amplitude for charged scalars.}\vskip 0.5cm
\end{table}
\vskip 2cm

\subsection{Pion amplitude}
We will use the $\pi^0\pi^+\rightarrow \pi^0\pi^+$ amplitude at $s_0=0.35$ using the S-matrix bootstrap \cite{pionworks}.  This amplitude has 2-channel symmetry with positive $a_\ell$'s for even spins. We will project on to even spins only to study EFT bounds. We can use the $a_\ell(s)$'s provided by the bootstrap in the formula eq.(\ref{Belldef}) to reconstruct $\mW_{pq}$'s. The results converge very fast in spin.
The $w_{pq}$ we obtained are recorded below (we quote up to 2 significant figures):
\be
w_{01}= 0.27\,,\quad w_{20}=0.13\,,\quad w_{21}=-0.0037\,,\quad w_{31}=-0.002\,,\quad w_{30}=0.020\,,\quad w_{40}=0.0039\,.
\ee


\begin{thebibliography}{99}
\bibitem{Aharonov}
Y.~Aharonov, A.~Komar and L.~Susskind,
Phys. Rev. \textbf{182}, 1400-1403 (1969)
\bibitem{Pham}
T.~N.~Pham and T.~N.~Truong,
Phys. Rev. D \textbf{31}, 3027 (1985)

\bibitem{anant}
B.~Ananthanarayan, D.~Toublan and G.~Wanders,
Phys. Rev. D \textbf{51}, 1093-1100 (1995)
[arXiv:hep-ph/9410302 [hep-ph]].

\bibitem{Adams}
A.~Adams, N.~Arkani-Hamed, S.~Dubovsky, A.~Nicolis and R.~Rattazzi,
JHEP \textbf{10}, 014 (2006)
[arXiv:hep-th/0602178 [hep-th]]
\bibitem{RMTZ}
C.~de Rham, S.~Melville, A.~J.~Tolley and S.~Y.~Zhou,
``Positivity bounds for scalar field theories,''
Phys. Rev. D \textbf{96}, no.8, 081702 (2017)
[arXiv:1702.06134 [hep-th]].


\bibitem{tolley}
A.~J.~Tolley, Z.~Y.~Wang and S.~Y.~Zhou,
``New positivity bounds from full crossing symmetry,''
[arXiv:2011.02400 [hep-th]].


   \bibitem{rattazzi}
B.~Bellazzini, J.~Elias Mir\'o, R.~Rattazzi, M.~Riembau and F.~Riva,
``Positive Moments for Scattering Amplitudes,''
[arXiv:2011.00037 [hep-th]].
   
\bibitem{SCH} 
  S.~Caron-Huot and V.~Van Duong,
  ``Extremal Effective Field Theories,''
  JHEP {\bf 2105}, 280 (2021)
  [arXiv:2011.02957 [hep-th]].

\bibitem{SCH2} 
  S.~Caron-Huot, D.~Mazac, L.~Rastelli and D.~Simmons-Duffin,
  ``Sharp Boundaries for the Swampland,''
  arXiv:2102.08951 [hep-th].

\bibitem{sasha}
M.~Correia, A.~Sever and A.~Zhiboedov,
``An Analytical Toolkit for the S-matrix Bootstrap,''
[arXiv:2006.08221 [hep-th]].

\bibitem{bern}
Z.~Bern, D.~Kosmopoulos and A.~Zhiboedov,
``Gravitational Effective Field Theory Islands, Low-Spin Dominance, and the Four-Graviton Amplitude,''
[arXiv:2103.12728 [hep-th]].

\bibitem{meltzer} D.~Meltzer,
``Dispersion Formulas in QFTs, CFTs, and Holography,''
JHEP \textbf{05}, 098 (2021)
[arXiv:2103.15839 [hep-th]].

   \bibitem{Miro:2021rof}
J.~E.~Mir\'o and A.~Guerrieri,
``Dual EFT Bootstrap: QCD flux tubes,''
[arXiv:2106.07957 [hep-th]].
   
\bibitem{nima}
N.~Arkani-Hamed, T.~C.~Huang and Y.~T.~Huang,
``The EFT-Hedron,''
[arXiv:2012.15849 [hep-th]].

\bibitem{HSZ}
P.~Haldar, A.~Sinha and A.~Zahed,
``Quantum field theory and the Bieberbach conjecture,'' SciPost Phys. {\textbf {11}}, 002 (2021)
[arXiv:2103.12108 [hep-th]].

\bibitem{ASAZ}
A.~Sinha and A.~Zahed,
``Crossing Symmetric Dispersion Relations in QFTs,''
Phys. Rev. Lett. \textbf{126}, no.18, 181601 (2021)
[arXiv:2012.04877 [hep-th]].
%


\bibitem{AK}
G.~Auberson and N.~N.~Khuri,
``Rigorous parametric dispersion representation with three-channel symmetry,''
Phys. Rev. D \textbf{6}, 2953-2966 (1972).

\bibitem{khuriuni}
N.~N.~Khuri and T.~Kinoshita,
``Forward Scattering Amplitude and Univalent Functions,''
Phys. Rev. \textbf{140}, B706-B720 (1965).

\bibitem{saso}
S.~Grozdanov, 
``Bounds on Transport from Univalence and Pole-Skipping,"
Phys. Rev. Lett. \textbf{126}, 051601 (2021) [arXiv:2008.00888 [hep-th]].

\bibitem{hebbar}
J.~Elias Mir\'o, A.~L.~Guerrieri, A.~Hebbar, J.~Penedones and P.~Vieira,
``Flux Tube S-matrix Bootstrap,''
Phys. Rev. Lett. \textbf{123}, no.22, 221602 (2019)
[arXiv:1906.08098 [hep-th]].

  \bibitem{Goluzin} 
  G.M.~Goluzin,
  ``On typically real functions,''
Rec. Math. (Math Sbronik) N.S. vol. 27(69)(1950) pp.201-218.
 
\bibitem {rogo}
W.~Rogosinski, ``\"{U}ber positive harmonische Entwicklungen and typisch-reelle Potenzreihen," Math. Z., 35, 93-121 (1932). Translated version by P.~Haldar, P.~Raman and A.~Zahed available on request.

\bibitem{yutin}
Y.~t.~Huang, J.~Y.~Liu, L.~Rodina and Y.~Wang,
``Carving out the Space of Open-String S-matrix,''
JHEP \textbf{04}, 195 (2021)
[arXiv:2008.02293 [hep-th]].
\bibitem{NS}
Z.~Nehari and B.~Schwarz, ``On the coefficients of univalent Laurent series,'' Proc. Amer. Math. Soc. {\bf 5} (1954), 212-217.

\bibitem{gribov}
V.~N.~Gribov, ``The theory of Complex Angular Momenta,'' Cambridge University Press, 2003. 

\bibitem{yutin2}
L.~Y.~Chiang, Y.~t.~Huang, W.~Li, L.~Rodina and H.~C.~Weng,
``Into the EFThedron and UV constraints from IR consistency,''
[arXiv:2105.02862 [hep-th]].

\bibitem{komatu}
Y.~Komatu, ``The coefficients of typically-real Laurent series,'' Kodai Math. Sem. Rep. 9 (1957), no. 1, 42--48.

\bibitem{Goodman} 
  A.~W.~Goodman,
  ``Functions typically real and meromorphic in the unit circle,''
  Annals of Math. 81(1956),92-105.
  
\bibitem{Wigner} 
  E.~P.~Wigner,
  ``On a class of analytic functions from the quantum theory of collisions ,''
  Annals of Math. vol. 53(1951) pp.36-67. 
 
 \bibitem{WignerEisenbud} 
  E.~P.~Wigner and L.~Eisenbud,
  ``Higher angular momentum and long-range interaction in resonance reactions ,''
  Phys.Rev.Lett., \textbf{72}, 1 (1947).
 
  \bibitem{Robertson} 
  M.~S.~Robertson,
  ``On coefficients of a typically real function,''
 Bull. Amer.Math. Soc.  vol. 41(1935) pp.565-572.
  
\bibitem{Markov1}
V.~A.~Markov,
 "On Functions of Least Deviation from Zero in a Given Interval,"(in Russian)
1892.
\bibitem{Markov2}
W.~Markoff, J.~Grossman ``\"{U}ber Polynome, die in einem gegebenen Intervalle mglichst wenig von Null abweichen,"
Math. Ann. 77, 213-258 (1916).
\bibitem{ST} 
  S.~S~Tiwari and V.~Vikramaditya, 
  to appear.
  
\bibitem{pionworks}
A.~L.~Guerrieri, J.~Penedones and P.~Vieira,
``Bootstrapping QCD Using Pion Scattering Amplitudes,''
Phys. Rev. Lett. \textbf{122}, no.24, 241604 (2019)
[arXiv:1810.12849 [hep-th]].\\
A.~Bose, P.~Haldar, A.~Sinha, P.~Sinha and S.~S.~Tiwari,
``Relative entropy in scattering and the S-matrix bootstrap,''
SciPost Phys. \textbf{9}, 081 (2020)
[arXiv:2006.12213 [hep-th]].\\
A.~Bose, A.~Sinha and S.~S.~Tiwari,
``Selection rules for the S-Matrix bootstrap,''
SciPost Phys. \textbf{10}, 122 (2021)
[arXiv:2011.07944 [hep-th]].
 
 \bibitem{CGHRS} 
 S.~D~Chowdhury, K.~Ghosh, P.~Haldar , P.~Raman and A.~Sinha
  to appear.


\bibitem{elbert}
A.~ Elbert and A.~ Laforgia, ``Upper bounds for the zeros of ultraspherical polynomials,''  J.
Approx. Theory 61 (1990) 88-97.

\bibitem{McKay:2021ejv}
J.~McKay and Y.~H.~He,
``Kashiwa Lectures on ''New Approaches to the Monster'',''
[arXiv:2106.01162 [math.HO]].

\bibitem{khare}
A.~Belton, D.~Guillot, A.~Khare and M.~Putinar,
``A panorama of positivity,'' arXiv:1812.05482 [math.CA].

\bibitem{pinkus}
 Allan Pinkus,
``Totally Positive Matrices,''
Cambridge Tracts in Mathematics, 2009, 
Cambridge University Press.

\bibitem{shohat}
J.~A.~Shohat and J.~D.~Tamarkin
``The problem of moments,"
Mathematical Surveys and Monographs
Volume: 1; 1943,
https://doi.org/http://dx.doi.org/10.1090/surv/001
Bulletein of American mathematical society.

\bibitem{schmudgen}
K.~Schm\"{u}dgen, 
``The Moment Problem,"
Graduate Texts Mathematics, 2017,
doi:10.1007/978-3-319-64546-9,
Springer.

\bibitem{smat3}
M.~F.~Paulos, J.~Penedones, J.~Toledo, B.~C.~van Rees and P.~Vieira,
``The S-matrix bootstrap. Part III: higher dimensional amplitudes,''
JHEP \textbf{12}, 040 (2019)
doi:10.1007/JHEP12(2019)040
[arXiv:1708.06765 [hep-th]].

\bibitem{AZ}
A.~Zahed,
``Positivity and Geometric Function Theory Constraints on Pion Scattering,''
[arXiv:2108.10355 [hep-th]]

\bibitem{nimayutin}
N.~Arkani-Hamed, T.~C.~Huang and Y.~t.~Huang,
``Scattering Amplitudes For All Masses and Spins,''
[arXiv:1709.04891 [hep-th]].
\bibitem{Kravchuk:2021kwe}
P.~Kravchuk, J.~Qiao and S.~Rychkov,
``Distributions in CFT II. Minkowski Space,''
[arXiv:2104.02090 [hep-th]].

\bibitem{GSZ}
R.~Gopakumar, A.~Sinha and A.~Zahed,
``Crossing Symmetric Dispersion Relations for Mellin Amplitudes,''
Phys. Rev. Lett. \textbf{126}, no.21, 211602 (2021)
[arXiv:2101.09017 [hep-th]].

\bibitem{kundu}
S.~Kundu,
``Swampland Conditions for Higher Derivative Couplings from CFT,''
[arXiv:2104.11238 [hep-th]].

\bibitem{rastelli2021}
S.~Caron-Huot, D.~Mazac, L.~Rastelli and D.~Simmons-Duffin,
``AdS Bulk Locality from Sharp CFT Bounds,''
[arXiv:2106.10274 [hep-th]].

 \bibitem{miguel}
M.~F.~Paulos,
``Dispersion relations and exact bounds on CFT correlators,''
[arXiv:2012.10454 [hep-th]].

\bibitem{anna}
A. ~ Tatarczak,
"Properties of Orthogonal polynomials and typically real functions related to the generalised Koebe function,"
Thesis,
Jagellonian university.


\bibitem{brickman}
L.~Brickman, T.~H.~MacGregor and D.~R.~Wilken,
``Convex hulls of some classical families of univalent function,''
Volume 156, May 1971,
Transaction of the American Mathematical Society.

\bibitem{green}
M.~B.~Green and C.~Wen,
``Superstring amplitudes, unitarily, and Hankel determinants of multiple zeta values,''
JHEP \textbf{11}, 079 (2019)
[arXiv:1908.08426 [hep-th]].


\bibitem{polyaszego}
G.~ P\'{o}lya and G. Szeg\"{o},
``Problems and Theorems in Analysis II,'' Springer-Verlag, New York, 1976.


\bibitem{powers}
V.~ Powers and B.~  Reznick,
``Polynomials that are positive on an interval,''
TRANSACTIONS OF THE AMERICAN MATHEMATICAL SOCIETY
 Volume 352, Number 10, 4677.

\bibitem{putinar}
N.~H.~A.~ Mai and V.~ Magron,
``On the complexity of Putinar-Vasilescu's Positivstellensatz,'' 
 arXiv:2104.11606v2.

\bibitem{massless}
A.~L.~Guerrieri, J.~Penedones and P.~Vieira,
``S-matrix bootstrap for effective field theories: massless pions,''
JHEP \textbf{06}, 088 (2021)
[arXiv:2011.02802 [hep-th]].

\bibitem{r4}
A.~Guerrieri, J.~Penedones and P.~Vieira,
``Where is String Theory?,''
[arXiv:2102.02847 [hep-th]].

\bibitem{bg}
R.~Blankenbecler and M.~L.~Goldberger,
``Behavior of scattering amplitudes at high energies, bound states, and resonances,''
Phys. Rev. \textbf{126}, 766-786 (1962)

\bibitem{Simon1}
B.~Simon,
``Coupling constant analyticity of the Anharmonic oscillator,''
Annals of Physics \textbf{58}, 76 (1970)

\bibitem{Simon2}
B.~Simon,
``The Anharmonic oscillator:a singular perturbation theory,''
Carg\'{e}se Lect.Phys \textbf{5}, 383-414 (1972)

\bibitem{BenderWu1}
C.~Bender,~T.~T~.Wu,
``Analytic structure of energy levels in a field theory model,''
Phys.Rev.Lett., \textbf{21}, 406(1968)

\bibitem{BenderWu2}
C.~Bender,~T.~T~.Wu,
``Anharmonic Oscillator,''
Phys.Rev.Lett., \textbf{184}, 1231 (1969)

\bibitem{Benderorzag}
C.M~Bender,~ S.~A~.Orszag,
``Advanced Mathematical Methods for Scientists and Engineers:
Asymptotic methods and perturbation theory,''
McGraw-Hill Book company

\bibitem{Wuomura}
T.~Y~Wu, T~.Ohmura,
``Quantum theory of scattering,''
Prentice-hall International series in physics, 1962

\bibitem{Jinmartin1}
Y.~S.~Jin,~A~.Martin,
``Number of subtractions in fixed transfer dispersion relations,"
Phys.Rev.Lett., \textbf{135}, 6B 1375-1377 (1964)

\bibitem{Jinmartin2}
Y.~S.~Jin,~A~.Martin,
``Connection between the asymptotic behaviour and the sign of the discontinuity in the one-dimensional dispersion relation,"
Phys.Rev.Lett., \textbf{135}, 6B  1369-1374(1964)

\bibitem{Mickens}
Ronald.~E.~Mickens,
``Mathematical properties of the vacuum polarisation function,''
Lectures in Mathematical Physics \textbf{2}, 343-347 (1978)
\end{thebibliography}
\end{document}